\documentclass[11pt,a4paper]{article}
\pdfoutput=1
\usepackage{jstyle}
\usepackage{amsfonts, amssymb}
\usepackage{graphicx, hyperref, color, verbatim}
\usepackage[dvipsnames]{xcolor}
\usepackage{float}
\usepackage{caption}
\usepackage{subcaption}
\usepackage{dsfont}

\usepackage{amsthm}

\theoremstyle{definition}
\newtheorem{exercise}{Exercise}[section]

\usepackage[most]{tcolorbox}

\tcolorboxenvironment{exercise}{
  colback=gray!10,
  colframe=gray!50,
  boxrule=0.5pt,
  arc=2mm,
  before skip=10pt,
  after skip=10pt
}

\usepackage{tikz}
\usetikzlibrary{shadings}

\makeatletter
\DeclareFontFamily{OMX}{MnSymbolE}{}
\DeclareSymbolFont{MnLargeSymbols}{OMX}{MnSymbolE}{m}{n}
\SetSymbolFont{MnLargeSymbols}{bold}{OMX}{MnSymbolE}{b}{n}
\DeclareFontShape{OMX}{MnSymbolE}{m}{n}{
    <-6>  MnSymbolE5
   <6-7>  MnSymbolE6
   <7-8>  MnSymbolE7
   <8-9>  MnSymbolE8
   <9-10> MnSymbolE9
  <10-12> MnSymbolE10
  <12->   MnSymbolE12
}{}
\DeclareFontShape{OMX}{MnSymbolE}{b}{n}{
    <-6>  MnSymbolE-Bold5
   <6-7>  MnSymbolE-Bold6
   <7-8>  MnSymbolE-Bold7
   <8-9>  MnSymbolE-Bold8
   <9-10> MnSymbolE-Bold9
  <10-12> MnSymbolE-Bold10
  <12->   MnSymbolE-Bold12
}{}

\let\llangle\@undefined
\let\rrangle\@undefined
\DeclareMathDelimiter{\llangle}{\mathopen}%
                     {MnLargeSymbols}{'164}{MnLargeSymbols}{'164}
\DeclareMathDelimiter{\rrangle}{\mathclose}%
                     {MnLargeSymbols}{'171}{MnLargeSymbols}{'171}
\makeatother

\title{Pre-Strings Lectures on Holographic Correlators and Analytic Bootstrap}

\author{Xinan Zhou}

\affiliation{Kavli Institute for Theoretical Sciences, University of Chinese Academy of Sciences, Beijing 100190, China}
\emailAdd{xinan.zhou@ucas.ac.cn}

\abstract{Over the past decade, the bootstrap strategy has transformed the computation of holographic correlators and revealed structures suggesting an emerging scattering amplitude program in AdS. These notes, which are an extended version of the five lectures delivered at the Pre-Strings 2026 School, give a pedagogical introduction and synthesis of these developments. We start with a quick reminder of the essentials of CFT and a brief review of AdS perturbation theory. We then demonstrate in detail  the bootstrap strategy for computing holographic correlators in the canonical example 4d $\mathcal{N}=4$ super Yang-Mills theory, at the level of four-point functions and in the regime dual to classical Type IIB supergravity. Several extensions, including higher-point functions, quantum corrections and stringy corrections, are discussed at varying levels of detail. We also explain how the bootstrap strategy can be extended to holographic CFTs with conformal defects, covering a diverse array of different setups.

}

\begin{document}
\maketitle
\tableofcontents

\newpage

\section{Introduction}\label{Sec:introduction}

In flat space scattering amplitudes are important physical quantities  connecting theory and experiments. In principle, as one learns in an introductory QFT course, these amplitudes can be computed as a sum of Feynman diagrams. However, after simplifying the expression one often finds the final answer to be much simpler than the naive sum. This is most evident in the Parke-Taylor formula \cite{Parke:1986gb} for the tree-level maximally helicity violating amplitudes of gluons
\begin{equation}
\mathcal{A}(1^+\ldots i^-\ldots j^-\ldots n^+)\sim \frac{\langle ij\rangle^4}{\langle 12\rangle \langle 23\rangle\ldots \langle n 1\rangle}\;,
\end{equation}
written in the spinor-helicity variables. Such a single-term result, valid for any number of gluons, would require summing over millions of diagrams to reproduce already at relatively small values of $n$. However, the existence of remarkable hidden simplicity is not the only lesson that the flat-space amplitudes can teach us. They also provide us with deep insights into the nature of gravity and QFT. A good example is the double copy relation \cite{Bern:2010ue} which allows one to construct graviton  amplitudes from gluon amplitudes in a ``product'' form. For example, the tree-level amplitude of four gluons can be written as 
\begin{equation}
\mathcal{A}_{\rm gluon}=\frac{c_s n_s}{s}+\frac{c_t n_t}{t}+\frac{c_u n_u}{u}\;,
\end{equation}
where $c_s$, $c_t$, $c_u$ are color structures satisfying the Jacobi identity $c_s+c_t+c_u=0$ and $n_s$, $n_t$, $n_u$ are kinematic factors satisfying the same algebraic relation. The graviton amplitude is obtained by simply replacing color factors $c_{s,t,u}$ by kinematic factors $n_{s,t,u}$
\begin{equation}
\mathcal{A}_{\rm graviton}=\frac{n_s^2}{s}+\frac{n_t^2}{t}+\frac{n_u^2}{u}\;.
\end{equation}
Such a relation is not a coincidence at four points and is valid more generally. It provides a tantalizing new perspective that gravity may be viewed as the ``square'' of Yang-Mills theory. These are some of the many remarkable properties of the flat-space amplitudes, the study of which has led to an extremely fruitful program. On the other hand, it is natural to ask the following: are the lessons from flat-space amplitudes truly robust? Do these properties also extend more generally to curved spacetime?

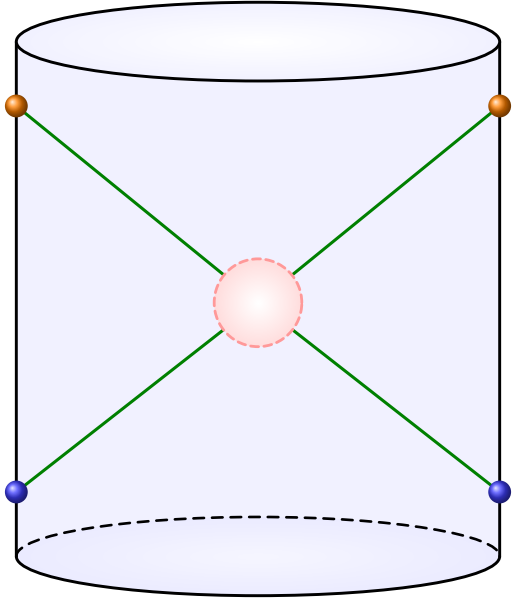
\begin{figure}[t]
\centering
\begin{tikzpicture}[
    x=1cm,
    y=1cm,
    line cap=round,
    line join=round]
\def\R{3.2}
\def\E{0.52}
\def\H{6.8}
\coordinate (I) at (0,3.35);
\coordinate (TL) at (-\R,5.95);
\coordinate (TR) at ( \R,5.95);
\coordinate (BL) at (-\R,0.85);
\coordinate (BR) at ( \R,0.85);
\shade[
    left color=blue!15,
    middle color=white,
    right color=blue!15,
    shading angle=0,
    opacity=0.35,
    draw=none
]
(-\R,\H)
--
(-\R,0)
arc[
    start angle=180,
    end angle=360,
    x radius=\R,
    y radius=\E
]
--
(\R,\H)
--
cycle;
\shade[
    inner color=white,
    outer color=blue!12,
    opacity=0.35,
    draw=none
]
(0,0) ellipse [
    x radius=\R,
    y radius=\E
];
\draw[
    green!50!black,
    line width=1.15pt]
(TL) -- (I);
\draw[green!50!black,
    line width=1.15pt]
(TR) -- (I);
\draw[
    green!50!black,
    line width=1.15pt
]
(BL) -- (I);
\draw[
    green!50!black,
    line width=1.15pt
]
(BR) -- (I);
\draw[black,line width=1.1pt]
(-\R,0) -- (-\R,\H);
\draw[black,line width=1.1pt]
(\R,0) -- (\R,\H);
\draw[
    black,
    line width=1pt,
    dashed,
    dash pattern=on 4pt off 3pt]
(-\R,0)
arc[
    start angle=180,
    end angle=0,
    x radius=\R,
    y radius=\E];
\shade[
    inner color=white,
    outer color=pink!55,
    draw=none]
(I) circle (0.58);
\draw[
    pink!80!red,
    line width=1.05pt,
    dashed,
    dash pattern=on 3pt off 2pt]
(I) circle (0.58);
\draw[
    black,
    line width=1.1pt]
(\R,0)
arc[
    start angle=0,
    end angle=-180,
    x radius=\R,
    y radius=\E];
\fill[
    white,
    draw=none]
(0,\H) ellipse [
    x radius=\R,
    y radius=\E];
\shade[
    inner color=white,
    outer color=blue!12,
    opacity=0.45,
    draw=none]
(0,\H) ellipse [
    x radius=\R,
    y radius=\E];
\draw[
    black,
    line width=1.1pt]
(0,\H) ellipse [
    x radius=\R,
    y radius=\E];
\shade[ball color=orange]
(TL) circle (4.3pt);
\shade[ball color=orange]
(TR) circle (4.3pt);
\shade[ball color=blue!75]
(BL) circle (4.3pt);
\shade[ball color=blue!75]
(BR) circle (4.3pt);
\end{tikzpicture}
\caption{Correlation functions on the boundary of AdS as scattering amplitudes in the bulk. The cylinder represents the bulk of AdS and operators are inserted on its boundary. Time runs upwards vertically. States were prepared in the past, scattered in the middle, and measured in the future.}
\label{fig:adscylinder}
\end{figure}

Clearly, studying scattering in general curved spacetime backgrounds is enormously complicated, if possible at all. The best starting point is spacetime which has maximal symmetry, just as in flat space. Here we will focus on the anti de Sitter (AdS) space. The AdS space is a background with constant negative curvature and has a time-like boundary.\footnote{The de Sitter (dS) space is also maximally symmetric. However, it has a space-like future boundary and therefore corresponds to a different observable problem.} The time-like boundary is especially useful because it allows for insertions which can be used to prepare and detect states. The qualitative picture is depicted in Figure \ref{fig:adscylinder}. The insertions on the boundary source wave-packets which travel into the bulk of AdS. After scattering, the outgoing wave-packets are measured again by the boundary insertions. In fact, this intuitive picture can be turned into a sharper statement: the correlation functions of the operator insertions at the boundary of AdS {\it are} the amplitudes in AdS. Using this definition, we can concretely explore the answer to our question. Even more ambitiously, one may try to establish a scattering amplitude program in AdS similar to the one in flat space. 

Aside from this amplitude motivation, the study of these holographic correlators of course also has the original incentives from the AdS/CFT correspondence \cite{Maldacena:1997re,Witten:1998qj,Gubser:1998bc}. The holographic correlators are the basic observables which one can use to explore and exploit the correspondence. In particular, in the regime where the strongly coupled boundary CFT is described as a weakly coupled gravitational theory in the bulk, one can in principle compute these correlators using curved space perturbation theory and obtain exact analytic results at strong coupling. Moreover, the AdS/CFT correspondence can also be used as an approach to study Quantum Gravity in a box. These holographic correlators are then the cleanest gauge invariant observables for such studies.

However, the computation of holographic correlators is quite challenging. One could try to use the same diagrammatic method as in flat space. However, we already know it might not be the smartest idea due to the proliferation of diagrams. This is further compounded by the curved nature of the background spacetime -- now each individual diagram is more complicated than in flat space. Therefore, new efficient techniques are needed to handle the computation of holographic correlators, and this is precisely what we will talk about in these lectures. On the one hand, we will introduce useful kinematic frameworks, such as the Mellin space formalism \cite{Mack:2009mi,Penedones:2010ue}, that manifest the scattering amplitude nature of holographic correlators and simplify their analytic structure. On the other hand, which is more important, we will try to understand the underlying organizing principles of the holographic correlators. It turns out that instead of following precisely the diagrammatic prescription, the most efficient way is to ``bootstrap'' the holographic correlators directly using symmetries and consistency conditions \cite{Rastelli:2016nze,Rastelli:2017udc}, without reference to the details of the effective Lagrangian. This is similar to the strategy of the on-shell scattering amplitude program in flat space where one focuses only on the on-shell observables, namely the amplitudes. We will explain in these lectures how this can be systematically done for the holographic correlators, as well as provide the necessary background.

So far, our discussion has only focused on local operators in CFT which correspond to particle-like excitations in AdS. Their correlation functions are interpreted as scattering amplitudes in AdS. But there are also nonlocal objects such as Wilson loops, surface defects, boundaries, interfaces, etc, which we will all collectively refer to as defects. These are dual to extended objects in AdS. We can consider correlation functions of local operators in the presence of such defects and they correspond to scattering form factors in AdS where particles scatter off extended objects. We will see that the same bootstrap techniques for correlators of local operators can be extended to these defect systems as well. In fact, the category of defects is quite large and can cover a wide range of different systems. In addition to the conventional defects which have $p\geq 1$ dimensions, it also makes sense to consider conformal defects with $0$ and $-1$ dimensions. Such unconventional defects can be realized, for example, as heavy local operators and putting the theory on real projective space respectively. The bootstrap techniques which we will develop in these lectures will also allow us to perform efficient holographic calculations in these systems as well.
 
The rest of these lecture notes is organized as follows. In Section \ref{Sec:basicsofCFT}, we review the basics of CFTs in higher dimensions which will be needed in these lectures. In Section \ref{Sec:AdSpertth}, we discuss the standard perturbation theory in AdS with a focus on tree-level diagrams and also introduce the Mellin space formalism. The preparations in this section will help with developing the necessary intuitions for holographic correlators. The discussion of the bootstrap strategy starts in Section \ref{Sec:bootstrap}, where we use 4d $\mathcal{N}=4$ SYM, which is dual to Type IIB superstring theory in AdS$_5\times$S$^5$, as the prototypical example. We show how holographic correlators in the supergravity limit can be computed by using only symmetries and consistency conditions. Then in Section \ref{Sec:extensions},  we discuss various extensions of the bootstrap strategy, including correlators at higher points, quantum loop corrections, and stringy corrections. In Section \ref{Sec:defects}, we switch to the discussion of holographic conformal defects. We will see how various bootstrap techniques developed for defect-free CFTs can be extended to correlators of local operators in the presence of defects. Finally, we conclude these lectures in Section \ref{Sec:outlook} with a brief outlook for open questions.

\vspace{1cm}

\noindent{\bf Useful references:} There already exist a number of great reviews which are related to the topics discussed in these lecture notes. In particular, the lecture notes by Rychkov \cite{Rychkov:2016iqz} and Simmons-Duffin \cite{Simmons-Duffin:2016gjk} give an excellent introduction to CFTs in higher dimensions. For a pedagogical introduction to AdS/CFT, the lectures by Penedones are a very good source \cite{Penedones:2016voo}, and see also the classic review by D'Hoker and Freedman 
\cite{DHoker:2002nbb} for more details about the traditional method for computing holographic correlators. The bootstrap methods for computing holographic correlators were pedagogically discussed in \cite{Bissi:2022mrs} for the defect-free case, where other more general developments in the analytic conformal bootstrap were also reviewed. Some parts of these lectures are based on these reviews. Other useful references include \cite{Aharony:1999ti,Poland:2018epd,Chester:2019wfx,Heslop:2022xgp,Rychkov:2023wsd}.

\section{Basics of CFT and AdS}\label{Sec:basicsofCFT}\label{Sec:CFTbasics}
In this section, we recall the CFT essentials needed for these lectures. The brief review emphasizes in particular the use of the embedding space formalism which linearizes the conformal symmetry. For a more detailed and pedagogical introduction to CFT in higher dimensions, we refer to \cite{Rychkov:2016iqz,Simmons-Duffin:2016gjk}.

\subsection{Conformal group and embedding space}
Let us start by recalling the definition of conformal transformations. In most of these lectures, we will work with the Euclidean space where the space is $\mathbb{R}^d$. Conformal transformations are all the coordinate transformations that preserve the form of the metric up to a local scaling factor
\begin{equation}
\delta_{\mu\nu}\frac{d x'^{\mu}}{dx^\alpha}\frac{d x'^{\nu}}{dx^\beta}=\Omega^2(x)\delta_{\alpha\beta}\;.
\end{equation}
Geometrically, this means that the angle between any two lines at the intersection point is preserved. This is an extension of the usual Poincar\'e symmetry consisting of translations and rotations
\begin{equation}
x^\mu\to x^\mu+a^\mu\;,\quad \quad x^\mu \to R^\mu{}_\nu x^\nu\;,
\end{equation}
of which the generators we denote as $P_\mu$ and $M_{\mu\nu}$. These transformations all have $\Omega(x)=1$. The new transformations are the dilatation 
\begin{equation}
x^\mu\to \Lambda x^\mu\;,
\end{equation}
generated by the operator $D$, and the special conformal transformation 
\begin{equation}
x^\mu\to \frac{x^\mu-b^\mu x^2}{1-2b\cdot x+b^2x^2}\;,
\end{equation}
generated by the operator $K_\mu$. Both transformations have factors of $\Omega(x)$ which are not unit. Here we have written down the finite transformations but it is straightforward to obtain the infinitesimal versions. Let us also unpack the special conformal transformation by noting it corresponds to conjugating translation with the conformal inversion
\begin{equation}
x^\mu \xrightarrow{I} \frac{x^\mu}{x^2} \xrightarrow{\text{translation by $-b^\mu$}} \frac{x^\mu}{x^2}-b^\mu \xrightarrow{I}\frac{x^\mu-b^\mu x^2}{1-2b\cdot x+b^2x^2}\;,
\end{equation}
where the conformal inversion $I$ acts as $x^\mu\to x^\mu/x^2$. 

It is straightforward to work out the commutation relations of these conformal transformation generators. Moreover, by taking the combinations 
\begin{equation}
\begin{split}
 {}&L_{\mu\nu}=M_{\mu\nu}\;,\quad L_{-1,0}=D\;,\\
 {}&L_{0,\mu}=\frac{1}{2}(P_\mu+K_\mu)\;,\quad L_{-1,\mu}=\frac{1}{2}(P_\mu-K_\mu)\;,
\end{split}
\end{equation}
it is not too difficult to recognize that the conformal algebra is isomorphic to $SO(d+1,1)$ where we denote the generators by $L_{AB}$. Here the indices take values $A,B=-1,0,1,\ldots, d$ and the $-1$ direction is time-like. Note that $SO(d+1,1)$ is just the rotation group in $\mathbb{R}^{d+1,1}$ which acts linearly. This motivates us to consider working with this embedding space to simplify the nonlinearity of the original conformal symmetry. 

\begin{figure}[htbp]
    \centering
        \includegraphics[width=0.45\textwidth]{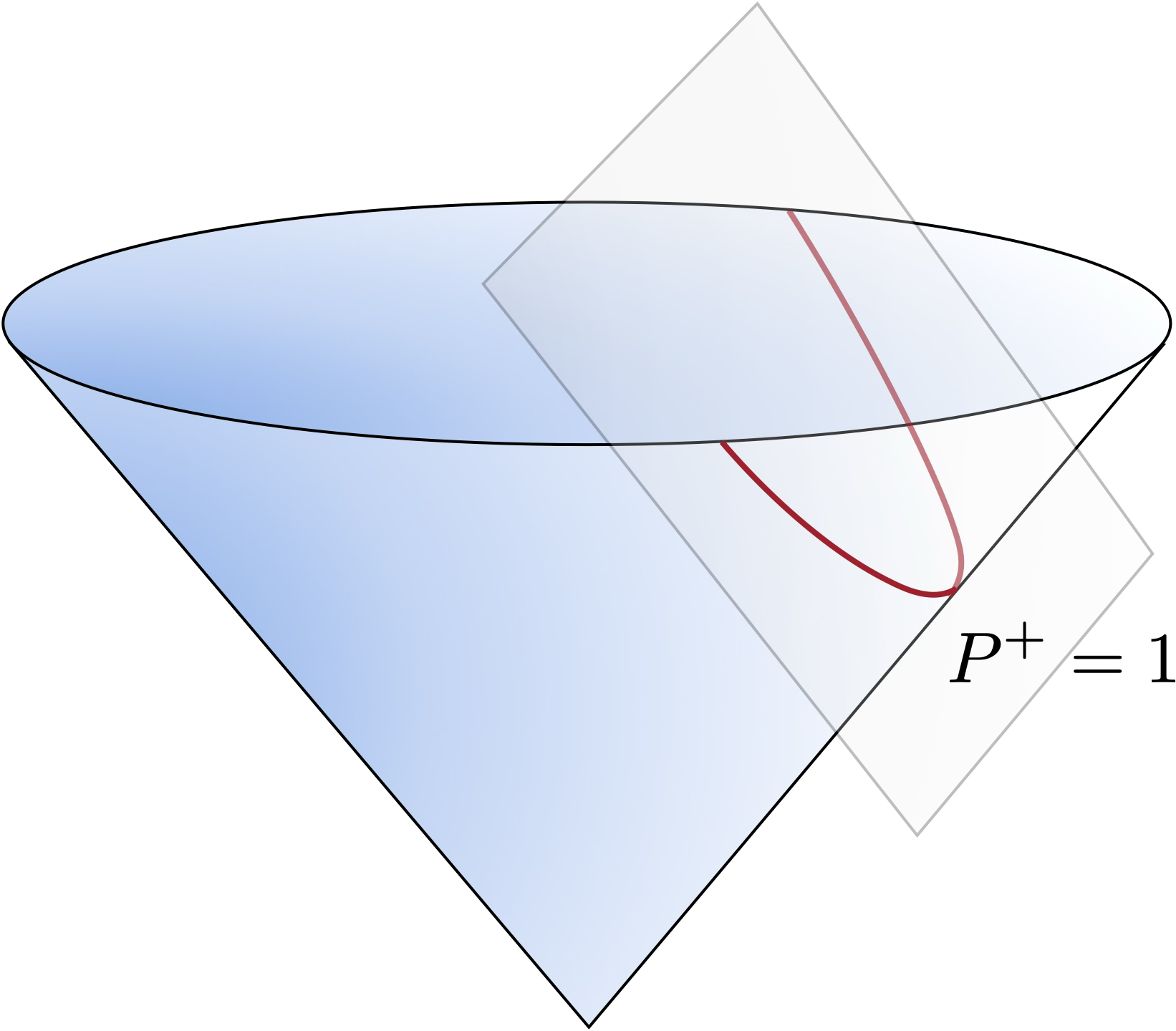}
    \caption{Embedding space. Each light ray on the null cone corresponds to a point $x$ in the physical space. We can fix the rescaling equivalence of the light ray by choosing the slice $P^+=1$ which corresponds to the red curve.}
    \label{fig:embeddingspace}
\end{figure}

For this to work, we first need to match the number of degrees of freedom. This can be achieved by identifying a point $x^\mu\in \mathbb{R}^d$ in the physical space with a null ray in the embedding space (see Figure \ref{fig:embeddingspace}). More precisely, for a vector $P^A\in\mathbb{R}^{d+1,1}$ we need to impose the null constraint
\begin{equation}\label{P2eq0}
P^A P_A=0\;,
\end{equation}
and demand equivalence under rescaling
\begin{equation}
P\sim \lambda P\;,\quad \lambda\neq 0\;.
\end{equation}
Each condition gets rid of one real degree of freedom, therefore giving the correct counting. To further get an explicit embedding, we can choose a gauge for the rescaling freedom by requiring 
\begin{equation}\label{Ppluseq1gauge}
P^+\equiv P^{-1}+P^0=1\;.
\end{equation}
Then we can relate $x^\mu$ to $P^A$ via the last $d$ components
\begin{equation}
P^A=\left(\frac{1+x^2}{2},\frac{1-x^2}{2},x^\mu\right)\;.
\end{equation}
The conformal transformation acts linearly on the embedding space vector as a rotation
\begin{equation}
P\to gP\;,\quad g\in SO(d+1,1)\;.
\end{equation}
However, in general the new vector no longer obeys the gauge choice (\ref{Ppluseq1gauge}). In order to compare, we must restore the gauge by rescaling
\begin{equation}
P\to gP\to gP/(gP)^+\;.
\end{equation}
Then comparing with $x^\mu$ components allows us to read off the conformal transformation in the physical space. 

To see this prescription produces correctly the conformal transformation, let us consider an example where the rotation $g$ is given by the following matrix
\begin{equation}
g^A{}_B=\left(\begin{array}{ccc}\frac{1+\Lambda^2}{2\Lambda} & \frac{1-\Lambda^2}{2\Lambda}& 0 \\\frac{1-\lambda^2}{2\Lambda} & \frac{1+\Lambda^2}{2\Lambda}  & 0 \\ 0 & 0 & 1\end{array}\right)\;.
\end{equation}
Acting it on the embedding space vector, we get
\begin{equation}
P\to gP=\left(\frac{\Lambda^{-1}+\Lambda x^2}{2},\frac{\Lambda^{-1}-\Lambda x^2}{2}, x^\mu\right)\;.
\end{equation}
This breaks the gauge because $(gP)^+=\Lambda^{-1}$. We need to rescale by $\Lambda$ to restore the gauge, and this gives
\begin{equation}
gP\to \Lambda gP= \left(\frac{1+\Lambda^2 x^2}{2},\frac{1-\Lambda^2 x^2}{2}, \Lambda x^\mu\right)\;.
\end{equation}
It is easy to see that this rotation in embedding space corresponds to the dilatation transformation $x^\mu \to \Lambda x^\mu$.
\begin{exercise}
Find the embedding space rotation matrices $g$ that produce all the other conformal transformations.  
\end{exercise}

\begin{figure}[htbp]
    \centering
        \includegraphics[width=0.5\textwidth]{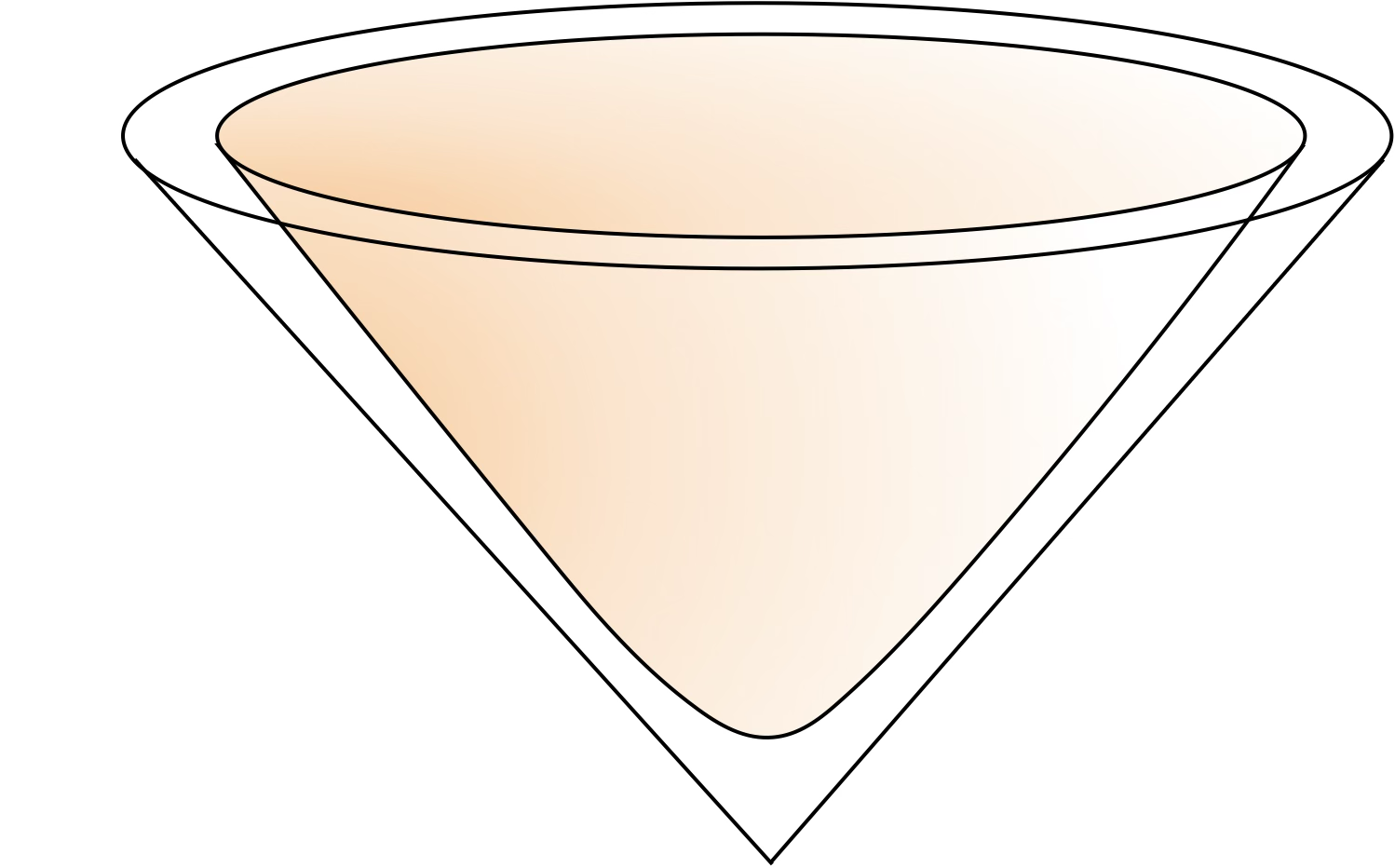}
    \caption{AdS space in embedding space. The AdS space asymptotes to the null cone, showing the boundary of AdS$_{d+1}$ is $\mathbb{R}^d$.}
    \label{fig:AdSinembedding}
\end{figure}

Note that in (\ref{P2eq0}) we have singled out the lightcone in the embedding space. We can also consider the surface 
\begin{equation}\label{X2eqm1}
X^A X_A=-R^2\;,
\end{equation}
which is also rotation invariant in embedding space. As is illustrated in Figure \ref{fig:AdSinembedding}, the time-like vector $X^A$ corresponds to a point in the Euclidean AdS space with radius $R$ (which later we will often set to 1). More precisely, we can parameterize $X$ as
\begin{equation}
X^A=R\left(\frac{1+z_0^2+\vec{z}^2}{2z_0},\frac{1-z_0^2-\vec{z}^2}{2z_0},\frac{\vec{z}}{z_0}\right)\;,
\end{equation}
where $\vec{z}=z^\mu$ and $z_0>0$. This corresponds to the Poincar\'e coordinates with the metric
\begin{equation}
ds^2=R^2\frac{dz_0^2+d\vec{z}^2}{z_0^2}\;.
\end{equation}
As we mentioned, the boundary of  AdS$_{d+1}$  is $\mathbb{R}^d$ and we can easily see it here. Note that in the limit $z_0\to 0$, we have
\begin{equation}
\lim_{z_0\to0}z_0 X^A\to P^A
\end{equation}
 We also note that in both (\ref{P2eq0}) and (\ref{X2eqm1}), the fixed locus is invariant under the same rotation group in $d+2$ dimensions. Therefore, conformal symmetry in $\mathbb{R}^d$ corresponds to isometry of AdS$_{d+1}$. 
 
The embedding space is not only useful for representing the full conformal symmetry group, but also convenient for characterizing conformal symmetry breaking by inserting defects. For example, we can consider a $p$-dimensional conformal defect occupying subspace $\mathbb{R}^p$ inside $\mathbb{R}^d$. In embedding space, this corresponds to defining a projector in $\mathbb{R}^{d+1,1}$ that separate into $\mathbb{R}^{p+1,1}$ and $\mathbb{R}^{d-p}$. The first factor $\mathbb{R}^{p+1,1}$ corresponds to the embedding space $\mathbb{R}^p$ and $\mathbb{R}^{d-p}$ is the space transverse to the defect. The preserved conformal symmetry is therefore $SO(p+1,1)\times SO(d-p)$ which is the rotation subgroup leaving these two subspace factors invariant in the embedding space. We will say more about defects in the last part of these lectures.

\subsection{Correlation functions}\label{subsec:corrfun}
In a CFT, local operators are divided into primary operators and descendant operators. When inserted at the origin, primary operators are defined by the property that they are annihilated by the generators of the special conformal transformations
\begin{equation}
[K_\mu, \mathcal{O}(0)]=0\;.
\end{equation}
Moreover, they are also eigenvectors of the dilatation generator and transform in irreducible representations of $SO(d)$
\begin{equation}
[D, \mathcal{O}(0)]=\Delta \mathcal{O}(0)\;,\quad [M_{\mu\nu}, \mathcal{O}_I(0)]=\left[M_{\mu\nu}\right]^J_I \mathcal{O}_J(0)\;.
\end{equation}
The descendant operators are obtained from the primary operators by repeated actions of $P_\mu$. Operators inserted at origin can then be moved away to an arbitrary point by using the translation operators $e^{x\cdot P}$. 

The correlators of primary operators are constrained by conformal symmetry to satisfy the Ward identity
\begin{equation}
\langle \mathcal{O}_1(x_1)\ldots \mathcal{O}_n(x_n)\rangle=\Omega(x'_1)^{\Delta_1}\ldots \Omega(x'_n)^{\Delta_n}\langle \mathcal{O}_1(x'_1)\ldots \mathcal{O}_n(x'_n)\rangle\;.
\end{equation}
where we recall a conformal transformation locally looks like a rescaling and a rotation
\begin{equation}
\frac{\partial x'^\mu}{\partial x^\nu}=\Omega(x') R^\mu{}_\nu(x')\;.
\end{equation}
The Ward identity therefore imposes nontrivial constraints on the form of the conformal correlators. 

To see the consequence of conformal symmetry on correlators, it is most convenient to use the embedding space. For simplicity, we will only focus on scalar operators. The primary operators admit a particularly simple characterization in embedding space. They are defined as functions on the null ray with a specific scaling behavior
\begin{equation}\label{defOembed}
\mathcal{O}(\lambda P)=\lambda^{-\Delta}\mathcal{O}(P)\;,\quad \mathcal{O}(P)\big|_{P^+=1}=\mathcal{O}(x)\;.
\end{equation}
Here it is more convenient to view it as we are extending the definition of the operator $\mathcal{O}$ from the physical space slice $P^+=1$ to the whole null cone.
\begin{exercise}
Prove this definition gives rise to the operator transformation
\begin{equation}
U_g\mathcal{O}(x)U_g^{-1}=\Omega(x')^\Delta \mathcal{O}(x')\;,
\end{equation}
in the Ward identity, where $U_g$ is the unitary operator corresponding to the conformal transformation $g$. 
\end{exercise}

Writing down possible structures of correlators in embedding space now becomes very straightforward thanks to the linearized symmetry. Correlators are constrained by two conditions
\begin{itemize}
\item Conformal invariance: this translates to rotation invariance in embedding space. Therefore, objects should be constructed from inner products of embedding space vectors. 
\item Scaling property: We can scale each operator independently $\mathcal{O}_i(\lambda_i P_i)=\lambda^{-\Delta_i}\mathcal{O}_i(P_i)$. The correlator must have the correct scaling behavior $\langle \mathcal{O}_1(\lambda_1 P_1)\ldots \mathcal{O}_n(\lambda_n P_n) \rangle=\lambda_1^{-\Delta_1}\ldots \lambda_n^{-\Delta_n}\langle \mathcal{O}_1( P_1)\ldots \mathcal{O}_n(P_n) \rangle$. 
\end{itemize}
It is not difficult to show that these conditions determine two-point functions uniquely
\begin{equation}
\langle \mathcal{O}_{\Delta_1}(P_1)\mathcal{O}_{\Delta_2}(P_2)\rangle=\frac{\delta_{\Delta_1\Delta_2}}{(-2P_1\cdot P_2)^{\Delta_1}}\;.
\end{equation}
Here $-2P_i\cdot P_j=x_{ij}^2$. The independent scalings of the two operators require that the conformal dimensions $\Delta_1$ and $\Delta_2$ must be equal. We have also set the coefficient to be one as a normalization for the operators. Similarly, three-point functions are fixed up to overall coefficients
\begin{equation}
\langle \mathcal{O}_{\Delta_1}(P_1)\mathcal{O}_{\Delta_2}(P_2)\mathcal{O}_{\Delta_3}(P_3)\rangle=\frac{C_{123}}{(-2P_1\cdot P_2)^{\frac{\Delta_1+\Delta_2-\Delta_3}{2}}(-2P_1\cdot P_3)^{\frac{\Delta_1+\Delta_3-\Delta_2}{2}}(-2P_2\cdot P_3)^{\frac{\Delta_2+\Delta_3-\Delta_1}{2}}}\;.
\end{equation}
Because we have already fixed the normalizations in two-point functions, the coefficient $C_{123}$ can no longer be eliminated and therefore contains physical information.

Going beyond three-point functions, the form of correlators can no longer be fully determined by conformal symmetry. For example, for four points we can construct two independent conformal cross ratios 
\begin{equation}
U=\frac{(-2P_1\cdot P_2)(-2P_3\cdot P_4)}{(-2P_1\cdot P_3)(-2P_2\cdot P_4)}\;,\quad V=\frac{(-2P_1\cdot P_4)(-2P_2\cdot P_3)}{(-2P_1\cdot P_3)(-2P_2\cdot P_4)}\;.
\end{equation}
It is straightforward to check that these combinations are invariant under both conditions. Therefore, we can only determine the four-point function up to an arbitrary function of $U$ and $V$. 

We can also consider correlators with $n>4$ operators. If we assume the cross ratios also take the form $\prod_{i<j} (-2 P_i\cdot P_j)^{n_{ij}}$, then imposing rescaling invariance gives $\frac{n(n-3)}{2}$ solutions. However, we note that when $n>d+2$ we also have the relation $\det (-2P_i\cdot P_j)=0$ where the matrix inside the determinant is $n$ by $n$. Therefore, not all these cross ratios are independent. To count the cross ratios more carefully as a function of $n$, we have the following exercise.
\begin{exercise}\label{countcrossratios}
Show that the number of independent conformal cross ratios for $n$ points is given by
\begin{equation}
\#\text{ (independent cross ratios)}=
\begin{cases}
nd-\dfrac{(d+2)(d+1)}{2}\;, & n>d+2\;,\\[0.5em]
\dfrac{n(n-3)}{2}\;, & n\le d+2\;.
\end{cases}
\end{equation}

\vspace{0.3cm}

{\bf Hint:} one can use generators of the conformal symmetry group to move points to specific locations. It is also important to consider the stability group which leaves the configuration unchanged.  
\end{exercise}

\subsection{Operator product expansion and conformal blocks}\label{subsec:OPEandconfblocks}
 An important concept in CFT is the Operator Product Expansion (OPE). The OPE tells us that we can replace the product of two local operators with an infinite sum of local operators
\begin{equation}\label{OPE122}
\mathcal{O}_i(x_1)\mathcal{O}_j(x_2)=\sum_k C_{ijk}\mathbb{D}(x_{12},\partial_2)\mathcal{O}_k(x_2)\;,
\end{equation}
where $C_{ijk}$ are the three-point function coefficients and $\mathbb{D}(x_{12},\partial_2)$ are differential operators that are completely determined by conformal symmetry (see Exercise \ref{ex:OPEdiffop} below). The OPE (\ref{OPE122}) is not asymptotic, but has a finite radius of convergence in CFT. It is convergent as long as we can draw a sphere to include both $\mathcal{O}_i$, $\mathcal{O}_j$ from the left side and the new operators $\mathcal{O}_k$ from the right side (which does not even need to be expanded around $x_2$), with no other operators inserted inside the sphere. These statements can be proven using the radial quantization. 
\begin{exercise} \label{ex:OPEdiffop} Prove for a scalar internal operator, the differential operator can be expanded as
\begin{equation}
\mathbb{D}(x,\partial)= |x|^{\Delta_k-\Delta_i-\Delta_j}(1+\alpha_1 x^\mu\partial_\mu+\alpha_2 x^\mu x^\nu \partial_\mu\partial_\nu+\alpha_3 x^2\partial^2+\ldots)\;.
\end{equation}
By computing the three-point function $\langle\mathcal{O}_i(x_1)\mathcal{O}_j(x_2)\mathcal{O}_k(x_3)\rangle$ using OPE and comparing the answer with the exact three-point function, show the first few expansion coefficients are
\begin{equation}\label{OPEdiffexpcoeff}
\alpha_1=\frac{1}{2}\;,\quad \alpha_2=\frac{\Delta+2}{8(\Delta+1)}\;,\quad \alpha_3=-\frac{\Delta}{16(\Delta+1)(\Delta-\frac{d-2}{2})}\;.
\end{equation}
Here we have assumed $\Delta_i=\Delta_j$ and both $\mathcal{O}_i$ and $\mathcal{O}_j$ are scalar operators.

\vspace{0.3cm}
\noindent{\bf Additional question:} the coefficient $\alpha_3$ in (\ref{OPEdiffexpcoeff}) has a pole at $\Delta=(d-2)/2$. Is this a problem?
\end{exercise}
The OPE is very useful because it allows us to reduce $n$-point functions to $(n-1)$-point functions. This reduction process continues all the way until we reach three-point functions which are fixed by symmetry. Note that the only parameters in the OPE are the conformal dimensions $\Delta_i$ and the three-point function coefficients $C_{ijk}$. Therefore, if we know all these parameters we can in principle compute all $n$-point correlators. This is why $\{\Delta_i,C_{ijk}\}$ are often called the CFT data -- these data define the CFT. However, in practice we do not have access to all the CFT data. Even if we do, the infinite resummations over operators are still very difficult.

The OPE leads to another important notion which is the conformal block. For definiteness, let us consider four-point functions of identical scalar operators $\langle \mathcal{O}(x_1) \mathcal{O}(x_2) \mathcal{O}(x_3) \mathcal{O}(x_4) \rangle$. We can apply the OPE in the 12 channel to reduce the correlator to three-point functions. The resulting three-point functions with 3 and 4 can only be nonzero when the operators in the 12 OPE are also present in the 34 OPE (here it is trivially true because the external operators are identical). So we can view this way of computing the four-point function as summing over operator exchanges in the 12-34 channel. The total contributions from the conformal family of an intermediate operator $\mathcal{O}_{\Delta,\ell}$, i.e., the primary and all its descendants, are defined to be the conformal block
\begin{equation}\label{defg}
\frac{C_{\mathcal{O}\mathcal{O}\mathcal{O}_{\Delta,\ell}}}{x_{12}^{2\Delta_{\mathcal{O}}}x_{34}^{2\Delta_{\mathcal{O}}}}g_{\Delta,\ell}(U,V)=\mathbb{D}(x_{12},\partial_2)\langle \mathcal{O}_{\Delta,\ell}(x_2) \mathcal{O}(x_3) \mathcal{O}(x_4)\rangle\;. 
\end{equation}
Here we have extracted some factor of $x_{ij}^2$ to make manifest the fact that the conformal block is a function of the conformal cross ratios. The three-point function coefficients will also cancel out in both sides so that the conformal blocks are unit normalized. The sum over descendants are encoded in the subleading terms in the differential operator $\mathbb{D}(x_{12},\partial_2)$. In this way, the four-point function can be decomposed as 
\begin{equation}
\langle \mathcal{O}(x_1) \mathcal{O}(x_2) \mathcal{O}(x_3) \mathcal{O}(x_4) \rangle=\frac{1}{x_{12}^{2\Delta_{\mathcal{O}}}x_{34}^{2\Delta_{\mathcal{O}}}}\sum_{\mathcal{O}_{\Delta,\ell}} C_{\mathcal{O}\mathcal{O}\mathcal{O}_{\Delta,\ell}}^2 g_{\Delta,\ell}(U,V)\;.
\end{equation}
The way in which we have applied the OPE is not the only one. We can also use the OPE in the 14 channel and it gives a different but equivalent decomposition. This leads to the crossing equation
\begin{equation}
\sum_{\mathcal{O}_{\Delta,\ell}} C_{\mathcal{O}\mathcal{O}\mathcal{O}_{\Delta,\ell}}^2 g_{\Delta,\ell}(U,V)=\left(\frac{U}{V}\right)^{\Delta_{\mathcal{O}}}\sum_{\mathcal{O}_{\Delta,\ell}} C_{\mathcal{O}\mathcal{O}\mathcal{O}_{\Delta,\ell}}^2 g_{\Delta,\ell}(V,U)\;,
\end{equation}
which places nontrivial constraints on the CFT data. The numerical conformal bootstrap exploits this condition together with unitarity, and obtains rigorous bounds for the CFT data, see \cite{Rychkov:2016iqz,Simmons-Duffin:2016gjk} for details. 

The conformal blocks in principle can be computed from (\ref{defg}) using the explicit form of $\mathbb{D}(x_{12},\partial_2)$. In practice, such a computation is quite complicated and the best way to obtain the conformal blocks is by solving the conformal Casimir equation \cite{Dolan:2003hv}. This is essentially exploiting the fact that all the exchanged operators, primary and descendant, have the same eigenvalue with respect to the quadratic conformal Casimir. From this, one obtains a second order partial differential equation for the conformal block which can then be solved with appropriate boundary conditions. Later in Section \ref{Subsec:confblockdecom}, we will explain how this works from a closely related perspective when we talk about exchange diagrams in AdS.  For the moment, let us mention that the conformal blocks can be written down in a closed form in even spacetime dimensions as ${}_2F_1$ hypergeometric functions (in odd dimensions it is hard). For example, in $d=4$ dimensions we have
\begin{equation}\label{confblock4d}
g^{4d}_{\Delta,\ell}(z,\bar{z})\propto \frac{z\bar{z}}{z-\bar{z}}(k_{\Delta+\ell}(z)k_{\Delta-\ell-2}(\bar{z})-k_{\Delta-\ell-2}(z)k_{\Delta+\ell}(\bar{z}))\;,
\end{equation}
where we have the change of variables
\begin{equation}\label{UVtozzb}
U=z\bar{z}\;,\quad V=(1-z)(1-\bar{z})\;,
\end{equation}
and
\begin{equation}
k_{2h}(z)=z^h{}_2F_1(h,h;2h;z)\;.
\end{equation}

\begin{exercise}
The new cross ratios $z$ and $\bar{z}$ in (\ref{UVtozzb}) have a clear geometric meaning. Show that you can use conformal symmetry to put all four points on a 2d plane, and then move three points at $0$, $1$ and $\infty$. The remaining point is no longer free to move around, and has complex coordinates $(z,\bar{z})$ on the plane.
\end{exercise}

\subsection{Free field in AdS}
We have seen that conformal symmetry in $\mathbb{R}^d$ is manifested in AdS$_{d+1}$ as the isometry of the background. Now let us take one step further and understand how local CFT operators are related to fields in AdS. Here we consider a free scalar field $\phi$ for simplicity, and we will follow the discussion in \cite{Penedones:2016voo}. 

The scalar field in AdS obeys the Klein-Gordon equation
\begin{equation}
\square\phi=m^2\phi\;.
\end{equation}
But let us start by considering the action of the quadratic Casimir of the isometry generators
\begin{equation}
-\frac{1}{2}L_{AB}L^{AB}\phi=\left(-X^2\partial_X^2+X\cdot\partial_X(d+X\cdot \partial_X)\right)\phi\;.
\end{equation}
Here we recall
\begin{equation}
L_{AB}=X_A\frac{\partial}{\partial X^B}-X_B\frac{\partial}{\partial X^A}\;,
\end{equation}
and we are formally extending the definition of $\phi$ from the fixed hyper surface $X^2=-R^2$. But the extension does not affect the action of the Casimir because $L_{AB}$ are interior to AdS, i.e., $[L_{AB}, X^2+R^2]=0$. We can foliate the embedding space into AdS slices with different radii $R$, then we get
\begin{equation}
\partial_X^2=-\frac{1}{R^{d+1}}\frac{\partial}{\partial R}R^{d+1}\frac{\partial}{\partial R}+\square\;.
\end{equation}
Moreover, because $X\cdot \partial_X$ generates overall scaling, we also have the relation $X\cdot \partial_X=R\partial_R$. Using this, we get
\begin{equation}\label{LAdSsquared}
-\frac{1}{2}L_{AB}L^{AB}\phi=R^2 \square \phi=m^2R^2\phi\;.
\end{equation}
This tells us that we should identify $m^2R^2$ with the eigenvalue of the Casimir. 

Let us now consider individual conformal generators, written in terms of the AdS coordinates. It is useful to use the global coordinates\footnote{Here we have restored time on the boundary so that AdS is Lorentzian
\begin{equation}
-\left(X_{-1}\right)^2-\left(X_0\right)^2+\left(X_1\right)^2+\ldots \left(X_d\right)^2=-R^2\;.
\end{equation}
}
\begin{equation}
\begin{split}
X_{-1}={}&R\cos t\cosh \rho\;,\\
X_0={}&-R\sin t\cosh\rho\;,\\
X_\mu={}&R\Omega^\mu\sinh\rho\;,
\end{split}
\end{equation}
where $\Omega^\mu$ with $\mu=1,\ldots,d$ is a unit vector parameterizing  a $d-1$ dimensional sphere. In terms of these coordinates, the conformal generators are 
\begin{equation}
\begin{split}
D={}& i\frac{\partial}{\partial t}\;,\\
M_{\mu\nu}={}&-i\left(\Omega_\mu\frac{\partial}{\partial \Omega_\nu}-\Omega_\nu\frac{\partial}{\partial \Omega_\mu}\right)\;,\\
P_\mu={}&-ie^{-it}\left(\Omega_\mu(\partial_\rho-i\tanh\rho\partial_t)+\frac{1}{\tanh\rho}\nabla_\mu\right)\;,\\
K_\mu={}&ie^{it}\left(\Omega_\mu(-\partial_\rho-i\tanh\rho\partial_t)-\frac{1}{\tanh\rho}\nabla_\mu\right)\;,
\end{split}
\end{equation}
where
\begin{equation}
\nabla_\mu=\frac{\partial}{\partial\Omega^\mu}-\Omega_\mu\Omega^\nu\frac{\partial}{\partial\Omega^\nu}\;.
\end{equation}
Imposing the conditions for the primary operator
\begin{equation}
D\phi=\Delta\phi\;,\quad K_\mu\phi=0\;,
\end{equation}
we find the wavefunction is
\begin{equation}
\phi\propto \left(\frac{e^{-it}}{\cosh\rho}\right)^\Delta\;.
\end{equation}
We can then explicitly compute the eigenvalue of the Casimir and find
\begin{equation}
\Delta(\Delta-d)=m^2R^2\;,
\end{equation}
which is the well known mass relation in AdS/CFT. We can also obtain the wavefunctions of the descendant operators  by acting with $P_\mu$\;.

\section{Perturbation Theory in AdS}\label{Sec:AdSpertth}
At the end of the previous section, we proved the mass relation $\Delta(\Delta-d)=m^2R^2$ which gives us the first entry in the AdS/CFT dictionary. We will now start exploring a richer class of physical observables, namely correlation functions of local operators in the CFT which are mapped to scattering amplitudes in the AdS theory. Note that the conformal boundary of AdS is time-like and this makes it difficult to define asymptotic states in the same way as in flat space. The AdS space behaves like a box and particles in AdS will bounce back and forth indefinitely. However, we can imagine inserting sources at the boundary and measure them again on the boundary after the wavepackets have scattered once in the bulk. More concretely, when the bulk theory is weakly coupled supergravity or QFT, which is the regime these lectures will focus on, we can describe the AdS amplitudes by using the standard perturbation theory adapted to AdS space. In fact, the AdS perturbation theory is very much similar to the flat-space perturbation theory in position space. The meaning of amplitudes in AdS can be defined precisely as a sum of AdS Feynman diagrams, i.e., Witten diagrams. 

\subsection{Tree-level Witten diagrams}\label{subsec:WDinpositionspace}

\begin{figure}[htbp]
    \centering
    \begin{subfigure}[b]{0.25\textwidth}
        \centering
        \includegraphics[width=\textwidth]{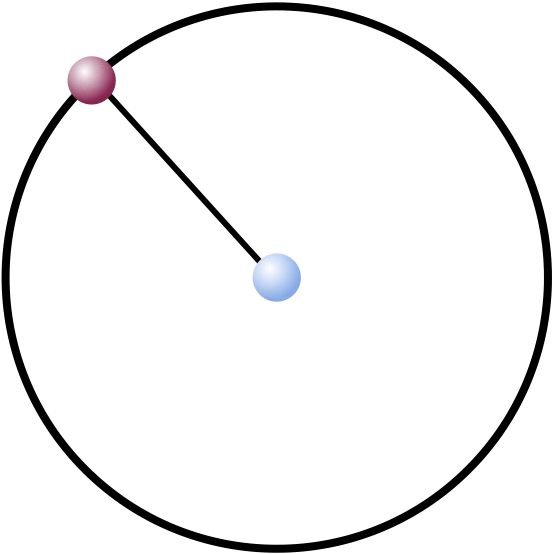}
        \caption{bulk-to-boundary}
        \label{fig:bulktoboundary}
    \end{subfigure}
    \hspace{0.1\textwidth}
    \begin{subfigure}[b]{0.25\textwidth}
        \centering
        \includegraphics[width=\textwidth]{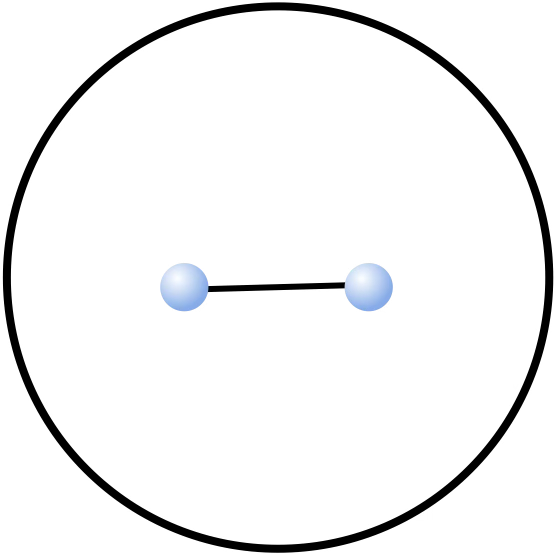}
        \caption{bulk-to-bulk}
        \label{fig:bulktobulk}
    \end{subfigure}
    \caption{Two types of propagators in AdS.}
    \label{fig:propagators}
\end{figure}

Like in flat space, Witten diagrams are constructed from propagators and vertices. There is nothing special about the vertices compared to flat space. But for the propagators, depending on where the end points are located, we should distinguish two cases: bulk-to-bulk and bulk-to-boundary (see Figure \ref{fig:propagators}). The bulk-to-bulk propagator has both points in the bulk of AdS and is the Green's function for the equation 
\begin{equation}\label{eomGBB}
(\square_Z-m^2)G_{BB}^\Delta(Z,W)=\delta(Z,W)\;,
\end{equation}
where we have set the AdS radius to $R=1$. The isometry of AdS tells us that the propagator can only be a function of the chordal distance
\begin{equation}
u=(Z-W)^2=\frac{(z-w)^2}{z_0w_0}\;.
\end{equation}
Appropriate boundary conditions select out the solution 
\begin{equation}
G_{BB}^\Delta(Z,W)=\mathcal{C}_\Delta u^{-\Delta}{}_2F_1\left(\Delta,\Delta-\frac{d}{2}+\frac{1}{2};2\Delta-d+1;-4u^{-1}\right)\;,
\end{equation}
where
\begin{equation}
\mathcal{C}_\Delta=\frac{\Gamma(\Delta)}{2\pi^{\frac{d}{2}}\Gamma(\Delta-\frac{d}{2}+1)}\;.
\end{equation}
The bulk-to-boundary propagator can be obtained by taking the limit where one bulk point approaches the boundary
\begin{equation}
\lim_{\lambda\to\infty} \lambda^\Delta G_{BB}^\Delta(Z,W=\lambda P+\ldots)\sim \frac{1}{(-2P\cdot Z)^\Delta}\;.
\end{equation}
We will choose the unit normalization and define the bulk-to-boundary propagator to be 
\begin{equation}\label{GBpartial}
G_{B\partial}^\Delta(Z,P)=\frac{1}{(-2Z\cdot P)^\Delta}=\left(\frac{z_0}{z_0^2+(\vec{z}-\vec{x})^2}\right)^\Delta\;.
\end{equation}
Note that the bulk-to-boundary propagator can also be determined by symmetry. We can view it as a ``two-point'' function where one point is an ``off-shell'' field inside the bulk AdS and the other point is an ``on-shell'' CFT operator on the boundary. The CFT operator has the scaling property (\ref{defOembed}). Up to the normalization, the combination (\ref{GBpartial}) is the only object which we can write down in terms of the invariant $Z\cdot P$ and satisfies the correct scaling under $P\to \lambda P$.

We will focus on tree-level Witten diagrams in this section. These can further be divided into two groups: contact Witten diagrams and exchange Witten diagrams. We will explain how these diagrams can be computed and discuss their properties. This is useful for developing intuitions for holographic correlators and also clarifies in which sense these objects are computed.

\subsubsection{Contact Witten diagrams}

\begin{figure}[htbp]
    \centering
        \includegraphics[width=0.25\textwidth]{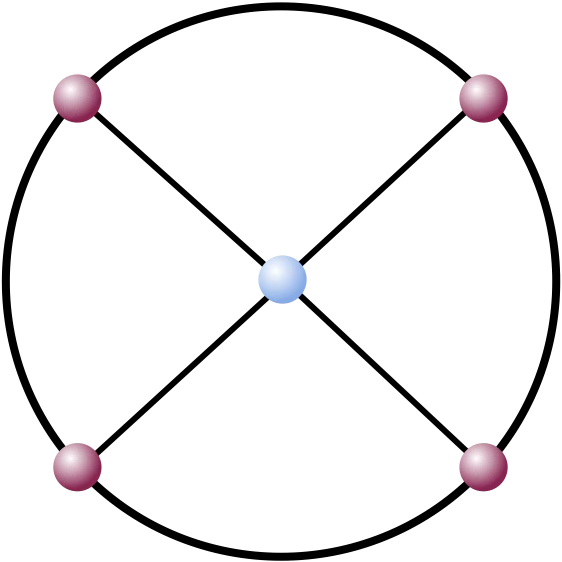}
    \caption{A contact Witten diagram.}
    \label{fig:contactWD}
\end{figure}

Let us start with the simpler contact Witten diagrams which are shown in Figure \ref{fig:contactWD}. They are defined as the product of bulk-to-boundary propagators and then integrated over the common bulk point
\begin{equation}\label{Wcon0der}
W_{\rm con}=D_{\Delta_1\ldots\Delta_n}=\int dZ \prod_{i=1}^n G_{B\partial}^{\Delta_i}(Z,P_i)=\int \frac{dz_0d^d\vec{z}}{z_0^{d+1}}\prod_{i=1}^n\left(\frac{z_0}{z_0^2+(\vec{z}-\vec{x}_i)^2}\right)^{\Delta_i}\;.
\end{equation}
In the literature these contact Witten diagrams are also known as the $D$-functions. To deal with the AdS integral, a convenient trick is to use the Schwinger parameterization
\begin{equation}
A^{-\Delta}=\frac{1}{\Gamma(\Delta)}\int_0^\infty dt t^{\Delta-1}e^{-t A}\;,
\end{equation}
to move the denominators into the exponent. This makes the integrations over the coordinates $\vec{z}$ along the boundary Gaussian integrals which can be easily performed. We can also perform the integral with respect to $z_0$. This gives us an integral representation in terms of the $n$ Schwinger parameters 
\begin{equation}\label{Dfunintrep}
D_{\Delta_1\ldots \Delta_n}=\frac{\pi^{\frac{d}{2}}\Gamma(\frac{1}{2}\Sigma-\frac{1}{2}d)\Gamma(\frac{1}{2}\Sigma)}{2\prod_i \Gamma(\Delta_i)}\int_0^\infty \prod_{i=1}^n dt_i t_i^{\Delta_i-1}  \frac{\delta(\sum_it_i-1)}{(\sum_{i<j}t_it_jx_{ij}^2)^{\frac{\Sigma}{2}}}\;,
\end{equation}
where $\Sigma=\sum_i \Delta_i$. This representation makes manifest the following differential recursion relation
\begin{equation}\label{Dfundiffrecur}
D_{\Delta_1\ldots \Delta_i+1\ldots \Delta_j+1\ldots \Delta_n}(x_i)=\frac{d-\Sigma}{2\Delta_i\Delta_j}\frac{\partial}{\partial x_{ij}^2}D_{\Delta_1\ldots\Delta_n}(x_i)\;.
\end{equation}
Another feature made manifest by this representation is that the dependence on the spacetime dimension $d$ only enters as a numerical overall factor. In other words, contact Witten diagrams in any AdS$_{d+1}$ have the same functional dependence on the cross ratios. In particular, for the four-point functions which we will primarily focus on in these lectures, we can make this property manifest by defining the $\bar{D}$-functions
\begin{equation}\label{defDbar}
\frac{2\prod_{i=1}^4\Gamma(\Delta_i)}{\pi^{\frac{d}{2}}\Gamma(\frac{\Sigma-d}{2})}D_{\Delta_1\Delta_2\Delta_3\Delta_4}(x_i)=\frac{(x_{14}^2)^{\frac{\Sigma}{2}-\Delta_1-\Delta_4}(x_{34}^2)^{\frac{\Sigma}{2}-\Delta_3-\Delta_4}}{(x_{13}^2)^{\frac{\Sigma}{2}-\Delta_4}(x_{24}^2)^{\Delta_2}}\bar{D}_{\Delta_1\Delta_2\Delta_3\Delta_4}(U,V)\;.
\end{equation}
The $\bar{D}$-functions are functions of the cross ratios $U$ and $V$, and are independent of $d$.

\begin{exercise}
Prove the integral representation (\ref{Dfunintrep}) and the differential recursion relation (\ref{Dfundiffrecur}).
\end{exercise}

On the other hand, we can consider the following type of conformal integrals in flat space $\mathbb{R}^D$ which have been considered already in the 70's
\begin{equation}\label{confintIn}
I_{\Delta_1\ldots\Delta_n}=\int d^Dx_0 \prod_{i=1}^n \frac{1}{(x_0-x_i)^{2\Delta_i}}\;.
\end{equation}
It is not difficult to show that these integrals are only conformal invariant provided the conformal dimensions and the spacetime dimension satisfy $\sum_i\Delta_i=D$. By using the same Schwinger parameterization trick, we can evaluate these integrals and it turns out that they are the same as the $D$-functions up to an overall numerical factor
\begin{equation}\label{IproptoD}
D_{\Delta_1\ldots \Delta_n}\propto I_{\Delta_1\ldots\Delta_n}\;.
\end{equation}
\begin{exercise}
Prove the relation (\ref{IproptoD}).
\end{exercise}
An important special case is $n=4$ and $\Delta_i=1$.  This flat-space conformal integral is nothing but the scalar one-loop box diagram in four dimensions. The equivalence with the one-loop Feynman diagram can be seen by defining momentum variables  $p_i=x_i-x_{i-1}$ (note momentum conservation $p_1+p_2+p_3+p_4=0$ is automatic) and writing $x_0=-l+x_1-p_1$. This gives 
\begin{equation}
I_{1111}=\int d^4x_0\frac{1}{x_{01}^2x_{02}^2x_{03}^2x_{04}^2}=\int d^4l\frac{1}{l^2(l+p_1)^2(l+p_1+p_2)^2(l+p_1+p_2+p_3)^2}\;,
\end{equation}
This one-loop integral has been explicitly computed in the literature \cite{Usyukina:1992jd}. Using the relation between the conformal integral and $D$-functions, we have the following explicit expression for $D_{1111}$  
\begin{equation}\label{Dbar1111aspolylogs}
\bar{D}_{1111}\equiv\Phi(z,\bar{z})= \frac{1}{z-\bar{z}}\left(2{\rm Li}_2(z)-2{\rm Li}_2(\bar{z})+\log (z\bar{z})\log\left(\frac{1-z}{1-\bar{z}}\right)\right)\;.
\end{equation}
Using the differential recursion relations (\ref{Dfundiffrecur}), it is clear that other $D$-functions whose weights can be obtained from $\{1,1,1,1\}$ by repeatedly shifting a pair of weights by 1 at a time (e.g., $D_{1212}$, $D_{2222}$), can also be written in terms of polylogarithms with transcendental weights no higher than 2. Later we will see that these are precisely the type of $D$-functions which we encounter in the holographic correlators in AdS$_5\times$S$^5$. 

In (\ref{Wcon0der}) we have focused on the simplest case where the contact vertex has no derivatives. More generally, we can also consider vertices which have $2L$ contracted derivatives.\footnote{In odd dimensional AdS space, it is also possible to have contact Witten diagrams with odd number of derivatives constructed using the Levi-Civita tensor. These contact Witten diagrams can also be written in terms of $D$-functions, see \cite{Rastelli:2019gtj}.} These contact Witten diagrams with derivatives can also be expressed in terms of $D$-functions. See the following exercise. 
\begin{exercise}
Prove the following identity for bulk-to-boundary propagators
\begin{equation}\label{idforGBpartialwder}
\nabla^\mu G_{B\partial}^{\Delta_1}\nabla_\mu G_{B\partial}^{\Delta_2}=\Delta_1\Delta_2(G_{B\partial}^{\Delta_1}G_{B\partial}^{\Delta_2}-2x_{12}^2G_{B\partial}^{\Delta_1+1}G_{B\partial}^{\Delta_2+1})\;,
\end{equation}
and then use it to show the contact Witten diagrams with derivatives can be expressed in terms of $D$-functions. 
\end{exercise}

Let us conclude the discussion of contact Witten diagrams by making a few comments. These are interesting facts but are not really needed for the rest of these lectures. Therefore, they can be safely skipped in a first reading. First of all, it is easy to notice that in the repeated applications of the differential recursion relations (\ref{Dfundiffrecur}), there is more than one way to go from one initial weight configuration to the same final one. For example, from $D_{\ldots \Delta_i\ldots \Delta_j\ldots \Delta_k \ldots \Delta_l\ldots}$ to $D_{\ldots \Delta_i+1\ldots \Delta_j+1\ldots \Delta_k+1 \ldots \Delta_l+1\ldots}$, where the other dimensions denoted by $\ldots$ are kept the same, we can proceed in three different ways. We can first increase the pair $\Delta_i$, $\Delta_j$ by 1, and then increase the $\Delta_k$, $\Delta_l$ pair by 1. Alternatively, we can increase first $\Delta_i$, $\Delta_k$ and then $\Delta_j$, $\Delta_l$, or first $\Delta_i$, $\Delta_l$ and then $\Delta_k$, $\Delta_j$. In terms of (\ref{Dfundiffrecur}), the consistency of these different ways to shift the weights translates to the following differential equations for the $D$-functions
\begin{equation}
\begin{split}
(\partial_{x_{ij}^2}\partial_{x_{kl}^2}-\partial_{x_{ik}^2}\partial_{x_{jl}^2})D_{\Delta_1\ldots\Delta_n}={}&0\;,\\
(\partial_{x_{ij}^2}\partial_{x_{kl}^2}-\partial_{x_{il}^2}\partial_{x_{jk}^2})D_{\Delta_1\ldots\Delta_n}={}&0\;.
\end{split}
\end{equation}
Quite interestingly, these consistency conditions also have an integrability meaning: the $D$-functions, or equivalently the conformal integrals $I_{\Delta_1\ldots\Delta_n}$ are invariant under a bigger symmetry known as the conformal Yangian \cite{Loebbert:2019vcj,Loebbert:2020hxk,Loebbert:2020glj}; the consistency conditions of the differential recursion relations turn out to be equivalent to the constraints imposed by the Yangian invariance \cite{Rigatos:2022eos}.

\begin{exercise}
The differential recursion relations actually also allow us to compute $D_{1111}$ without evaluating the integral. Rewrite the conditions
\begin{equation}
\begin{split}
(\partial_{x_{12}^2}\partial_{x_{34}^2}-\partial_{x_{14}^2}\partial_{x_{23}^2})D_{1111}={}&0\;,\\
(\partial_{x_{12}^2}\partial_{x_{34}^2}-\partial_{x_{13}^2}\partial_{x_{24}^2})D_{1111}={}&0\;,
\end{split}
\end{equation}
as partial differential equations with respect to $z$ and $\bar{z}$. Then show (\ref{Dbar1111aspolylogs}) can be obtained from solving these equations.

\vspace{0.3cm}

{\bf Hint:} see \cite{Loebbert:2019vcj} for details.
\end{exercise}

\begin{exercise}
$D$-functions also satisfy various other identities. For example, show the $\bar{D}$-functions defined in (\ref{defDbar}) satisfy
\begin{equation}
\Delta_4 \bar{D}_{\Delta_1\Delta_2\Delta_3\Delta_4}= \bar{D}_{\Delta_1+1\Delta_2\Delta_3\Delta_4+1}+\bar{D}_{\Delta_1\Delta_2+1\Delta_3\Delta_4+1}+\bar{D}_{\Delta_1\Delta_2\Delta_3+1\Delta_4+1}\;.
\end{equation}
\end{exercise}
Another comment is that the $D$-functions have a natural geometric meaning. This can be quite tantalizing because one may wonder if the AdS amplitudes also have underlying geometric structures similar to those of the flat-space amplitudes. Here the observation is still very preliminary. The statement is that the function $D_{1111}$ can be interpreted as the volume of an ideal tetrahedron in Euclidean AdS$_3$
\begin{equation}\label{D1111asvolume}
D_{1111}(P_i)=\frac{\pi}{2}\frac{{\rm vol}(T)}{\sqrt{-\det(P_i\cdot P_j)}}\;,
\end{equation}  
where the tetrahedron $T$ has vertices $P_i$. In what follows, we will outline a simple proof for this fact. Let $\alpha_i\geq 0$ and define
\begin{equation}
\widehat{Q}(\alpha)=\frac{Q}{\sqrt{-Q^2}}\;,
\end{equation}
with $Q=\sum_i\alpha_iP_i$. Then we have $\widehat{Q}^2=-1$ and $\widehat{Q}(\alpha)$ parametrizes the ideal tetrahedron in AdS. We can choose representatives for the projective null vectors $P_i$. Since $\widehat{Q}(\alpha)$ is invariant under a common rescaling of all $\alpha_i$, we may further fix this redundancy by imposing $\sum_i\alpha_i=1$. The $D_{1111}$ integral can be written using Feynman parameters as 
\begin{equation}
D_{1111}=\int_{H^3}d{\rm vol}(X) \prod_{i=1}^4\frac{1}{-2P_i\cdot X}=6\int_{\Delta_3} d^3\alpha \int_{H^3}\frac{d{\rm vol}(X)}{(-2Q\cdot X)^4}\;,
\end{equation}
where $\Delta_3$ denotes the integration region bounded by the constraints $\{\alpha_i\geq0,\sum_i\alpha_i=1\}$. This integral can be performed by choosing a frame for the time-like vector $Q=(q,0,0,0)$ with $q=\sqrt{-Q^2}$, and parameterizing the AdS coordinates as $X=(\cosh r,\sinh r \hat{n})$. We get
\begin{equation}
D_{1111}=\frac{\pi}{2}\int_{\Delta_3}\frac{d^3\alpha}{(-Q^2)^2}\;.
\end{equation}
On the other hand, the invariant volume form is
\begin{equation}
d{\rm vol}_{H^3}=|\epsilon(\widehat{Q},\partial_1\widehat{Q},\partial_2\widehat{Q},\partial_3\widehat{Q})|d^3\alpha\;,
\end{equation}
after rewriting it in the $\alpha_i$ coordinates. It is easy to show 
\begin{equation}
\epsilon(\widehat{Q},\partial_1\widehat{Q},\partial_2\widehat{Q},\partial_3\widehat{Q})=\frac{1}{(-Q^2)^2}\epsilon(Q,\partial_1Q,\partial_2Q,\partial_3Q)\;.
\end{equation}
Using the explicit parameterization $Q=P_4+\alpha_1(P_1-P_4)+\alpha_2(P_2-P_4)+\alpha_3(P_3-P_4)$, we get 
\begin{equation}
\epsilon(Q,\partial_1Q,\partial_2Q,\partial_3Q)=-\epsilon(P_1,P_2,P_3,P_4)\;.
\end{equation}
It then follows
\begin{equation}
d{\rm vol}_{H^3}=\frac{|\epsilon(P_1,P_2,P_3,P_4)|}{(-Q^2)^2}d^3\alpha=\frac{\sqrt{-\det(P_i\cdot P_j)}}{(-Q^2)^2}d^3\alpha\;,
\end{equation}
where we have used $\det(P_i\cdot P_j)=-[\epsilon(P_1,P_2,P_3,P_4)]^2$. Comparing these two expressions, we prove (\ref{D1111asvolume}). 

\subsubsection{Exchange Witten diagrams}\label{Subsubsec:exchinposi}

\begin{figure}[htbp]
    \centering
        \includegraphics[width=0.25\textwidth]{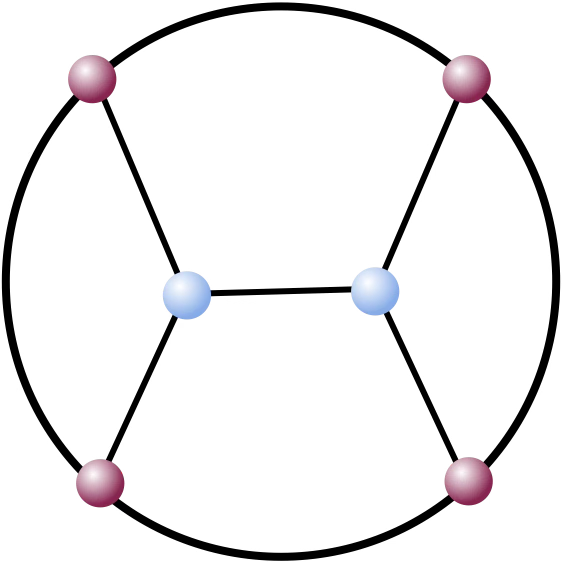}
    \caption{An exchange Witten diagram.}
    \label{fig:exchangeWD}
\end{figure}

Now let us consider the exchange Witten diagrams shown in Figure \ref{fig:exchangeWD}. For concreteness and simplicity, we will focus on four points and consider only the scalar exchange diagrams. The exchange Witten diagram is defined by the integral  
\begin{equation}\label{defWexscalar}
W_{\Delta,0}(P_i)=\int dZ dW G_{B\partial}^{\Delta_1}(P_1,Z)G_{B\partial}^{\Delta_2}(P_2,Z) G_{BB}^\Delta(Z,W) G_{B\partial}^{\Delta_3}(P_3,W)G_{B\partial}^{\Delta_4}(P_4,W)\;,
\end{equation}
where we need to perform two AdS integrals. Unlike the contact case, we will not compute the exchange Witten diagram by directly evaluating the integrals because this is technically very complicated. Instead, we will first find the differential equation satisfied by the exchange Witten diagram and then compute the diagram by solving the differential equation \cite{DHoker:1999mqo}. 

\begin{figure}[h]
    \centering
        \includegraphics[width=0.26\textwidth]{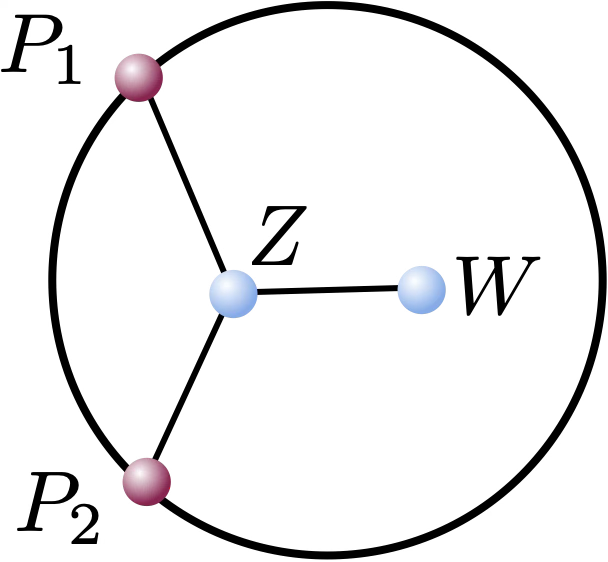}
    \caption{The ``three-point function'' part of the exchange Witten diagram.}
    \label{fig:3ptint}
\end{figure}

To do that, we first focus on a part of the diagram which involves just one of the integrals (Figure \ref{fig:3ptint})
\begin{equation}\label{def3ptAint}
A_{\Delta,0}(P_1,P_2,W)=\int dZ G_{B\partial}^{\Delta_1}(P_1,Z)G_{B\partial}^{\Delta_2}(P_2,Z) G_{BB}^\Delta(Z,W)\;.
\end{equation}
This can be viewed as a ``three-point function'' where we have two ``on-shell'' CFT local operators and one ``off-shell'' AdS field. As is clear from this definition, we have the rescaling property
\begin{equation}\label{A3ptscaling}
A_{\Delta,0}(\lambda_1P_1,\lambda_2P_2,W)=\lambda_1^{-\Delta_1}\lambda_2^{-\Delta_2}A_{\Delta,0}(P_1,P_2,W)\;.
\end{equation}
But for the bulk point $W$ we do not impose any rescaling constraint. We can proceed in the same way as we analyzed correlators of local operators in embedding space. The three-point integral (\ref{def3ptAint}) is fixed up to a function of a single cross ratio
\begin{equation}\label{deffgamma}
A_{\Delta,0}=\frac{1}{(-2P_1\cdot W)^{\Delta_1}(-2P_2\cdot W)^{\Delta_2}}f(\gamma)\;.
\end{equation}
Here the extracted factor takes care of the scaling behavior in (\ref{A3ptscaling}), and $\gamma$ is the only ``cross ratio'' we can write down that is invariant under both conformal symmetry and rescaling
\begin{equation}
\gamma=\frac{(-2P_1\cdot P_2)}{(-2P_1\cdot W)(-2P_2\cdot W)}=\frac{(\vec{x}_1-\vec{x}_2)^2w_0^2}{(w_0^2+(\vec{w}-\vec{x_1})^2)(w_0^2+(\vec{w}-\vec{x_2})^2)}\;.
\end{equation}
The three-point function $A_{\Delta,0}$ is conformal invariant, which implies it is annihilated by the action of the conformal generators
\begin{equation}
(L^{(1)}_{AB}+L^{(2)}_{AB}+L^{\rm bulk}_{AB})A_{\Delta,0}=0\;.
\end{equation}
Here $L^{(i)}_{AB}$ are the conformal generators acting on boundary point $i$ and $L^{\rm bulk}_{AB}$ are the AdS isometry generators acting on the bulk point $W$. We can then square this relation and get the following identity
\begin{equation}
{\rm Cas}^{(12)} A_{\Delta,0}=-\frac{1}{2}(L^{(1)}_{AB}+L^{(2)}_{AB})^2A_{\Delta,0}=-\frac{1}{2} (L^{\rm bulk}_{AB})^2 A_{\Delta,0}=\square_W A_{\Delta,0}\;,
\end{equation}
where on the LHS we have a two-particle conformal Casimir and on the RHS we have used (\ref{LAdSsquared}) to write the action of the AdS isometry Casimir as the AdS Laplacian. Note that the Laplacian acts on the bulk point $W$ and in (\ref{def3ptAint}) $W$ appears only in the bulk-to-bulk propagator $G_{BB}^\Delta(Z,W)$. So we have
\begin{equation}\label{CaseqforA}
\begin{split}
({\rm Cas}^{(12)}-m^2)A={}&\int dZ G_{B\partial}^{\Delta_1}(P_1,Z)G_{B\partial}^{\Delta_2}(P_2,Z) (\square_W-m^2)G_{BB}^\Delta(Z,W)\\
={}&\frac{1}{(-2P_1\cdot W)^{\Delta_1}(-2P_2\cdot W)^{\Delta_2}}\;,
\end{split}
\end{equation}
where we have used the equation of motion for $G_{BB}^\Delta(Z,W)$ (\ref{eomGBB}) and the integral over the delta function just sets $Z=W$. We also recall the mass relation $m^2=\Delta(\Delta-d)$. Using (\ref{deffgamma}) and stripping off the inhomogeneous kinematic factor, we get a second order ODE for $f(\gamma)$ which reads
\begin{equation}\label{eomforf}
\begin{split}
{}&4\gamma^2(\gamma-1)f''(\gamma)+\gamma(4(\gamma-1)(\Delta_1+\Delta_2+1)+2d)f'(\gamma)\\
{}&+((\Delta-\Delta_1-\Delta_2)(-d+\Delta+\Delta_1+\Delta_2)+4\Delta_1\Delta_2\gamma)f(\gamma)=1\;.
\end{split}
\end{equation}
To determine $f(\gamma)$, we further need to impose two boundary conditions:
\begin{enumerate}
\item {\bf Behavior in the $\gamma\to 0$ limit.} The $\gamma\to0$ limit is the OPE limit where the external operators $\mathcal{O}_1$ and $\mathcal{O}_2$ come close. In this OPE, we in particular will exchange the scalar operator with dimension $\Delta$ dual to $G_{BB}^\Delta$. This gives rise to the behavior $f(\gamma)\to \gamma^{\frac{\Delta-\Delta_1-\Delta_2}{2}}$.
\item {\bf Regularity at $\gamma=1$.} The condition $\gamma=1$ geometrically means that the bulk point $W$ is sitting at the geodesic line connecting the two boundary points.\footnote{Geodesics in the Poincar\'e coordinates are semi-circles (or a straight line) which intersect with the conformal boundary perpendicularly. } But clearly from the definition (\ref{def3ptAint}) there is nothing singular about the integral at this configuration. 
\end{enumerate}
\begin{exercise}
Derive the ODE (\ref{eomforf}).
\end{exercise}
Let us start by looking for power-like solutions to (\ref{eomforf}). We make the assumption that $f(\gamma)$ takes the form
\begin{equation}
f(\gamma)=\sum_k a_k\gamma^k\;,
\end{equation}
where for the moment we do not specify the rank of the sum over $k$. Substituting this ansatz into the ODE, we find that different powers of $\gamma$ are shifted into each other and this gives the following recursion relation for the coefficients
\begin{equation}
a_{k-1}=\frac{(k-\frac{\Delta-\Delta_1-\Delta_2}{2})(k+\frac{\Delta+\Delta_1+\Delta_2}{2}-\frac{d}{2})}{(k+\Delta_1-1)(k+\Delta_2-1)}a_k\;.
\end{equation}
Moreover, the inhomogeneous term in (\ref{eomforf}) determines the upper bound of $k$ to be $k_{\rm max}=-1$ with the coefficient
\begin{equation}
a_{k_{\rm max}}=\frac{1}{4(\Delta_1-1)(\Delta_2-1)}\;.
\end{equation}
Here we note that if the conformal dimensions satisfy
\begin{equation}\label{trunccond}
\Delta_1+\Delta_2-\Delta\in 2\mathbb{Z}_{>0}\;,
\end{equation}
the series truncates into a polynomial with $k_{\rm min}=\frac{\Delta-\Delta_1-\Delta_2}{2}$. Note that such a polynomial solution for $f(\gamma)$ clearly satisfies both boundary conditions and therefore is a true solution. We then get
\begin{equation}\label{cubicinttocontact}
\begin{split}
A_{\Delta,0}(P_1,P_2,W){}&=\sum_{k_{\rm min}}^{k_{\rm max}}\frac{a_k (-2P_1\cdot P_2)^k}{(-2P_1\cdot W)^{\Delta_1+k}(-2P_2\cdot W)^{\Delta_2+k}}\\
{}&=\sum_{k_{\rm min}}^{k_{\rm max}}a_k x_{12}^{2k}G_{B\partial}^{\Delta_1+k}(W,P_1)G_{B\partial}^{\Delta_2+k}(W,P_2)\;,
\end{split}
\end{equation}
which is a sum of contact interactions with shifted dimensions for the propagators. Inserting it into the definition of the exchange Witten diagram (\ref{defWexscalar}), we find that it can be written as a finite sum of contact Witten diagrams
\begin{equation}
W_{\Delta,0}=\sum_{k_{\rm min}}^{k_{\rm max}}a_k x_{12}^{2k} D_{\Delta_1+k,\Delta_2+k,\Delta_3,\Delta_4}\;.
\end{equation}
Generically when the truncation condition (\ref{trunccond}) is not satisfied, the situation is a bit more complicated. The solution to $f(\gamma)$ is a linear combination of hypergeometric functions and the expansion in $\gamma$ contains infinitely many terms. However, each term in the series still corresponds to a product of bulk-to-boundary propagators. Therefore, the general exchange Witten diagram can still be written as the sum of infinitely many $D$-functions.  

Note that even without explicitly solving the three-point integral (\ref{def3ptAint}), we can use (\ref{CaseqforA}) in (\ref{defWexscalar}) and obtain the following relation
\begin{equation}\label{EoMWD}
({\rm Cas}^{(12)}-m^2)W_{\Delta,0}=D_{\Delta_1\Delta_2\Delta_3\Delta_4}\;.
\end{equation}
This is a useful identity that relates exchange and contact Witten diagrams, and is valid for arbitrary external and internal conformal dimensions.

Although here we have restricted ourselves to the scalar exchange case to illustrate the method, the techniques and results also apply more generally to exchange Witten diagrams where the internal line carries nonzero spins. For an exchange field with dimension $\Delta$ and $\ell$, the identity (\ref{EoMWD}) becomes
 \begin{equation}\label{EoMWDgen}
({\rm Cas}^{(12)}-C_{\Delta,\ell})W_{\Delta,\ell}=W_{\rm con}\;,
\end{equation}
where $C_{\Delta,\ell}$ is the quadratic conformal Casimir eigenvalue $C_{\Delta,\ell}=\Delta(\Delta-d)+\ell(\ell+d-2)$ and $W_{\rm con}$ is a collection of contact Witten diagrams with no more than $2\ell-2$ contracted derivatives. 
 
\subsection{Conformal blocks and geodesic Witten diagrams}\label{Subsec:confblockdecom}
\begin{figure}[h]
    \centering
        \includegraphics[width=0.25\textwidth]{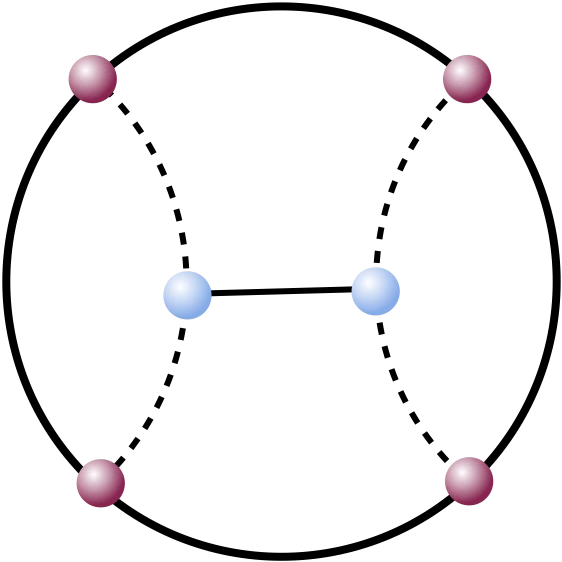}
    \caption{A geodesic Witten diagram. The dashed lines are the geodesics in AdS.}
    \label{fig:geoWD}
\end{figure}

To make more connections with our CFT discussion, let us go back to the conformal blocks introduced in Section \ref{subsec:OPEandconfblocks} by considering a special type of exchange Witten diagrams. These are the geodesic Witten diagrams and they are the holographic dual of conformal blocks \cite{Hijano:2015zsa}. Again for simplicity we will focus only on the scalar case. The geodesic Witten diagram is defined by AdS integrals similar to (\ref{defWexscalar})
\begin{equation}
W_{\Delta,0}^{\rm geo}(P_i)=\int_{\gamma_{12}} \int_{\gamma_{34}}dZ dW G_{B\partial}^{\Delta_1}(P_1,Z)G_{B\partial}^{\Delta_2}(P_2,Z) G_{BB}^\Delta(Z,W) G_{B\partial}^{\Delta_3}(P_3,W)G_{B\partial}^{\Delta_4}(P_4,W)\;,
\end{equation}
but now with the bulk integrations restricted to the two geodesics (see Figure \ref{fig:geoWD}). 

We can also consider the action of the operator in (\ref{CaseqforA}) which only concerns the three-point function with external points $P_1$, $P_2$ and $Z$. However, an important difference is that because the integration over $Z$ is restricted to the geodesic line $\gamma_{12}$, generically $Z$ will not coincide with $W$. Therefore, instead of an inhomogeneous term, we get zero on the RHS in (\ref{CaseqforA}). It then follows
\begin{equation}
({\rm Cas}-m^2)W_{\Delta,0}^{\rm geo}=0\;.
\end{equation}
This is precisely the same conformal Casimir equation satisfied by the scalar conformal block $g_{\Delta,0}$. Moreover, it is easy to show that both $W_{\Delta,0}^{\rm geo}$ and $g_{\Delta,0}$ satisfy the same boundary conditions. Therefore, the geodesic Witten diagram is just the conformal block up to normalizations\footnote{The geodesic Witten diagram representation of conformal blocks can also be extended to CFTs with defects, see, e.g., \cite{Rastelli:2017ecj} for the case of boundary CFT.}
\begin{equation}
W_{\Delta,0}^{\rm geo}\propto g_{\Delta,0}\;.
\end{equation}
This correspondence $W_{\Delta,\ell}^{\rm geo}\propto g_{\Delta,\ell}$ also extends to the case where the spin $\ell$ of the exchanged field is nonzero and we have
 \begin{equation}\label{confCaseqn}
({\rm Cas}^{(12)}-C_{\Delta,\ell})g_{\Delta,\ell}=0\;,
\end{equation}
Just like the equation of motion identity (\ref{EoMWDgen}) which allows us to solve the exchange Witten diagrams, the homogeneous conformal Casimir equations (\ref{confCaseqn}) also allows us to efficiently obtain the conformal blocks \cite{Dolan:2003hv}.

Let us now briefly discuss how the Witten diagrams are decomposed into conformal blocks. We first try to understand what types of operators are expected to arise in the decomposition. The tree-level Witten diagrams considered here can be viewed as the leading correction to the generalized free fields (GFF) which describes the theory at strictly infinite $N$. Here GFF just means that in the CFT all correlators are computed by Wick contractions of the basic fields and in the dual AdS description we have just free fields. Therefore, the Hilbert space, or equivalently the CFT spectrum, is just the Fock space constructed from the normal ordered products of the basic CFT operator. Consider for example the following four-point function in the GFF limit
\begin{equation}\label{GFFcorr}
\begin{split}
\langle \mathcal{O}(x_1)\mathcal{O}(x_2)\mathcal{O}(x_3)\mathcal{O}(x_4) \rangle_{\rm GFF} ={}& \langle \mathcal{O}(x_1)\mathcal{O}(x_2)\rangle \langle \mathcal{O}(x_3)\mathcal{O}(x_4) \rangle+\langle \mathcal{O}(x_1)\mathcal{O}(x_3)\rangle \langle \mathcal{O}(x_2)\mathcal{O}(x_4) \rangle\\
{}& + \langle \mathcal{O}(x_1)\mathcal{O}(x_4)\rangle \langle \mathcal{O}(x_2)\mathcal{O}(x_3) \rangle\\
={}&\frac{1}{x_{12}^{2\Delta}x_{34}^{2\Delta}}\left(1+U^\Delta+\left(\frac{U}{V}\right)^\Delta\right)\;.
\end{split}
\end{equation}
Decomposing this correlator into conformal blocks, one finds in addition to the identity operator also infinitely many ``double-trace''\footnote{This terminology is borrowed from 4d $\mathcal{N}=4$ SYM but is often used to large $N$ theories which do not have a gauge theory description. Basic operators in the gauge theory are the ``single-trace'' operators which are products of basic adjoint fields in the Lagrangian and then with a single trace taken in the color space to ensure gauge invariance. Double-trace operators, or more generally multi-trace operators, are defined as the normal ordered products of single-trace operators.} conformal blocks
\begin{equation}
\langle \mathcal{O}(x_1)\mathcal{O}(x_2)\mathcal{O}(x_3)\mathcal{O}(x_4) \rangle_{\rm GFF} \propto 1+ \sum_{n,\ell} a^{(0)}_{n,\ell}\, g_{2\Delta+2n+\ell,\ell}(z,\bar{z})\;.
\end{equation}
These double-trace operators are schematically
\begin{equation}
:\mathcal{O}_i\square^n\partial^\ell \mathcal{O}_j:\;.
\end{equation}
They are constructed from the single-trace operators $\mathcal{O}_i$ and  $\mathcal{O}_j$ and has conformal dimensions $\Delta^{ij}_{n,\ell}=\Delta_i+\Delta_j+2n+\ell$. In this example, we looked at the case where the operators $\mathcal{O}_i$ and  $\mathcal{O}_j$ are the same. 
\begin{exercise}
Use the conformal blocks given in (\ref{confblock4d}) to decompose the GFF correlator in 4d (\ref{GFFcorr}).  Confirm the structure that there are only the identity and the double-trace operators, and obtain the decomposition coefficients.

\vspace{0.3cm}

{\bf Hint:} Do this in Mathematica. First consider a small $U$ expansion, then for each power of $U$ consider an expansion in $1-V$.
\end{exercise}

The tree-level Witten diagrams correspond to the leading corrections after turning on the $1/N$ suppressed bulk interactions. Therefore, the double-trace operators which we observed at the previous leading order are expected to still persist at the tree level. For the contact Witten diagram $D_{\Delta_1\Delta_2\Delta_3\Delta_4}$, assuming $\Delta_i$ are generic, we have 
\begin{equation}\label{confblockdecDfun}
D_{\Delta_1\Delta_2\Delta_3\Delta_4}\propto \sum_{n=0}^\infty a_{n,0}^{12}\, g_{\Delta^{12}_{n,0}}+\sum_{n=0}^\infty a_{n,0}^{34}\, g_{\Delta^{34}_{n,0}}\;,
\end{equation}
where the two infinite towers of double-trace conformal blocks correspond to the operators $:\mathcal{O}_1\square^n\mathcal{O}_2:$ and $:\mathcal{O}_3\square^n\mathcal{O}_4:$. Note that the decomposition has only support on spin $\ell=0$. If we consider contact Witten diagrams with $2L$ derivatives, we generically expect support on spins up to $L$. 

For the exchange Witten diagram, considering the scalar case for concreteness, we have both the single-trace operator and double-trace operators due to exchange of the field in AdS
\begin{equation}
W_{\Delta,0}\propto A\, g_{\Delta,0}+\sum_{n=0}^\infty A_{n,0}^{12}\, g_{\Delta^{12}_{n,0}}+\sum_{n=0}^\infty A_{n,0}^{34}\, g_{\Delta^{34}_{n,0}}\;.
\end{equation}
Note that it also has a finite (in this case only $\ell=0$) support on spins. The equation of motion relation (\ref{EoMWD}) and the Casimir equation of conformal blocks together imply a useful relation between the coefficients
\begin{equation}
a^{ij}_{n,0}=(C_{\Delta,0}-C_{\Delta^{ij}_{n,0},0})A^{ij}_{n,0}\;.
\end{equation}

Here we have focused on the decomposition in the s-channel where we perform the OPE for $\mathcal{O}_1$ and $\mathcal{O}_2$. We can also consider the t-channel OPE where $\mathcal{O}_1$ and $\mathcal{O}_4$ come close, for the contact Witten diagram and the s-channel exchange Witten diagram. In the case of the contact Witten diagram, the t-channel decomposition is similar to the s-channel and one only gets double-trace operators
\begin{equation}
D_{\Delta_1\Delta_2\Delta_3\Delta_4}\propto \sum_{n=0}^\infty a_{n,0}^{14}\, g^{(t)}_{\Delta^{14}_{n,0}}+\sum_{n=0}^\infty a_{n,0}^{23}\, g^{(t)}_{\Delta^{23}_{n,0}}\;.
\end{equation}
Here we added a superscript for the conformal blocks to remind us that they are in the t-channel. Clearly, the t-channel decomposition coefficients are given by the s-channel coefficients by permuting the operator labels.

For the exchange Witten diagram, we also get only double-trace operators in the t-channel
\begin{equation}\label{Wscalarintchannel}
W_{\Delta,0}\propto \sum_{n=0}^\infty \sum_{\ell=0}^\infty B_{n,\ell}^{14}\, g^{(t)}_{\Delta^{14}_{n,\ell}}+\sum_{n=0}^\infty\sum_{\ell=0}^\infty B_{n,\ell}^{23}\, g^{(t)}_{\Delta^{23}_{n,\ell}}\;,
\end{equation}
but an important difference to notice is that the double-trace conformal blocks now have infinite support on spins. To efficiently compute these decomposition coefficients, the differential operator in the Casimir equation is again very useful. Acting it on (\ref{Wscalarintchannel}), we get from the LHS a $D$-function which we already know how to decompose. On the RHS, the action on each t-channel conformal block also admits nice recursion relations which allow us to express it as a finite sum of  t-channel conformal block with shifted dimensions and spins. This leads to recursion relations for the t-channel decomposition coefficients which make solving the coefficients a tractable problem. More details of the conformal block decomposition of Witten diagrams can be found in \cite{Zhou:2018sfz}.

Let us conclude the discussions in this subsection with two comments. The first comment is about the case where the double-trace towers become degenerate. This happens for example when the external operators are all identical. In addition to the double-trace conformal blocks, we also get their derivatives with respect to the conformal dimension. The derivative conformal blocks arise because the decomposition coefficients in the two towers become divergent but with opposite signs. Taking the limit then leads to derivatives on the conformal blocks. See, e.g., \cite{Zhou:2018sfz} for details. The second comment is that these exchange Witten diagrams are later understood to be essentially the unitary blocks proposed by Polyakov in the `70s \cite{Polyakov:1974gs} well before the advent of AdS/CFT, see \cite{Gopakumar:2016wkt,Gopakumar:2016cpb,Gopakumar:2018xqi}.  This connection is exploited in the Polyakov style bootstrap in Mellin space \cite{Gopakumar:2016wkt,Gopakumar:2016cpb,Gopakumar:2018xqi} and is also very useful in the analytic functional approach to the conformal bootstrap \cite{Mazac:2016qev,Mazac:2018mdx,Mazac:2018ycv,Kaviraj:2018tfd,Mazac:2018biw,Mazac:2019shk}. We will not go into further details of these development, but the interested reader is welcome to have a look at the pedagogical discussions in the review \cite{Bissi:2022mrs} to get a rough flavor.

\subsection{Mellin amplitudes}\label{Subsec:Mellinrepresentation}
In flat space, scattering amplitudes are usually considered in momentum space because their analytic structure is simpler there. This simplification has a symmetry reason, namely the plane waves in the Fourier transform are the eigenfunctions of the translation operators. In AdS, however, the translations do not commute and it is difficult to directly define a useful momentum space. However, we can exploit the dilatation symmetry and consider the Mellin transform
\begin{equation}
\varphi(s)=\int_0^\infty x^{s-1}f(x) dx\;,\quad f(x)=\frac{1}{2\pi i}\int_{c-i\infty}^{c+i\infty} x^{-s}\varphi(s) ds\;,
\end{equation}
which relates the power in $x^{-a}$ to a pole at $s=a$. This is very natural in CFT correlators because correlators can be expanded into powers of cross ratios with powers controlled by the conformal dimensions of the operators appearing in the OPE.  

Let us consider an $n$-point function scalar operators. We write the correlator as a multi-dimensional inverse Mellin transform \cite{Mack:2009mi,Penedones:2010ue}
\begin{equation}\label{defnptMellin}
\langle \mathcal{O}_1(x_1)\ldots \mathcal{O}_n(x_n)\rangle=\int [d\delta_{ij}]\left(\prod_{i<j}(-2P_i\cdot P_j)^{-\delta_{ij}}\Gamma(\delta_{ij})\right)\mathcal{M}(\delta_{ij})\;.
\end{equation}
For the integration variables $\delta_{ij}$, we can extend their definition from $i<j$ to $i\geq j$ by setting 
\begin{equation}
\delta_{ij}=\delta_{ji}\;, \quad \delta_{ii}=-\Delta_i\;.
\end{equation}
Note that in the second equality, no summation over the index $i$ is performed. Requiring the correlator to have the correct behavior under individual rescaling $P_i \to \lambda_i P_i$ then imposes the conditions
\begin{equation}
\sum_{j=1}^n\delta_{ij}=0\;.
\end{equation}
Finally, we integrate over independent $\delta_{ij}$ along the imaginary axes in (\ref{defnptMellin}). All the information in the $n$-point correlator is now transferred to the function $\mathcal{M}(\delta_{ij})$ which is called the Mellin amplitude.

\subsubsection{Why Mellin amplitudes look like amplitudes}

To solve the constraints and to see why $\mathcal{M}(\delta_{ij})$ can be viewed as an ``amplitude'', we introduce fictitious flat-space momenta and write $\delta_{ij}$ as 
\begin{equation}
\delta_{ij}=\vec{p}_i\cdot \vec{p}_j\;.
\end{equation}
The constraints for $\delta_{ij}$ are automatically solved if the momenta are conserved and on-shell
\begin{equation}\label{ficpconstraints}
\sum_i \vec{p}_i=0\;,\quad\quad \vec{p}_i^2=-\Delta_i\;,
\end{equation}
where the role of the squared mass $m^2$ is played by the conformal dimension $\Delta$. The dual variables $\delta_{ij}$ are just the Mandelstam variables of the fictitious momenta. This motivates us to think of correlators in CFT as some kind of scattering amplitudes in higher dimensions (for why higher dimensions, see the following exercise). 

\begin{exercise}
Count the number of independent Mandelstam variables $\delta_{ij}$ as a function of $n$ subject to the constraints (\ref{ficpconstraints}). Then compare the result with the counting of cross ratios in a $d$-dimensional CFT in Exercise \ref{countcrossratios}. Assuming the fictitious flat space is $D$-dimensional, show the countings match provided $D=d+1$. 
\end{exercise}

We now consider in more detail at how the poles in the Mellin transformed correlator are related to the operators appearing in the OPE. Recall that in the small $x_{12}^2$ limit, the OPE has the form 
\begin{equation}
\mathcal{O}_1(x_1)\mathcal{O}_2(x_2)=\sum_k C_{12k} (x_{12}^2)^{\frac{\Delta_k-\Delta_1-\Delta_2}{2}}(\mathcal{O}_k(x_2)+c x_{12}^2\partial^2\mathcal{O}_k(x_2)+\ldots)\;,
\end{equation}
where on the RHS we considered only scalar operators for simplicity. It is clear that we have a series of powers $(x_{12}^2)^{\frac{\Delta_k-\Delta_1-\Delta_2}{2}+n}$ with $n=0,1,2,\ldots$. To reproduce these powers from the Mellin representation, we can close the $\delta_{12}$ contour and pick up the poles. This tells us that the integrand should have poles at 
\begin{equation}\label{poleindelta12}
\delta_{12}=\frac{\Delta_1+\Delta_2-\Delta_k}{2}-n\;.
\end{equation}

We can also understand the meaning more clearly in four-point functions which we will mostly focus on in later parts of these lectures. The general definition (\ref{defnptMellin}) becomes
\begin{equation}\label{defMellin4pt}
\begin{split}
\langle \mathcal{O}_1(x_1)\ldots \mathcal{O}_4(x_4)\rangle={}&\frac{1}{(x_{12}^2)^{\frac{\Delta_1+\Delta_2}{2}}(x_{34}^2)^{\frac{\Delta_3+\Delta_4}{2}}}\left(\frac{x_{14}^2}{x_{24}^2}\right)^{\frac{\Delta_2-\Delta_1}{2}}\left(\frac{x_{14}^2}{x_{13}^2}\right)^{\frac{\Delta_3-\Delta_4}{2}}\\
{}&\times \int_{-i\infty}^{i\infty}\frac{dsdt}{(4\pi i)^2}U^{\frac{s}{2}}V^{\frac{t}{2}-\frac{\Delta_3+\Delta_4}{2}}\mathcal{M}(s,t)\,\Gamma_{\Delta_1\Delta_2\Delta_3\Delta_4}(s,t,u)\;,
\end{split}
\end{equation}
where various $\delta_{ij}$ can be written in terms of the $s$, $t$, $u$ Mandelstam variables, e.g.,
\begin{equation}
\delta_{12}=\frac{\Delta_1+\Delta_2-s}{2}\;,
\end{equation}
and $s$, $t$, $u$ are subject to the constraint
\begin{equation}
s+t+u=\Delta_1+\Delta_2+\Delta_3+\Delta_4\;.
\end{equation}
We have also defined the Gamma function factor
\begin{equation}
\begin{split}
\Gamma_{\Delta_1\Delta_2\Delta_3\Delta_4}(s,t,u)={}&\Gamma\left(\frac{\Delta_1+\Delta_2-s}{2}\right)\Gamma\left(\frac{\Delta_3+\Delta_4-s}{2}\right)\Gamma\left(\frac{\Delta_1+\Delta_4-t}{2}\right)\\
{}&\times \Gamma\left(\frac{\Delta_2+\Delta_3-t}{2}\right) \Gamma\left(\frac{\Delta_1+\Delta_3-u}{2}\right)\Gamma\left(\frac{\Delta_2+\Delta_4-u}{2}\right)\;.
\end{split}
\end{equation}
The poles (\ref{poleindelta12}) in $\delta_{12}$ coming from the 12 OPE then translate to the following poles in $s$
\begin{equation}
s=\Delta_k+2n\;.
\end{equation}
This is similar to flat-space scattering where an exchanged particle of squared mass $m^2$ shows up in the amplitude as a pole at $s=m^2$.  The important difference here is that we now have infinitely many poles for each conformal family because there are infinitely many conformal descendants. For spinning operators in the OPE, the structure is the same and the conformal dimension $\Delta$ is replaced by the conformal twist $\tau=\Delta-\ell$\;.

\subsubsection{Witten diagrams in Mellin space}
To appreciate the usefulness of this formalism, let us consider again the tree-level Witten diagrams but now in Mellin space. Compared to position space, the analytic structure of the Witten diagrams becomes much simpler and the scattering nature becomes much more manifest. 

We start by considering the contact Witten diagrams. As we have seen in Section \ref{subsec:WDinpositionspace}, these contact Witten diagrams are proportional to  the conformal integrals $I_n$ (\ref{confintIn}) in flat space. Such integrals have been considered in the literature and admit a Mellin representation via the Symanzik formula \cite{Symanzik:1972wj}
\begin{equation}
I_n=\frac{\pi^\frac{D}{2}}{\prod_i\Gamma(\Delta_i)}(2\pi i)^{\frac{n(n-3)}{2}}\int [d\delta_{ij}]\prod_{i<j}\Gamma(\delta_{ij})x_{ij}^{-2\delta_{ij}}\;.
\end{equation}
This is precisely in the form of the Mellin representation (\ref{defnptMellin}). After taking into account the proportional constant, it follows that the Mellin amplitude of the contact Witten diagram with no derivatives is just the constant 
\begin{equation}
\mathcal{M}_{\rm con}=\frac{\pi^{\frac{d}{2}}\Gamma(\frac{\Sigma-d}{2})}{\prod_i\Gamma(\Delta_i)}\;.
\end{equation}
We can also try to understand this result in more detail. Recall in (\ref{confblockdecDfun}) the contact Witten diagram decomposes into only double-trace conformal blocks of which the twists are $\Delta_i+\Delta_j+2m$. This is true for any channel. Therefore, the Mellin integrand should have poles only at
\begin{equation}
\delta_{ij}=\frac{\Delta_i+\Delta_j-(\Delta_i+\Delta_j+2m)}{2}-n=-m-n\in \mathbb{Z}_{\leq 0}\;.
\end{equation}
The $\Gamma(\delta_{ij})$ factor in the definition precisely has these poles and therefore no additional poles are needed in the Mellin amplitude $\mathcal{M}(\delta_{ij})$. This is a general feature of the Mellin representation, namely the Gamma factor separates the  ubiquitous large $N$ multi-trace poles from the single-trace poles which are dynamically more interesting. More generally, we can also consider the Mellin amplitude of contact Witten diagrams with derivatives. If the number of contracted derivatives in the contact vertex is $2L$, the Mellin amplitude is a degree $L$ polynomial. 
\begin{exercise}
Explicitly show this is the case for the contact Witten diagram with two derivatives. 

\vspace{0.3cm}

{\bf Hint:} First use (\ref{idforGBpartialwder}) to write the contact Witten diagram with two derivatives in terms of $D$-functions. Then use the Mellin representation of $D$-functions to show the Mellin amplitude of the contact Witten diagram is a linear function of the Mellin-Mandelstam variables.
\end{exercise}

Let us now consider exchange Witten diagrams. In addition to the double-trace operators which are already captured by the Gamma function factor, there should also be a series of single-trace poles corresponding to the exchanged field. Let us consider the exchange in the s-channel and assume the exchanged field has dimension $\Delta$ and spin $\ell$. The four-point exchange Mellin amplitude has the following general form
\begin{equation}\label{Mellin4ptexchange}
\mathcal{M}_{\Delta,\ell}=\sum_{m=0}^\infty \frac{Q_{\ell,m}(t)}{s-(\Delta-\ell)-2m}+P_{\ell-1}(s,t)\;,
\end{equation}
where $Q_{\ell,m}(t)$ are degree-$\ell$ polynomials of $t$ and $P_{\ell-1}(s,t)$ is a degree-$(\ell-1)$ polynomial of $s$, $t$. The polynomials $Q_{\ell,m}(t)$ are called the Mack polynomials. Note that if we take the residues of the Mellin integral at these single-trace poles, we just obtain the conformal block $g_{\Delta,\ell}$ in position space.\footnote{If we take the residues at the Gamma function poles, we get the contributions from the double-trace conformal blocks.} In other words, conformal blocks and exchange Witten diagrams have the same poles and residues in Mellin space. This is a very useful property because it allows us to straightforwardly obtain Witten diagram recursion relations from those of conformal blocks \cite{Zhou:2020ptb}.

The computation of exchange Witten diagrams also becomes simpler in Mellin space thanks to the simpler analytic structure. In position space, we computed the exchange Witten diagrams by using the equation of motion of the bulk-to-bulk propagator which led to the identity 
\begin{equation}\label{EoMWDposition}
({\rm Cas}-C_{\Delta,\ell})W_{\Delta,\ell}=W_{\rm con}\;.
\end{equation}
We can use the same relation as the starting point in Mellin space. If we strip away the kinematic factor of $x_{ij}^2$ in (\ref{defMellin4pt}), the exchange Witten diagram becomes a function of the cross ratios $U$ and $V$. Then the Casimir operator also becomes a second order differential operator in these cross ratios that can be built out of the following basic operators
\begin{equation}
U\partial_U\;,\quad V\partial_V\;,\quad U^mV^n\;.
\end{equation}
These operators have simple interpretations when acting on the Mellin representation 
\begin{equation}\label{defMellingenab}
\int \frac{ds dt}{2\pi i} U^{\frac{s}{2}+a}V^{\frac{t}{2}+b}\mathcal{M}(s,t)\Gamma_{\Delta_1\Delta_2\Delta_3\Delta_4}(s,t,u)\;.
\end{equation}
Each translates into a different operator
\begin{equation}
U\partial_U\to \widehat{U\partial_U}\;,\quad V\partial_V \to \widehat{V\partial_V}\;,\quad U^mV^n\to \widehat{U^mV^n}\;,
\end{equation}
with the action
\begin{equation}\label{diffbbaction}
\begin{split}
{}&\widehat{U\partial_U}\circ \mathcal{M}(s,t)=  \left(\frac{s}{2}+a\right)\times \mathcal{M}(s,t) \;,\\
{}&\widehat{V\partial_V }\circ \mathcal{M}(s,t)= \left(\frac{t}{2}+b\right)\times \mathcal{M}(s,t)\;,\\
{}& \widehat{U^mV^n}\circ\mathcal{M}(s,t)=\mathcal{M}(s-2m,t-2n)\frac{\Gamma_{\Delta_1\Delta_2\Delta_3\Delta_4}(s-2m,t-2n,u+2m+2n)}{\Gamma_{\Delta_1\Delta_2\Delta_3\Delta_4}(s,t,u)}\;.
\end{split}
\end{equation}
In particular, to get the last expression, we absorb the multiplicative power $U^mV^n$ by shifting $s$ and $t$. But we should also take into account the action of the shifts on the Gamma factor which gives rise to the ratio. Using these rules, we can turn the differential equation (\ref{EoMWDposition}) into a difference equation for the Mellin amplitude
\begin{equation}\label{EoMWDMellin}
(\widehat{{\rm Cas}}-C_{\Delta,\ell})\mathcal{M}_{\Delta,\ell}=\mathcal{M}_{\rm con}\;.
\end{equation}
This equation can be solved pole by pole. Note that the contact Mellin amplitude on the RHS is a polynomial with no poles. On the LHS, the exchange Mellin amplitude has the structure of (\ref{Mellin4ptexchange}) but the poles get shifted into each other by the difference operator. Requiring the poles to vanish gives rise to recursion relations for the numerator polynomials $Q_{\ell,m}(t)$ which allow us to solve them efficiently. The key strategy here, namely translating the position space differential operators into Mellin space difference operators, is a trick which we will use again later in these lectures. 

An important comment we should make is that although the sum over poles in (\ref{Mellin4ptexchange}) is generically infinite, for special cases the series truncates and the exchange Mellin amplitude becomes a rational function. This happens precisely when the truncation conditions which we have seen in position space are met, i.e., when $\Delta_1+\Delta_2-\Delta+\ell$ or $\Delta_3+\Delta_4-\Delta+\ell$ are positive even integers. These conditions on the spectrum are satisfied for example in the supergravity limit of 4d $\mathcal{N}=4$ SYM and the 6d $\mathcal{N}=(2,0)$ theory, but are not satisfied in 3d $\mathcal{N}=8$ ABJM. The tree-level truncation can be viewed just as a fact of the computation outcome. But it can also be understood as a consistency condition of the large $N$ expansion, see \cite{Rastelli:2017udc} for the argument. 

\begin{exercise}
First obtain the Casimir operator as a differential operator and then translate it into Mellin space. Try to solve (\ref{EoMWDMellin}) for the case of scalar field exchange. In this case, there is no regular term $P$ and the numerators $Q_{0,m}$ are just constants.
\end{exercise}

From these discussions, it should be evident that the analytic structure of Mellin amplitudes is very similar to that of tree-level amplitudes in flat space. In a certain sense, we can regard the Mellin space as the ``momentum space'' in AdS and import many of our intuitions from flat space. For example, an important property of  amplitudes in flat space is factorization and it is the statement that the residues at the poles of amplitudes factorize into lower-point amplitudes. This property has a natural analogue in Mellin space and is essentially dictated by the OPE. If we take the residue of the Mellin amplitude at a pole, the residue can also be expressed in terms of the lower-point Mellin amplitudes in a way that is schematically a product. See \cite{Goncalves:2014rfa} for details. 

Another interesting feature of the Mellin amplitude is the flat-space limit. If the scattering process in AdS has high enough energy, we can expect that the curvature effect is negligible and we therefore should recover the flat-space scattering amplitude. This can be made precise and it is particularly simple to do it in Mellin space where the flat-space limit is just the high energy limit of the Mellin amplitude. The relation turns out to further involve a Borel transform \cite{Penedones:2010ue}
\begin{equation}\label{Mellinflatspacelim}
\mathcal{M}(\delta_{ij})\propto \int_0^\infty d\beta \beta^{\frac{1}{2}\sum_i\Delta_i-\frac{d}{2}-1}e^{-\beta}\mathcal{A}_{\rm flat}\left(S_{ij}=\frac{2\beta}{R^2}\delta_{ij}\right)\;,\quad R\to\infty\;,
\end{equation}
where we have omitted the precise coefficient for simplicity. For simple Mellin amplitudes such as the tree-level exchange amplitudes, the Borel transform is quite trivial and just multiplies the high energy limit of the Mellin amplitude by a number.

\section{Bootstrapping Holographic Correlators}\label{Sec:bootstrap}

Starting from this section, we will discuss how to use the bootstrap strategy to compute the holographic correlators. We will mainly focus on the prototypical example of AdS/CFT, namely four dimensional $\mathcal{N}=4$ SYM dual to Type IIB superstring theory in AdS$_5\times$S$^5$. However, before we go into the details of the bootstrap strategy, we begin with a quick review of the duality setup and explain why computing holographic correlators by brute force is an intimidating task. 

\subsection{Review of the duality and why brute force computation is difficult}

On one side of the duality, we have 4d $\mathcal{N}=4$ SYM.  A convenient way to obtain its Lagrangian is to dimensionally reduce the 10d $\mathcal{N}=1$ SYM to $\mathbb{R}^4$. Explicitly, the Lagrangian contains three basic fields $A_\mu$, $\Psi_a$, $\Phi^I$ and takes the form
\begin{equation}
\begin{split}
L={}&\frac{1}{g_{YM}^2}{\rm tr}\bigg[\frac{1}{4}F^{\mu\nu}F_{\mu\nu}+\frac{1}{2}(D^\mu\Phi^I)^2+\bar{\Psi}^a\sigma^\mu D_\mu \Psi_a\\
{}&\quad\quad\quad-\frac{1}{4}[\Phi^I,\Phi^J]^2-C^{ab}_I\Psi_a[\Phi^I,\Psi_b]-\bar{C}_{Iab}\bar{\Psi}^a[\Phi^I,\bar{\Psi}^b]\bigg]\;.
\end{split}
\end{equation}
All the fields are in the adjoint representation with respect to the $SU(N)$ gauge group. The theory has an $SO(6)$ R-symmetry group. In particular, the six scalars $\Phi^I$ are rotated into each other as a vector by $SO(6)$. The theory is not only maximally supersymmetric, meaning it has sixteen supercharges, but also has a zero beta function. This makes the theory superconformal. 

The symmetry is matched by the dual side where the bulk theory has the geometry of AdS$_5\times$S$^5$. In particular, for the bosonic symmetries the conformal group $SO(4,2)$ symmetry corresponds to the isometry of AdS$_5$ as we have already seen, and the $SO(6)$ R-symmetry is naturally associated to the S$^5$ isometry. 

\begin{figure}[h]
    \centering
    \begin{subfigure}[b]{0.23\textwidth}
        \centering
        \includegraphics[width=\textwidth]{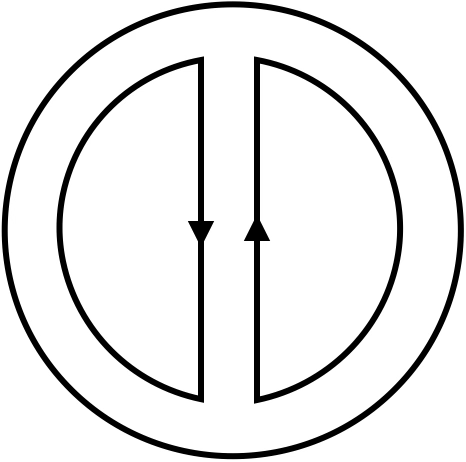}
        \caption{A planar diagram}
        \label{fig:planardiagram}
    \end{subfigure}
    \hspace{0.1\textwidth}
    \begin{subfigure}[b]{0.3\textwidth}
        \centering
        \includegraphics[width=\textwidth]{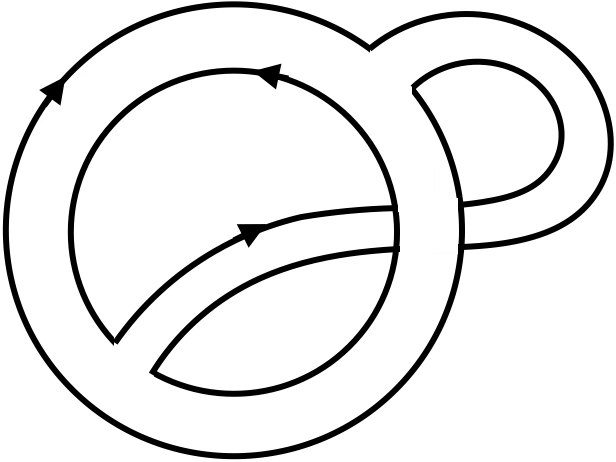}
        \caption{A non-planar diagram}
        \label{fig:nonplanardiagram}
    \end{subfigure}
    \caption{Feynman diagrams in the 't Hooft limit are organized by their topology. The non-planar diagram \ref{fig:nonplanardiagram} is suppressed by $1/N^2$ compared to the planar diagram \ref{fig:planardiagram}.}
    \label{fig:planarandnonplanar}
\end{figure}

In the field theory, a useful organization principle is the large $N$ expansion of 't Hooft. We define the 't Hooft coupling $\lambda=g_{YM}^2N$. In terms of the Feynman diagram expansion, we can draw propagators and vertices in the double line notation for the color indices where each propagator is
\begin{equation}
\langle \Phi^i_j \Phi^k_l\rangle\propto \frac{\lambda}{N}\delta^i_l\delta^k_j\;,
\end{equation}
and each vertex contributes a factor of $N/\lambda$. For a diagram with $V$ vertices, $E$ propagators and $F$ loops, we get a factor
\begin{equation}
\left(\frac{N}{\lambda}\right)^V\left(\frac{\lambda}{N}\right)^EN^F=N^\chi\lambda^{E-V}\;,
\end{equation}
where $\chi=V+F-E=2-2g$. Here we take the limit $N\to\infty$ while keeping $\lambda$ finite. It is easy to see that the large $N$ limit organizes the diagrams by topology and the theory is dominated by the subset of diagrams with the smallest $g$, see Figure \ref{fig:planarandnonplanar}.

The large $N$ simplification can also be seen on the string theory side. The string theory parameters are related to the field theory ones by 
\begin{equation}\label{parameterrelation}
g_{YM}^2=4\pi g_s\;,\quad \frac{R^4}{\ell_s^4}=g_{YM}^2N=\lambda\;,
\end{equation} 
where $g_s$ is the string coupling, $\ell_s$ is the string length and $R$ is the radius of AdS$_5$ (also the radius of S$^5$). In the 't Hooft limit, $g_s$ is very small. Therefore, in the planar limit we only need to consider tree-level processes in AdS and can ignore the string loops. 

In this large $N$ limit, we can further consider different regimes of the 't Hooft coupling $\lambda$. When $\lambda$ is small, observables can in principle be computed by summing over planar Feynman diagrams in the field theory. The large $\lambda$ regime is nonperturbative and is difficult for the field theory. However, on the dual AdS side, large $\lambda$ corresponds to $\ell_s\ll R$ which means excited string states are very heavy. Therefore, in the $\lambda\to\infty$ limit we can ignore all the higher string states and Type IIB superstring theory reduces to weakly coupled Type IIB supergravity. This is the limit which we will focus on in computing the holographic correlators. 

Since the AdS theory is weakly coupled and the leading contribution consists of only tree-level Witten diagrams, it might seem that the AdS perturbation theory discussed in Section \ref{Sec:AdSpertth} has already equipped us with all the necessary techniques to compute the holographic correlators. However, going through the literature, one finds that explicit results of holographic correlators computed using the standard diagrammatic algorithm remained extremely limited even after two decades since the discovery of AdS/CFT \cite{DHoker:1999pj,Arutyunov:2000py,Arutyunov:2002ff,Arutyunov:2002fh,Arutyunov:2003ae,Dolan:2006ec,Berdichevsky:2007xd,Uruchurtu:2008kp,Uruchurtu:2011wh}. This was certainly not due to a lack of trying. But then why are the computations of holographic correlators so difficult?

There are several sources of difficulties. A major reason is that in this theory we have infinitely many particles in AdS. These infinitely many particles come from the Kaluza-Klein (KK) reduction of the 10d supergravity field fluctuations on the internal S$^5$.\footnote{It is tempting to ask if one can do computations directly in ten dimensions. This has not been possible so far, but we will see hints of higher dimensional structures later in this section.} As a result, we have infinitely many correlators to compute even after we fix the number of operators in the correlator. The interactions among the infinitely many particles are conceivably very complicated. In a heroic effort, the complete cubic and quartic vertices were worked out explicitly for AdS$_5\times$S$^5$ Type IIB supergravity in \cite{Arutyunov:1999fb}, which are needed for computing general four-point functions at tree level. But the result for these vertices filled 14 pages! This is definitely bad news for the diagrammatic algorithm since the Feynman rules are only the starting point. Moreover, even though for a four-point correlator with fixed external operators the number of Witten diagrams is finite, the diagrams proliferate quickly as we increase the dimensions of the external operators. This happens because the conformal dimensions are correlated with the KK levels on S$^5$ which are essentially the angular momenta (or R-symmetry charge). As we consider heavier external operators, more intermediate channels for angular momentum become available and consequently we have increasingly many exchange Witten diagrams. Also the contact interactions become more complicated because there are more allowed R-symmetry structures. These lead to increasingly lengthy expressions in terms of $D$-functions which are highly nontrivial to simplify. 

\subsection{Bootstrap using symmetries and consistency}
A better strategy for computing these holographic correlators is bootstrap: we will obtain them directly from symmetries and consistency conditions, bypassing altogether all the complicated details of the effective Lagrangian. 

Before explaining any details of the bootstrap strategy, let us first discuss briefly why it is not unreasonable to expect such an approach to work. First, we are inspired by the similar success in flat space. An important lesson from flat space is that on-shell observables are simple while off-shell objects can be very complicated. The vertices, or the effective action, are off-shell objects. They are not even uniquely determined because their form changes under field redefinitions. Even in bosonic gravity in flat space, expanding the Einstein-Hilbert action leads to increasingly many terms. The complexity of the action however does not prevent the flat-space graviton amplitudes, which are the on-shell observables, from having a simple form. As we have argued in the introduction and seen in explicit details in Section \ref{Sec:AdSpertth}, holographic correlators are the natural analogue of on-shell scattering amplitudes in AdS. Therefore, despite the 14 pages of complicated vertices, the holographic correlators may still have nontrivial hidden simplicity which allows them to be efficiently constructed. Second, $\mathcal{N}=4$ SYM is a theory with the maximal amount of allowed supersymmetry and is commonly believed to be unique up to the choice of the gauge group. This uniqueness suggests that we might be able to  fix the correlators entirely from the symmetries of the theory and consistency conditions, without needing to go through the conventional workflow via the action. Both points turn out to be true as we will see. However, the success of the bootstrap approach is not restricted to $\mathcal{N}=4$ SYM and extends to other theories as well. 

In what follows, we will set the stage for the bootstrap calculation by first discussing the kinematics of the holographic correlators in $\mathcal{N}=4$ SYM. Then in the rest of this section, we will show how to bootstrap the holographic correlators in three different ways. Each method has its own pros and cons as we will point out. 

We are going to focus on the following $\frac{1}{2}$-BPS operators\footnote{These operators are called $\frac{1}{2}$-BPS because they are annihilated by half of the supercharges which is also the maximal amount possible. Another subtlety which we will ignore for pedagogical reasons is that the supergravity modes are not really dual to the single-trace operators (\ref{defOkop}) but there are also $1/N$ suppressed multi-trace operators. This amounts to a change of basis and it affects the near extremal correlators. See \cite{Arutyunov:1999en,Arutyunov:2000ima,Aprile:2019rep,Alday:2019nin,Aprile:2020uxk} for details.}
\begin{equation}\label{defOkop}
\mathcal{O}_k^{I_1\ldots I_k}(x)={\rm tr}(\Phi^{\{I_1}(x)\ldots \Phi^{I_k\}}(x))\;.
\end{equation}
These are scalar operators with protected conformal dimension $k$ and transform in the rank-$k$ symmetric traceless representation of the $SO(6)$ R-symmetry. In the supergravity limit, the supergravity KK modes are organized by superconformal symmetry into short $\frac{1}{2}$-BPS multiplets. These $\frac{1}{2}$-BPS operators $\mathcal{O}_k^{I_1\ldots I_k}(x)$ are the superconformal primaries of the multiplets and they are dual to scalar fields $s_k$ in AdS$_5$ which we will refer to as supergravitons\footnote{This is to distinguish from the real spin-2 gravitons and their higher KK analogues.}. The multiplets contain superconformal descendant operators which are dual to KK modes of other fields in AdS. In Table \ref{tab:kkexch} we list the component fields with their quantum numbers for those which will be relevant for the computation of holographic correlators.  

\begin{table}[ht]
    \centering
    \renewcommand{\arraystretch}{1.3}
    \begin{tabular}{c c c c c}
        \hline
        Field & spin ($ \ell $) & dimension ($\Delta$) & twist ($\Delta - \ell$) & $SU(4)_R$ representation \\
        \hline
        $s_k$ & $0$ & $k$ & $k$ & $[0,k,0]$ \\
        $A_{\mu,k}$ & $1$ & $k+1$ & $k$ & $[1,k-2,1]$ \\
        $C_{\mu,k}$ & $1$ & $k+3$ & $k+2$ & $[1,k-4,1]$ \\
        $\phi_k$ & $0$ & $k+2$ & $k+2$ & $[2,k-4,2]$ \\
        $t_k$ & $0$ & $k+4$ & $k+4$ & $[0,k-4,0]$ \\
        $\varphi_{\mu\nu,k}$ & $2$ & $k+2$ & $k$ & $[0,k-2,0]$ \\
        \hline
    \end{tabular}
    \caption{List of component fields which can appear in the exchanges and their quantum numbers. The R-symmetry representations are denoted by the $SU(4)$ Dynkin labels, and the component does not exist of any number in the Dynkin label is negative.}
    \label{tab:kkexch}
\end{table}

We will be interested in computing the correlation functions of the $\frac{1}{2}$-BPS operators (\ref{defOkop}). One reason for considering these operators over others is technical simplicity. The kinematics of correlators of scalars are in general simpler than spinning operators. Moreover, the $\frac{1}{2}$-BPS property of these operators also gives us more control from superconformal symmetry. However, the information captured by these correlators is nontrivial, and in fact for four-point functions all correlators of different superconformal descendants can be obtained from the super primary ones by using superconformal symmetry.

A nice trick to keep track of the many R-symmetry indices of $\mathcal{O}_k$ is to contract them with null polarization vectors
\begin{equation}\label{defOkwt}
\mathcal{O}_k(x,t)=\mathcal{O}_k^{I_1\ldots I_k}(x)t^{I_1}\ldots t^{I_k}={\rm tr}(\Phi\cdot t)^k\;,
\end{equation}
where $t\cdot t=0$. It is clear that anything multiplied with $t^{I_1}\ldots t^{I_k}$ gets symmetrized in the indices. Moreover, the null property makes sure that the traces are zero and therefore the operator gets projected correctly to the symmetric traceless representation. In this way, the $\frac{1}{2}$-BPS operators not only depends on the spacetime coordinates $x$ but also on the ``internal coordinates'' $t$. To restore the indices, we just need to take derivatives with respect to $t$. 

Our target in this section will be the four-point functions of $\mathcal{O}_k(x,t)$. For the simplicity of the demonstration, we will consider the equal weight case with $k_i=k$
\begin{equation}
G_{kkkk}(x_i,t_i)=\langle \mathcal{O}_k(x_1,t_1)\ldots \mathcal{O}_k(x_4,t_4) \rangle\;,
\end{equation}
although it is not difficult to extend to the case where all the operators have unequal weights. In Section \ref{subsec:corrfun} we have seen that conformal symmetry allows us to express a four-point function as a function of two conformal cross ratios. The same is true also for R-symmetry. More precisely, we need to extract a kinematic factor and then we can write the correlator as
\begin{equation}\label{defcalG}
G_{kkkk}(x_i,t_i)=\left(\frac{t_{12}}{x_{12}^2}\right)^k\left(\frac{t_{34}}{x_{34}^2}\right)^k\mathcal{G}_{kkkk}(U,V;\sigma,\tau)\;,
\end{equation}
where 
\begin{equation}
U=\frac{x_{12}^2x_{34}^2}{x_{13}^2x_{24}^2}\;,\quad V=\frac{x_{14}^2x_{23}^2}{x_{13}^2x_{24}^2}\;,\quad \sigma=\frac{t_{13}t_{24}}{t_{12}t_{34}}\;,\quad \tau=\frac{t_{14}t_{23}}{t_{12}t_{34}}\;,
\end{equation}
with $t_{ij}\equiv t_i\cdot t_j$. Note that the R-symmetry polarization tensors $t_i$ enter the operators (\ref{defOkwt}) in a multiplicative way. As a consequence, they can only appear in $G_{kkkk}$ as polynomials of $t_{ij}$ to ensure R-symmetry invariance, and each $t_i$ must appear precisely $k$ times. It follows that $\mathcal{G}_{kkkk}$ is a polynomial of $\sigma$ and $\tau$ of degree $k$. 

In addition to the bosonic symmetry subgroups which have allowed us to write the correlators in terms of cross ratios, the fermionic generators of the superconformal group impose further constraints. These constraints relate the dependence of the correlator on the conformal cross ratios with the R-symmetry dependence. It is convenient to make a change of variables
\begin{equation}
U=z\bar{z}\;,\quad V=(1-z)(1-\bar{z})\;,\quad \sigma=\alpha\bar{\alpha}\;,\quad \tau=(1-\alpha)(1-\bar{\alpha})\;.
\end{equation}
These additional constraints are known as the superconformal Ward identities \cite{Eden:2000bk,Nirschl:2004pa}
\begin{equation}\label{scfWardidNeq4}
(z\partial_z-\alpha\partial_\alpha)\mathcal{G}_{kkkk}(z,\bar{z};\alpha,\bar{\alpha})\big|_{\alpha=1/z}=0\;,
\end{equation}
where the differential operator acts on the correlator before $\alpha=1/z$ is imposed. These are all the kinetic constraints on the correlators that come from symmetries alone. In what follows, we will introduce three different methods to bootstrap these holographic correlators. 

\subsection{First method: position space}\label{Subsec:posispamethod}
The discussion of the superconformal kinematics in the previous subsection applies to four-point correlators of $\frac{1}{2}$-BPS operators at any value of the coupling. To perform bootstrap style calculations, we must also input at least some physical information so that we can specify we are working in the supergravity limit. In the first implementation of this idea, we will be less ambitious and stay close to the traditional diagrammatic method. We will still explicitly compute the Witten diagrams but we will ask the following important question: can the precise details of the KK mode interactions be determined entirely from symmetries? If this turns out to be possible, it would tell us that the bootstrap idea is working and we can avoid inputting the many pages of interaction vertices which are difficult to obtain in the first place. 

More precisely, our strategy is to write down an ansatz for the correlator in terms of all the possible Witten diagrams. However, the coefficients of these diagrams are not fixed. Instead of using the precise vertices of the effective action, our plan is to use superconformal symmetry to determine these coefficients. This is the position space method which was first proposed in \cite{Rastelli:2017udc}.

Let us demonstrate this algorithm with the simplest example where the $\frac{1}{2}$-BPS operator has the lowest KK level $k_i=2$ and corresponds to the stress tensor multiplet. We can split the ansatz into two parts, namely an exchange part and a contact part, in parallel with the structure of the Witten diagram expansion
\begin{equation}
A=A_{\rm exchange}+A_{\rm contact}\;.
\end{equation}
The exchange part can further be written as the sum of three channels
\begin{equation}
A_{\rm exchange}=A_{s}+A_{t}+A_{u}\;.
\end{equation}
Because the three channels are related by crossing, we can just focus on the s-channel. Furthermore, there are only three fields in the $k=2$ multiplet that can be exchanged. This follows more generally from two types of consistency conditions. The first is the R-symmetry selection rule, namely the exchanged field must carry R-symmetry representations which appear in the tensor product of the two external representations. Note this already guarantees that the number of exchanges in a given correlator is finite. The second is the non-extremality condition, requiring the twist of the exchanged field in the s-channel to satisfy $\Delta-\ell< \min\{\Delta_1+\Delta_2,\Delta_3+\Delta_4\}$. The reason for this condition is that the Witten diagram integrals for such extremal three-point functions (when the equality is taken) are divergent and the couplings must vanish for the effective action to be finite.\footnote{Alternatively, this can be understood as the requirement that the two-particle states should be orthogonal to the single-particle states.} 
Explicitly, we have
\begin{equation}\label{ansatzAs}
A_s=\lambda_s Y_s W_{2,0}+\lambda_V Y_V W_{3,1}+\lambda_\varphi Y_\varphi W_{4,2}\;.
\end{equation}
Here $W_{\Delta,\ell}$ are the exchange Witten diagrams, which are evaluated using the method used in Section \ref{Subsubsec:exchinposi} (after generalizing it to include exchanging spinning fields)
\begin{equation}
\begin{split}
W_{2,0}={}&\frac{1}{4x_{12}^2}D_{1122}\;,\\
W_{3,1}={}&\frac{1}{2x_{12}^2}(x_{13}^2D_{2132}-x_{14}^2D_{2123}+x_{24}^2D_{1223}-x_{34}^2D_{1232})\;,\\
W_{4,2}={}&-\frac{8}{3x_{12}^2}(D_{1122}-3(x_{13}^2D_{2132}+x_{14}^2(D_{2123}-2x_{13}^2D_{3133})))\;.
\end{split}
\end{equation}
The factors $Y_s$, $Y_V$, $Y_\varphi$ are the analogues of exchange Witten diagrams on S$^5$, or more precisely, the analogues of conformal blocks. Each field carries a specific irrep under $SO(6)$ R-symmetry and its contribution in the exchange is captured by a specific polynomial of the R-symmetry cross ratios. These polynomials can be solved from the R-symmetry version of the Casimir equation and read
\begin{equation}
\begin{split}
Y_s={}&(t_1\cdot t_2)^2(t_3\cdot t_4)^2\left(\sigma+\tau-\frac{1}{3}\right)\;,\\
Y_V={}&(t_1\cdot t_2)^2(t_3\cdot t_4)^2(\sigma-\tau)\;,\\
Y_\varphi={}&(t_1\cdot t_2)^2(t_3\cdot t_4)^2\;.
\end{split}
\end{equation}
Finally, the coefficients $\lambda_s$, $\lambda_V$, $\lambda_\varphi$ for each exchange contribution are unknowns. For the contact part of the ansatz, we assume that the vertices either contain zero or two derivatives. Higher derivatives are forbidden because such contact Witten diagrams survive the flat-space limit and this would contradict the fact that supergravity is a two-derivative theory in flat space. Therefore, $A_{\rm contact}$ is constructed from $D_{2222}$ and $x_{34}^2 D_{2233}$ with its crossing images. On the other hand, we will assume the R-symmetry dependence to be completely general which means these $D$-functions are multiplied by general degree-2 polynomials of $\sigma$ and $\tau$ with unknown coefficients. 

The next step is to evaluate this ansatz. We have already proven in (\ref{Dfundiffrecur}) that the $D$-functions satisfy the weight-shifting relations
\begin{equation}
D_{\Delta_1\ldots \Delta_i+1\ldots \Delta_j+1\ldots \Delta_n}(x_i)=\frac{d-\Sigma}{2\Delta_i\Delta_j}\frac{\partial}{\partial x_{ij}^2}D_{\Delta_1\ldots\Delta_n}(x_i)\;.
\end{equation}
For our case, it is obvious that all the $D$-functions can be recursively generated from $D_{1111}$ in this way. On the other hand, we also recall from (\ref{Dbar1111aspolylogs}) that $D_{1111}$ can be explicitly evaluated in terms of polylogarithms as
\begin{equation}
x_{13}^2x_{24}^2 D_{1111}\propto \Phi(z,\bar{z})=\frac{1}{(z-\bar{z})}\left(2{\rm Li}_2(z)-2{\rm Li}_2(\bar{z})+\log (z\bar{z})\log\left(\frac{1-z}{1-\bar{z}}\right)\right)\;.
\end{equation}
Using this expression, one can easily prove the following recursion relations
\begin{equation}
\begin{split}
\partial_z\Phi={}&-\frac{\Phi}{z-\bar{z}}+\frac{\log U}{(z-1)(z-\bar{z})}-\frac{\log V}{z(z-\bar{z})}\;,\\
\partial_{\bar{z}}\Phi={}&\frac{\Phi}{z-\bar{z}}-\frac{\log U}{(\bar{z}-1)(z-\bar{z})}+\frac{\log V}{\bar{z}(z-\bar{z})}\;.
\end{split}
\end{equation}
These relations allow us to show that the ansatz, after extracting a kinematic factor as in (\ref{defcalG}), can be expanded in a basis spanned by $\Phi$, $\log U$, $\log V$ and $1$
\begin{equation}
\mathcal{A}=R_\Phi \Phi+ R_U \log U+R_V \log V+R_1\;.
\end{equation}
Here the coefficient functions are rational functions of $z$, $\bar{z}$, $\alpha$, $\bar{\alpha}$.

The last step is to impose the superconformal Ward identities. It is easy to see that the LHS of (\ref{scfWardidNeq4}) can also be expanded in the same basis
\begin{equation}
(z\partial_z-\alpha\partial_\alpha)\mathcal{A}\big|_{\alpha=1/z}=R'_\Phi \Phi+ R'_U \log U+R'_V \log V+R'_1\;.
\end{equation}
The superconformal Ward identities then require all the new rational coefficient functions $R'_\Phi$, $R'_U$, $R'_V$, $R'_1$ to be zero, which simply translates to a set of over-determined linear equations for the unknowns in the ansatz. These linear equations solve all the unknown coefficients up to an overall constant which can be fixed by any protected three-point function. In this way, we have completely determined the four-point function in position space. 

\begin{exercise}
Explicitly carry out the position space bootstrap calculation as outlined for the $k=2$ correlator. Then show that the holographic correlator becomes a holomorphic function depending only on $z$ and $\alpha$ after setting $\bar{\alpha}=1/\bar{z}$. The same also happens to the free theory correlator which gives identical holomorphic function.  The holomorphic function is in fact the correlator of a chiral algebra which is a closed sector of theory \cite{bllprv13}. This chiral algebra correlator is coupling independent and therefore can be used as another way to fix the overall constant. 
\end{exercise}

Let us conclude the discussion of the first bootstrap method by making a few comments. First of all, this position space method confirms that our bootstrap idea is valid. In particular, the complicated interaction details do not need to be inputted by hand, but rather can be fixed by superconformal symmetry. Second, the strategy of expanding holographic correlators into diagrams and then solving their coefficients using symmetry is clearly not specific to $\mathcal{N}=4$ SYM, and can be straightforwardly applied to other theories. In particular, it can be used in the 6d $\mathcal{N}=(2,0)$ theory in the large $N$ limit which is dual to 11d supergravity in AdS$_7\times$S$^4$ \cite{Rastelli:2017ymc} because there the exchange Witten diagrams also truncate to finitely many $D$-functions. However, it is difficult to apply the algorithm to the 3d $\mathcal{N}=8$ ABJM theory which is dual to 11d supergravity in AdS$_4\times$S$^7$ because the spectrum of the theory does not lead to truncations. Finally, here we have demonstrated the method with the correlator of operators with the lowest KK level but the algorithm also works for higher KK levels $k_i$. However, the method also becomes cumbersome because the size of the ansatz expression grows quickly as we increase $k_i$.

\subsection{Second method: Mellin space}\label{Subsec:mellinspacemethod}
We have seen in Section \ref{Subsec:Mellinrepresentation} that the Mellin representation greatly simplifies the analytic structure of AdS Witten diagrams. Therefore, it is natural to ask if the position space bootstrap can be performed in Mellin space instead, which can potentially be much more efficient. Here we explain how this can be done, following the discussion in \cite{Zhou:2017zaw}. 

The structure of the algorithm requires no major changes. We start from the same ansatz which is a sum of all possible Witten diagrams with unfixed coefficients, but we replace the position space Witten diagrams with their Mellin amplitudes. The key difference is in the last step where we implement the superconformal constraints. We need to translate the position space superconformal Ward identities into Mellin space. To see why this translation appears to be challenging, we note that in the LHS of (\ref{scfWardidNeq4}) $z$ and $\bar{z}$ appear asymmetrically. On the other hand, in the Mellin representation (\ref{defMellingenab})
\begin{equation}
\int \frac{ds dt}{2\pi i} U^{\frac{s}{2}+a}V^{\frac{t}{2}+b}\mathcal{M}(s,t)\Gamma_{\Delta_1\Delta_2\Delta_3\Delta_4}(s,t,u)\;,
\end{equation}
only the variables $U$ and $V$ enter. Inside the cross ratios $U$ and $V$, $z$ and $\bar{z}$ are symmetric and the invariance under the $z\leftrightarrow\bar{z}$ exchange is a parameterization ambiguity. This appears to give rise to an obstruction for the translation. But this problem can be easily resolved by noticing the following trick. We can exploit the $z\leftrightarrow \bar{z}$ symmetry of the correlator to write down another copy of the superconformal Ward identity
\begin{equation}
(\bar{z}\partial_{\bar{z}}-\alpha\partial_\alpha)\mathcal{G}_{kkkk}(z,\bar{z};\alpha,\bar{\alpha})\big|_{\alpha=1/\bar{z}}=0\;.
\end{equation}
If we now take the sum or the difference (and further divide by $z-\bar{z}$) of these two identities, the $z\leftrightarrow \bar{z}$ symmetry is restored. This amounts to separating the even and odd parts. Even better, we find that the action of the differential operator can be written in terms of $U\partial_U$, $V\partial_V$ and the monomials $U^mV^n$ , see Exercise \ref{ex:sumdiff}. On the other hand, we recall from (\ref{diffbbaction}) that we already know how to interpret them as difference operators in Mellin space. This means that the superconformal Ward identities in position space become difference equations in Mellin space. Solving these difference equations is straightforward and allows us to fix the unknowns in the ansatz.

This Mellin space method turns out to be the more efficient strategy with wider applicability. The reason is that we can write down closed form Mellin expressions for the exchange Witten diagrams even when they fail to truncate into finitely many $D$-functions. An important case we mentioned is 3d $\mathcal{N}=8$ ABJM which is dual to 11d supergravity in AdS$_4\times$S$^7$ when $N$ is large. The superconformal Ward identities take the same form as in (\ref{scfWardidNeq4}) except that the relative coefficient of the two terms is different. This implies that we can use the same trick to translate the superconformal Ward identities. Finally, to be able to  compute all the correlators in the theory, one can also exploit certain simplifying limits related to special R-symmetry configurations to build a more efficient algorithm \cite{Alday:2020lbp,Alday:2020dtb}. Here we will not digress further with this development. But we refer to Section 11.3 of \cite{Bissi:2022mrs} for a pedagogical discussion and Exercise \ref{ex:MRV} to see the simplification in such a limit. 

\begin{exercise}\label{ex:sumdiff}
Show that by taking the sum and difference of the superconformal Ward identities, we only encounter the following combinations
\begin{equation}
\zeta^{(n)}_+=z^n+\bar{z}^n\;,\quad \zeta^{(n)}_-=\frac{z^n-\bar{z}^n}{z-\bar{z}}\;.
\end{equation} 
Then show these combinations always reduce to polynomials of $U$ and $V$.
\end{exercise}

\begin{exercise}\label{ex:Mellinkeq2}
Redo the $k_i=2$ bootstrap in Mellin space for $\mathcal{N}=4$ SYM and obtain the Mellin amplitude.
\end{exercise}

\begin{exercise}\label{ex:MRV}
Show that the $k_i=2$ Mellin amplitude simplifies significantly after setting $t_1=t_3$ for the R-symmetry polarization vectors. In particular, you should see there are no poles in $u$ and the poles in $s$ and $t$ are accompanied by zeros in $u$. 
\end{exercise}

\subsection{Third method: as an algebraic problem}\label{Subsec:algebraicbootstrap}
Quite remarkably, for 4d $\mathcal{N}=4$ SYM we can even obtain all the tree-level four-point holographic correlators without computing a single Witten diagram. This can be achieved by using the algebraic bootstrap method of \cite{Rastelli:2016nze,Rastelli:2017udc} and this is the last method which we will introduce in this section. 

In this approach we will take a different starting point. Instead of imposing the superconformal Ward identities as a final constraint, we start by solving them. The solution splits into two parts \cite{Eden:2000bk,Nirschl:2004pa}
\begin{equation}\label{solscfWardidNeq4}
G_{k_1k_2k_3k_4}=G_{{\rm free}, k_1k_2k_3k_4}+{\rm R} H_{k_1k_2k_3k_4}\;.
\end{equation}
Here $G_{{\rm free}, k_1k_2k_3k_4}$ is the correlator in the free theory, ${\rm R}$ is a factor determined by superconformal symmetry
\begin{equation}
{\rm R}=(t_1\cdot t_2)^2(t_3\cdot t_4)^2 x_{13}^4x_{24}^4(1-z\alpha)(1-z\bar{\alpha})(1-\bar{z}\alpha)(1-\bar{z}\bar{\alpha})\;,
\end{equation}
and $H_{k_1k_2k_3k_4}$ is a reduced correlator which captures all the dynamical information. We have restored the extracted kinematic factors to make crossing symmetry manifest. This form of the decomposition is also known in the literature as partial non-renormalization. An important feature of this solution is that $H_{k_1k_2k_3k_4}$ behaves just like an ordinary four-point correlator, except its conformal dimensions are shifted from $k_i$ to $k_i+2$, and R-symmetry charges are shifted from $k_i$ to $k_i-2$. 

We then want to translate the solution (\ref{solscfWardidNeq4}) into Mellin space. On the LHS, we have already defined the Mellin amplitude
\begin{equation}\label{defGgenMellin}
G_{k_1k_2k_3k_4}=\int_{-i\infty}^{i\infty}\frac{dsdt}{(4\pi i)^2} K(x_{ij}^2;s,t,u)\mathcal{M}_{k_1k_2k_3k_4}(s,t,u;t_{ij})\Gamma_{\{k_i\}}(s,t,u)\;,
\end{equation}
where 
\begin{equation}
K(x_{ij}^2;s,t,u)=(x_{12}^2)^{\frac{s-k_1-k_2}{2}}(x_{34}^2)^{\frac{s-k_3-k_4}{2}}(x_{14}^2)^{\frac{t-k_1-k_4}{2}}(x_{23}^2)^{\frac{t-k_2-k_3}{2}}(x_{13}^2)^{\frac{u-k_1-k_3}{2}}(x_{24}^2)^{\frac{u-k_2-k_4}{2}}\;,
\end{equation}
and 
\begin{equation}
\Gamma_{\{k_i\}}(s,t,u)=\Gamma(\tfrac{k_1+k_2-s}{2})\Gamma(\tfrac{k_3+k_4-s}{2})\Gamma(\tfrac{k_1+k_4-t}{2})\Gamma(\tfrac{k_2+k_3-t}{2})\Gamma(\tfrac{k_1+k_3-u}{2})\Gamma(\tfrac{k_2+k_4-u}{2})\;,
\end{equation}
with $s+t+u=k_1+k_2+k_3+k_4$. We can similarly define a reduced Mellin amplitude from $H_{k_1k_2k_3k_4}$
\begin{equation}\label{defHgenMellin}
H_{k_1k_2k_3k_4}=\int_{-i\infty}^{i\infty}\frac{dsdt}{(4\pi i)^2} K(x_{ij}^2;s,t,\tilde{u})\widetilde{\mathcal{M}}_{k_1k_2k_3k_4}(s,t,\tilde{u};t_{ij})\Gamma_{\{k_i\}}(s,t,\tilde{u})\;,
\end{equation}
where we have $s+t+\tilde{u}=k_1+k_2+k_3+k_4-4$ to account for the shift in weights caused by ${\rm R}$, and note that the new Gamma function factor is also different. On the other hand, the free correlator $G_{{\rm free},k_1k_2k_3k_4}$ is a rational function and does not have a well defined Mellin amplitude. It is most convenient to define its amplitude to be zero, and the free correlator then can be produced from $0\times \infty$ effect of the contour pinching. Comparing both sides, the multiplicative factor ${\rm R}$ can be treated similarly as a difference operator $\widehat{\rm R}$ as we have done before. This gives the following relation
\begin{equation}\label{MandMtilde}
\mathcal{M}_{k_1k_2k_3k_4}=\widehat{\rm R}\circ\widetilde{\mathcal{M}}_{k_1k_2k_3k_4}\;.
\end{equation}
But here we should notice that the Gamma function factors in (\ref{defGgenMellin}) and (\ref{defHgenMellin}) are different. Therefore, the action of the difference operator is slightly modified from (\ref{diffbbaction}). We also note that the relation (\ref{MandMtilde}) is quite similar to that of the super amplitude in flat space. There we extract a supercharge delta function $\delta^{(16)}(Q)$ to define the reduced amplitude, and here the role of the delta function is played by the difference operator $\widehat{\rm R}$.

Now we can formulate an algebraic bootstrap problem by imposing a set of constraints from symmetries and consistency conditions. These conditions are
\begin{enumerate}
\item {\bf Superconformal symmetry.} The consequence of superconformal symmetry has already been taken care of by the difference relation (\ref{MandMtilde}). 
\item {\bf Bose symmetry.} Because the external operators are scalars, the Mellin amplitude should have the usual Bose symmetry. More precisely, the Mellin amplitude is invariant under permuting all the particle labels.
\item {\bf Analytic structure.} The Mellin amplitude $\mathcal{M}_{k_1k_2k_3k_4}$ should only have simple poles with polynomial residues. This follows from the fact that the Mellin amplitude in principle can be computed from finitely many Witten diagrams and this analytic structure is satisfied by each diagram.
\item {\bf Asymptotic growth.} The Mellin amplitude $\mathcal{M}_{k_1k_2k_3k_4}$ should not grow more than linearly in the high energy limit. This is because the high energy limit of the Mellin amplitude is related to the flat-space supergravity amplitude via (\ref{Mellinflatspacelim}), and the growth behavior of the latter is known. 
\end{enumerate}
Note that importantly, unlike in the previous two methods where correlators were considered on a case-by-case basis, this algebraic problem is formulated for all different correlators at the same time. This is very useful because we can now obtain all infinitely many correlators in one go by solving a single problem.

The conditions in this algebraic bootstrap problem turn out to be very constraining and completely fix the answer. The reduced Mellin amplitude for general weights has the form 
\begin{equation}\label{Mtildegen}
\widetilde{\mathcal{M}}_{k_1k_2k_3k_4}=\prod_{i<j}t_{ij}^{\gamma_{ij}^0}(t_{12}t_{34})^{\mathcal{E}} \sum_{i,j}\frac{a_{ijk}\sigma^i\tau^j}{(s-s_M+2k)(t-t_M+2j)(\tilde{u}-u_M+2i)}\;,
\end{equation}
where $s_M$, $t_M$, $u_M$ and $\gamma_{ij}^0$ are fixed parameters depending on the weights $k_i$. The sum over all R-symmetry structures is constrained by $i+j+k=\mathcal{E}-2$ where $\mathcal{E}$ is called the extremality and is also determined by $k_i$ (e.g., $\mathcal{E}=k$ if all $k_i=k$). Here we refer to the original papers \cite{Rastelli:2016nze,Rastelli:2017udc} for details of these definitions. Importantly, the coefficients $a_{ijk}$ are uniquely fixed from the bootstrap constraints (up to an overall constant)
\begin{equation}
a_{ijk}=\frac{2\sqrt{k_1k_2k_3k_4}}{N^2 i!j!k!(i+\frac{\kappa_u}{2})!(j+\frac{\kappa_t}{2})!(k+\frac{\kappa_s}{2})!}\;,
\end{equation}
where $\kappa_s$, $\kappa_t$, $\kappa_u$ are also $k_i$ dependent constants. In this way, we solve all the tree-level four-point functions with arbitrary KK levels from just symmetries and consistency conditions. In particular, for the $k_i=2$ case, there is a single term in the sum and reduced Mellin amplitude reads
\begin{equation}
\widetilde{\mathcal{M}}_{2222}\propto\frac{1}{(s-2)(t-2)(\tilde{u}-2)}\;.
\end{equation}

\begin{exercise}
Work out $\widehat{\rm R}$ as a difference operator and check its action on the $k_i=2$ reduced Mellin amplitude reproduces correctly the full Mellin amplitude you obtained in Exercise \ref{ex:Mellinkeq2}.
\end{exercise}

Writing the holographic correlators in terms of the reduced correlators also helps to manifest a nice hidden structure in these objects. Note that due to the shifts in the quantum numbers, the lowest $k_i=2$ reduced correlator has conformal dimension 4 at each point and is an R-symmetry singlet. Therefore, $H_{2222}$ is a function of $x_{ij}^2$ only and has no $t_{ij}$ dependence. However, it turns out that the simplest reduced correlator can be promoted into a generating function if we perform the following replacement \cite{Caron-Huot:2018kta}
\begin{equation}\label{bfH}
{\bf H}(x_i;t_i)=H_{2222}(x_{ij}^2-2t_i\cdot t_j)\;,
\end{equation}
which introduces $t$ dependence. All the other higher KK reduced correlators can then be obtained from the generating function ${\bf H}(x_i;t_i)$ by Taylor expanding in $t_i\cdot t_j$ and collecting all the R-symmetry structures that can appear. Note that the replaced argument can be conveniently written as 
\begin{equation}\label{PPtoZZ}
x_{ij}^2=-2P_i\cdot P_j\to x_{ij}^2-2t_i\cdot t_j=-2Z_i\cdot Z_j\;,
\end{equation}
where $Z_i$ are 12 dimensional null vectors obtained by grouping $P_i$ and $t_i$ together
\begin{equation}
Z_i=(P_i,t_i)\;.
\end{equation}
These vectors can be viewed as conformal embedding space vectors of $\mathbb{R}^{10}$, and their appearance seems to indicate a hidden conformal symmetry in 10 dimensions. In other words, in obtaining the generating function (\ref{bfH}) we have just replaced the 4d distances by the 10d distances. The same higher dimensional hidden structure also appears at weak coupling \cite{Caron-Huot:2021usw}, as well as in other theories \cite{Rastelli:2019gtj,Alday:2021odx,Zhou:2021gnu,Huang:2024dxr,Du:2024xbd,Chen:2025yxg,Chen:2026ium}.  However, currently these hidden structures are only exploited as useful tools in various calculations. There has been no sharp understanding (and probably is also difficult to find one) about their physical nature or the precise mechanism that gives rise to such structures.

\section{Many Extensions}\label{Sec:extensions}
The bootstrap philosophy is very powerful and also very general. Using this philosophy, we can extend the study of holographic correlators in many different directions. In this section, we will review some of these extensions beyond the simplest cases considered in the previous section. We will mainly focus on outlining the ideas rather than spelling out all the technical details.

\subsection{Beyond four-point functions}
So far our discussions have focused on only four-point correlators which are the simplest correlators with nontrivial spacetime dependence. However, higher-point correlators contain new theory data which are not encoded in four-point functions. Moreover, they are also particularly important in the analogy with the flat-space on-shell scattering amplitude program as there one is often interested in the behavior at general multiplicity. 

\begin{figure}[h]
    \centering
    \begin{subfigure}[b]{0.28\textwidth}
        \centering
        \includegraphics[width=\textwidth]{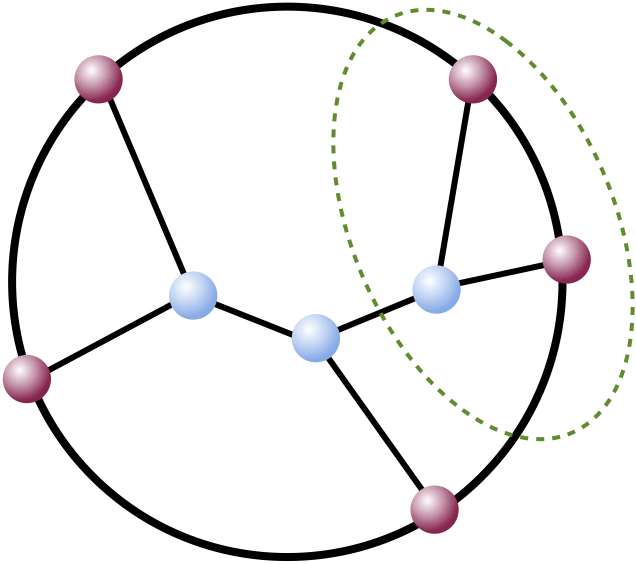}
        \caption{double exchange}
        \label{fig:5ptWD1}
    \end{subfigure}
     \hspace{0.05\textwidth}
    \begin{subfigure}[b]{0.28\textwidth}
        \centering
        \includegraphics[width=\textwidth]{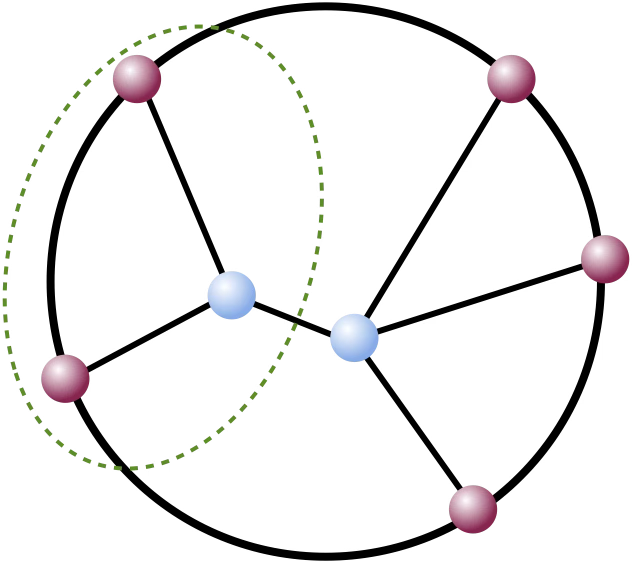}
        \caption{single exchange}
        \label{fig:5ptWD2}
    \end{subfigure}
     \hspace{0.05\textwidth}
    \begin{subfigure}[b]{0.26\textwidth}
        \centering
        \includegraphics[width=\textwidth]{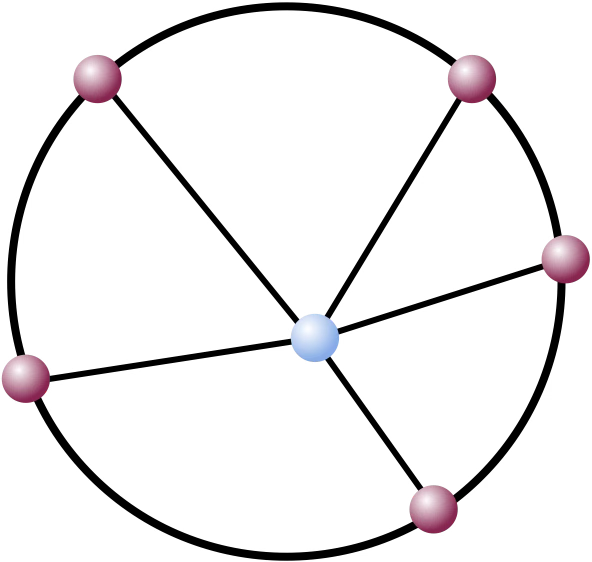}
        \caption{contact}
        \label{fig:5ptWD3}
    \end{subfigure}
    \caption{Three different topologies of five-point Witten diagrams at tree level. For the part of the diagram inside the green dashed circle, we can use the integrated vertex identity to collapse the bulk-to-bulk correlator. We can reduce from topology \ref{fig:5ptWD1} to \ref{fig:5ptWD2}, and then to the contact diagram topology \ref{fig:5ptWD3}.}
    \label{fig:5ptWD}
\end{figure}

To extend the bootstrap strategy beyond four points, a straightforward attempt is  to use the position space algorithm introduced in Section \ref{Subsec:posispamethod}. We first make an ansatz for the correlator as the sum of all diagrams with unknown coefficients and then solve the coefficients by using symmetries. This is possible for five-point functions and was implemented in \cite{Goncalves:2019znr,Alday:2022lkk}. The reason why this is feasible is that we can use the same trick used in evaluating exchange Witten diagrams in Section \ref{Subsubsec:exchinposi} to reduce five-point exchange Witten diagrams of all topologies to contact Witten diagrams. This is schematically illustrated in Figure \ref{fig:5ptWD}.  In these diagrams we can always find the bulk-to-bulk propagators being part of cubic integrals where the other two propagators are bulk-to-boundary, and then we can use (\ref{cubicinttocontact}) to collapse these internal propagators into contact interactions. For example,  in the Witten diagram in Figure \ref{fig:5ptWD1}, if we assume for simplicity all the lines are scalars of conformal dimension 2, we can use the identity twice to write the exchange Witten diagram as
\begin{equation}
W\propto \frac{1}{x_{12}^2x_{34}^2}D_{11112}\;.
\end{equation}
In fact, in $\mathcal{N}=4$ SYM all Witten diagrams including the contact diagrams can be reduced to the basic contact Witten diagram $D_{11112}$ via repeated actions of the differential recursion relations (\ref{Dfundiffrecur}). The basic function $D_{11112}$ is again a known function and can be reduced to the flat-space one-loop box diagram (\ref{Dbar1111aspolylogs}), see \cite{Goncalves:2019znr} for the explicit expression. By further imposing the constraints from superconformal symmetry, it is possible to determine all the unknowns in the ansatz from solving a set of overdetermined linear equations. However, this simple reduction-based implementation of the position space method ceases to work at six points because the reduction identity (\ref{cubicinttocontact}) requires cubic vertices. For six points or more, there are Witten diagrams, such as the three-to-three exchange Witten diagram, which cannot be reduced to contact Witten diagrams.

To efficiently go to higher points, the Mellin space formalism turns out to be the more convenient choice. This is partly due to the simpler analytic structure and is also because Mellin space makes manifest the scattering amplitude properties of holographic correlators which are similar to those of flat-space amplitudes. A particularly useful tool is the factorization property of Mellin amplitudes which allows us to express the Mellin pole residues in terms of lower-point Mellin amplitudes \cite{Goncalves:2014rfa}. Another useful property is the flat-space limit (\ref{Mellinflatspacelim}) which allows us to fix the leading terms of the Mellin amplitudes in the high energy limit. In fact the flat-space amplitude we get from AdS in (\ref{Mellinflatspacelim}) is not the most generic one but is in a simplifying limit. The $\frac{1}{2}$-BPS operators in $\mathcal{N}=4$ SYM are mapped to spinning gravitons in flat space and they have special polarizations which are orthogonal to all particle momenta. This orthogonality also makes the flat-space amplitude simpler. These properties have been exploited in e.g., \cite{Alday:2023kfm,Cao:2023cwa} to compute higher-point correlators which show certain recursive features. However, a truly recursive and on-shell method in the spirit of the flat-space Britto-Cachazo-Feng-Witten (BCFW) recursion relation still remains to be discovered.\footnote{There is an AdS version of the BCFW recursion relation in momentum space \cite{Raju:2010by}. However, the AdS relation is still quite complicated because the use of the momentum space obscures the conformal symmetry.}

\subsection{Gluon scattering in AdS}\label{Subsec:gluonscattering}
For the example of $\mathcal{N}=4$ SYM which we considered in great detail in the previous section, the holographic correlators in the theory correspond to scattering amplitudes of gravitons in AdS. But in the flat-space amplitude program, the simpler, also in a sense more fundamental, objects to study are the gluon amplitudes. In AdS, we can study gluon scattering as well, by considering a subsector of theories which have less superconformal symmetry \cite{Zhou:2018ofp,Alday:2021odx}. In the case of $\mathcal{N}=4$ SYM, the simplest modification leading to such less superconformal theories is to couple $\mathcal{N}=4$ SYM to additional fundamental $\mathcal{N}=2$ hypermultiplets. On the bulk side, this corresponds to adding $N_F$ D7 branes as probes with $N_F\ll N$ and this breaks half of the superconformal symmetry.\footnote{More precisely, such a theory is only conformal in the strict $N_F\ll N$ limit which is satisfied when $N_F$ is finite and $N$ is taken to infinity. One can also have another type of construction which is exactly conformal by considering D3 branes probing F-theory 7-brane singularities. At tree level, the gluon correlators are the same up to a choice of gauge groups. See, e.g., the discussions in \cite{Alday:2021odx} for more details.} The D7 branes are embedded in AdS$_5\times$S$^5$ as the subspace AdS$_5\times$S$^3$ and the low energy theory on the brane is just 8d $\mathcal{N}=1$ SYM with gauge group $SU(N_F)$. An important feature of this system is that at large $N$ the gluon degrees of freedom, which are localized to the D7 brane worldvolume, couple parametrically more strongly with themselves  than with the gravitons living in the full 10d bulk. More precisely, we have 
\begin{equation}\label{gravitongluonlargeNcount}
\langle \mathcal{O}_{\rm gluon}\mathcal{O}_{\rm gluon}\mathcal{O}_{\rm gluon}\rangle\sim 1/\sqrt{N}\;,\quad \langle \mathcal{O}_{\rm gluon}\mathcal{O}_{\rm gluon}\mathcal{O}_{\rm graviton}\rangle\sim 1/N\;.
\end{equation}
Here schematically the AdS gravitons have the form $\mathcal{O}_{\rm graviton}={\rm tr}(\Phi\ldots \Phi)$, and the AdS gluons  $\mathcal{O}_{\rm gluon}=q(\ldots\Phi\ldots)\bar{q}$ are constructed by further using the fundamental scalar quarks and anti-quarks from the $\mathcal{N}=2$ hypermultiplet matter. This $1/N$ hierarchy of the three-point functions also follows from the usual counting in the field theory: $\mathcal{O}_{\rm gluon}$ are meson states and $\mathcal{O}_{\rm graviton}$ are glue states; (\ref{gravitongluonlargeNcount}) is nothing but the standard $1/N$ counting of the large $N$ QCD. Importantly, the relation (\ref{gravitongluonlargeNcount}) shows that in the large $N$ limit we can decouple the bulk gravity and therefore talk about only gluon scattering on AdS$_5\times$S$^3$.

We now consider $\mathcal{N}=1$ SYM on AdS$_5\times$S$^3$ by itself. Like in AdS$_5\times$S$^5$, the KK reduction of the theory on S$^3$ also gives rise to an infinite tower of states which fit into $\frac{1}{2}$-BPS multiplets of 4d $\mathcal{N}=2$ superconformal symmetry. The superconformal primaries, which we will refer to as the supergluons, have the form
\begin{equation}
\mathcal{O}_k^I(x,v,\bar{v})=\mathcal{O}^{I;\alpha_1,\ldots,\alpha_k;\bar{\alpha}_1\ldots\bar{\alpha}_{k-2}}(x)v_{\alpha_1}\ldots v_{\alpha_k}\bar{v}_{\bar{\alpha}_1}\ldots \bar{v}_{\bar{\alpha}_{k-2}}\;.
\end{equation} 
Here $\alpha_i$ are the $SU(2)_R$ R-symmetry indices and $\bar{\alpha}_i$ are the indices of another $SU(2)_L$ global symmetry. These two $SU(2)$ factors together make up the $SO(4)=SU(2)_R\times SU(2)_L$ isometry of S$^3$. The supergluon operator $\mathcal{O}^I_k$ therefore transforms in the $(\frac{k}{2},\frac{k-2}{2})$ representation of $SU(2)_R\times SU(2)_L$, and also in the adjoint representation of $SU(N_F)$ with the index $I$. Just like the null R-symmetry vectors which we introduced for the supergravitons, here we have introduced the polarization spinors $v_\alpha$ and $\bar{v}_{\bar{\alpha}}$ to correspondingly keep track of the $SU(2)_R$ and $SU(2)_L$ indices. The simplest correlators with nontrivial insertion position dependence are the four-point functions  
\begin{equation}
G_{k_1k_2k_3k_4}^{I_1I_2I_3I_4}(x_i;v_i;\bar{v}_i)=\langle \mathcal{O}_{k_1}^{I_1}(x_1,v_1,\bar{v}_1)\ldots \mathcal{O}_{k_4}^{I_4}(x_4,v_4,\bar{v}_4) \rangle\;. 
\end{equation}
On the one hand, the superconformal kinematics is similar to that of the $\mathcal{N}=4$ correlators. After extracting a kinematic factor, the correlator can be written as a function of cross ratios
\begin{equation}
\mathcal{G}_{k_1k_2k_3k_4}^{I_1I_2I_3I_4}(z,\bar{z};\alpha,\beta)\;,
\end{equation}
where the new $SU(2)_R$ and $SU(2)_L$ cross ratios are defined as 
\begin{equation}
\alpha=\frac{(v_1\cdot v_3)(v_2\cdot v_4)}{(v_1\cdot v_2)(v_3\cdot v_4)}\;,\quad \beta=\frac{(\bar{v}_1\cdot\bar{v}_3)(\bar{v}_2\cdot\bar{v}_4)}{(\bar{v}_1\cdot\bar{v}_2)(\bar{v}_3\cdot\bar{v}_4)}\;,
\end{equation}
with $v_i\cdot v_j=\epsilon^{\gamma\delta}v_{i,\gamma}v_{j,\delta}$ and $\bar{v}_i\cdot \bar{v}_j=\epsilon^{\bar{\gamma}\bar{\delta}}\bar{v}_{i,\bar{\gamma}}\bar{v}_{j,\bar{\delta}}$. The fermionic supercharges of $\mathcal{N}=2$ superconformal symmetry further give rise to the superconformal Ward identities
\begin{equation}
(z\partial_z-\alpha\partial_\alpha)\mathcal{G}_{k_1k_2k_3k_4}^{I_1I_2I_3I_4}(z,\bar{z};\alpha,\beta)\big|_{\alpha=1/z}=0\;,
\end{equation}
together with another one with $z\leftrightarrow\bar{z}$. The superconformal Ward identities can be solved in general. This allows us to write the gluon correlators in a form similar to (\ref{solscfWardidNeq4})
\begin{equation}
\mathcal{G}_{k_1k_2k_3k_4}^{I_1I_2I_3I_4}=\mathcal{G}_{{\rm prot},k_1k_2k_3k_4}^{I_1I_2I_3I_4}+R \mathcal{H}_{k_1k_2k_3k_4}^{I_1I_2I_3I_4}\;,
\end{equation}
where $R=(1-z\alpha)(1-\bar{z}\alpha)$. Here $\mathcal{G}_{{\rm prot},k_1k_2k_3k_4}^{I_1I_2I_3I_4}$ is the protected part and $\mathcal{H}_{k_1k_2k_3k_4}^{I_1I_2I_3I_4}$ is the reduced correlator. On the other hand, the new feature which did not appear before is that the supergluon correlators now carry the color indices. The color structures can be handled by decomposing the correlators into different color channels using projectors. 

The bootstrap methods we introduced for $\mathcal{N}=4$ SYM also apply to these gluon correlators. We will not go into the details here to avoid being repetitive and also overly technical. We refer the reader to the original paper \cite{Alday:2021odx} for the implementation details. The final results, written in terms of the Mellin amplitudes of the reduced correlators, take the following compact form
\begin{equation}\label{Mtildegluon}
\widetilde{\mathcal{M}}^{\mathcal{N}=2}_{k_1k_2k_3k_4}\sim \sum_{i,j} a_{ijk}\sigma^i\tau^j \left(\frac{n_s^{i,j}c_s}{s-s_M+2k}+\frac{n_t^{i,j}c_t}{t-t_M+2j}+\frac{n_u^{i,j}c_u}{\tilde{u}-u_M+2i}\right)
\end{equation}
where we have omitted the kinematic factor and constant factors such as $\sqrt{k_1k_2k_3k_4}$. But the sum over $i$, $j$ is the same as in (\ref{Mtildegen}) with  the same coefficients $a_{ijk}$ and
\begin{equation}
\sigma=\alpha\beta\;,\quad \tau=(1-\alpha)(1-\beta)\;,
\end{equation}
except here the Mandelstam variables satisfy instead the relation $s+t+\tilde{u}=\sum_ik_i-2$. Moreover, we have defined the color factors
\begin{equation}
c_s=f^{I_1I_2J}f^{JI_3I_4}\;,\quad c_t=f^{I_1I_4J}f^{JI_2I_3}\;,\quad c_u=f^{I_1I_3J}f^{JI_2I_4}\;,
\end{equation}
in terms of the color group structure constants $f^{IJK}$ and they satisfy $c_s+c_t+c_u=0$. The kinematic factors are
\begin{equation}
\begin{split}
n_s^{ij}={}&\frac{1}{t-t_M+2j}-\frac{1}{\tilde{u}-u_M+2i}\;,\\
n_t^{ij}={}&\frac{1}{\tilde{u}-u_M+2i}-\frac{1}{s-s_M+2k}\;,\\
n_u^{ij}={}&\frac{1}{s-s_M+2k}-\frac{1}{t-t_M+2j}\;,
\end{split}
\end{equation}
which satisfy the same relation $n_s^{ij}+n_t^{ij}+n_u^{ij}=0$. Writing the supergluon amplitudes in this form, \cite{Zhou:2021gnu} pointed out that the result exhibits an AdS version of the double copy relation which is very similar to the flat-space relation we mentioned in Section \ref{Sec:introduction}. If we replace all $c_s$, $c_t$, $c_u$ in (\ref{Mtildegluon}) by the corresponding $n_s^{ij}$, $n_t^{ij}$, $n_u^{ij}$ for each $\sigma^i\tau^j$ structure, we get precisely the AdS$_5\times$S$^5$ supergraviton amplitudes (\ref{Mtildegen}). On the other hand, we can also perform a ``zero copy'' where we replace $n_s^{ij}$, $n_t^{ij}$, $n_u^{ij}$ by the color factors $c'_s$, $c'_t$, $c'_u$ of a different color group, we obtain the Mellin amplitudes of a bi-adjoint scalar field theory which is purely bosonic and couples conformally to AdS$_5\times$S$^1$. However, such a double copy relation for holographic correlators is currently restricted to only four-point functions. It would be very interesting to find analogous AdS double copy relations for general higher-point correlators. 

\subsection{Loops in AdS}\label{Subsec:loopsinAdS}
In the large $N$ expansion of the holographic correlators, so far we have only considered the leading connected contribution which comes from tree-level diagrams in AdS.  Higher orders terms in $1/N$ correspond to quantum corrections and are given by loop-level processes in AdS. These loop-level corrections are another front where the traditional method falls short. Not only do we need to know how to compute all the individual loop-level Witten diagrams, but we also need to sum over an infinite number of diagrams  because all KK modes run in the loops. However, we can still compute the loop corrections to correlators by using the AdS analogue of the unitarity method in flat space \cite{Aharony:2016dwx}.

\begin{figure}[htbp]
    \centering
        \includegraphics[width=0.9\textwidth]{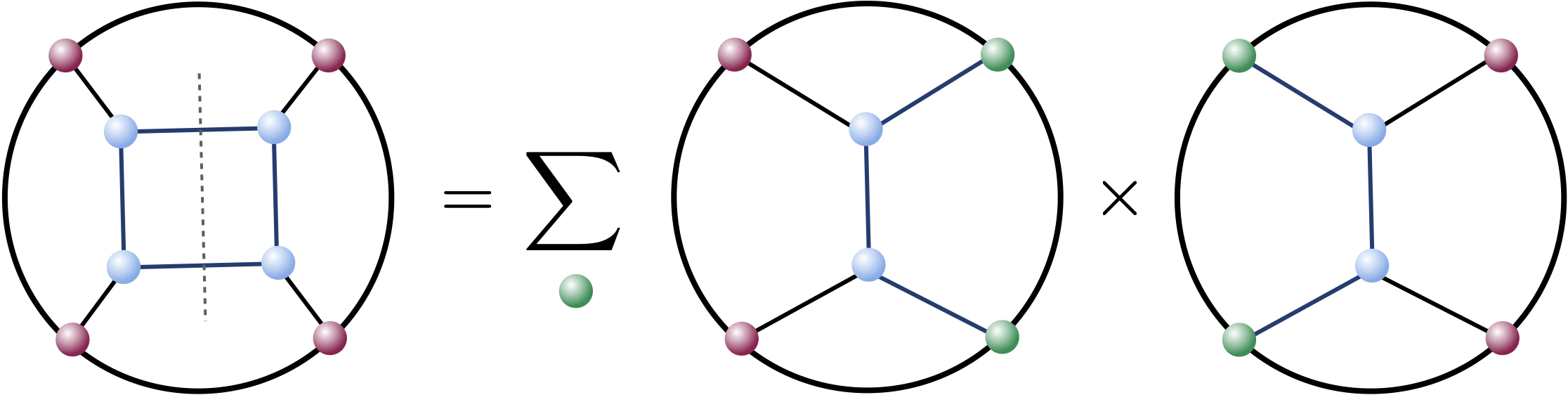}
    \caption{Unitarity method in AdS: obtaining loop-level correlators from gluing tree-level correlators. Here we also need to sum over all the intermediate KK modes which corresponds to summing over the glued external operators.}
    \label{fig:AdSunitarity}
\end{figure}

The intuitive idea for obtaining loop corrections is by gluing together tree-level correlators. We will focus on the one-loop correlator and this is schematically illustrated in Figure \ref{fig:AdSunitarity}. However, such a picture is only motivational. What really gets glued are certain logarithmic singularities which arise in special kinematic limits of the correlator. To understand how the singularities arise, let us consider the s-channel OPE. The conformal blocks have the following behavior in the small $U$ limit
\begin{equation}
g_{\Delta,\ell}(U,V)=U^{\tau/2}(f_0(V)+Uf_1(V)+U^2f_2(V)\ldots)\;.
\end{equation}
If the operator is not protected and has a small anomalous dimension $\Delta=\Delta^{(0)}+\frac{\gamma}{N^2}+\ldots$, we can expand with respect to the small parameter $1/N^2$. The $U^{\frac{\tau}{2}}$ term will give rise to logarithmic singularities
\begin{equation}
g_{\Delta,\ell}(U,V)\big|_{\frac{1}{N^2}}\supset \left(\frac{\gamma}{2}\log U\right) g_{\Delta^{(0)},\ell}(U,V)\;.
\end{equation}
We will say that a logarithmic singularity in the correlator is leading if it has the highest possible power at a given order of $1/N^2$. Note for such a leading logarithmic singularity, the power of the leading $\gamma$ is always equal to the power of $\log U$.\footnote{The anomalous dimension can also have higher order terms $\Delta=\Delta^{(0)}+\frac{\gamma}{N^2}+\frac{\gamma'}{N^4}+\ldots$. But terms with $\gamma'$ will not be part of the leading logarithmic singularities.} 

The double-particle operators we saw in the conformal block decomposition of Witten diagrams in Section \ref{Subsec:confblockdecom} are precisely such unprotected operators with leading corrections to conformal dimensions suppressed by $1/N^2$. Now let us consider the four-point function in $1/N^2$ expansion
\begin{equation}
\mathcal{G}(U,V)=\mathcal{G}^{(0)}(U,V)+\frac{1}{N^2}\mathcal{G}^{(1)}(U,V)+\frac{1}{N^4}\mathcal{G}^{(2)}(U,V)+\ldots\;.
\end{equation}

For simplicity, let us consider a toy example assuming there is only one kind of double-particle operators $:\mathcal{O}\square^n\partial^l\mathcal{O}:$. The leading order $\mathcal{G}^{(0)}(U,V)$ is just generalized free fields and these operators contribute to the decomposition as
\begin{equation}
\mathcal{G}^{(0)}(U,V)=\sum_{n,\ell} a_{n,\ell}^{(0)}\;g_{n,\ell}(U,V)\;,
\end{equation}
where we have slightly abused the notation to use $g_{n,\ell}(U,V)$ to denote the conformal block of an operator with dimension $2\Delta_\mathcal{O}+2n+\ell$ and spin $\ell$. At the next order, which corresponds to tree-level supergravity, we can allow at most one power of the tree-level anomalous dimension $\gamma$. Therefore, at this order we can have a $\log U$ singularity with the coefficients
\begin{equation}
\mathcal{G}^{(1)}(U,V)\big|_{\log U}=\sum_{n,\ell} \frac{1}{2}a_{n,\ell}^{(0)}\gamma_{n,\ell}^{(1)} \;g_{n,\ell}(U,V)\;.
\end{equation}
The next order in $1/N$ comes from supergravity at one loop. We can have at most two powers of tree-level $\gamma$. The leading logarithmic singularity is therefore $\log^2U$ with the coefficients
\begin{equation}
\mathcal{G}^{(2)}(U,V)\big|_{\log^2 U}=\sum_{n,\ell} \frac{1}{8}a_{n,\ell}^{(0)}(\gamma_{n,\ell}^{(1)})^2 \;g_{n,\ell}(U,V)\;.
\end{equation}
An important point to notice is that the $\log^2U$ singularity involves only the tree-level and GFF data. More explicitly, we can rewrite the coefficients as
\begin{equation}
\mathcal{G}^{(2)}(U,V)\big|_{\log^2 U}=\sum_{n,\ell}\frac{1}{2} \frac{(\frac{1}{2}a_{n,\ell}^{(0)}\gamma_{n,\ell}^{(1)})^2}{a_{n,\ell}^{(0)}} \;g_{n,\ell}(U,V)\;,
\end{equation}
which is the ratio of the squared tree-level coefficients and GFF coefficients. 

In this toy example, the rewriting of the coefficients might seem quite trivial and somewhat arbitrary. The full-fledged case of AdS$_5\times$S$^5$ supergravity is more complicated because the double-particle operators are degenerate at infinite $N$. For example, the following operators all have the same quantum number and can mix with each other
\begin{equation}
:\mathcal{O}_4\mathcal{O}_4:\;,\quad :\mathcal{O}_3\square\mathcal{O}_3:\;,\quad :\mathcal{O}_2\square^2\mathcal{O}_2:\;.
\end{equation} 
However, it turns out that we can still write the $\log^2U$ coefficients of the one-loop correlator in such a squared form in terms of the tree-level $\log U$ coefficients. But to achieve that, in particular for the one-loop correlator $\langle\mathcal{O}_2\mathcal{O}_2\mathcal{O}_2\mathcal{O}_2\rangle$ which we will focus on, we will need all the $\langle\mathcal{O}_2\mathcal{O}_2\mathcal{O}_p\mathcal{O}_p\rangle$  correlators at tree level \cite{Alday:2017xua,Aprile:2017bgs,Aprile:2017xsp}.\footnote{We also need all the $\langle\mathcal{O}_p\mathcal{O}_p\mathcal{O}_p\mathcal{O}_p\rangle$ correlators at the GFF order in the denominator.} This is consistent with our schematic picture: all the KK modes run in the loop. To summarize this somewhat technical discussion so far, let us reiterate that the main point is the leading logarithmic singularity at one loop can be obtained via a gluing procedure at the level of conformal blocks using data from lower orders. In fact, the explicit computation further shows that the $\log^2U$ coefficient has the following structure where the $\log V$ singularities at small $V$ are manifested
\begin{equation}\label{LLS}
\mathcal{H}^{(2)}(U,V)\big|_{\log^2 U}=F_2(U,V)\log^2V+F_1(U,V)\log V+F_0(U,V)\;,
\end{equation}
and $F_i(U,V)$ all have regular expansions at small $U$ and $V$. Here we have also switched from the full correlator to the reduced correlator because the latter contains all the dynamical information and is more convenient to work with. 

The next step is to complete the leading logarithmic singularity (\ref{LLS}) in the full reduced one-loop correlator. We have two choices to proceed: working either in position space \cite{Aprile:2017bgs,Aprile:2017qoy,Aprile:2019rep} or in Mellin space \cite{Alday:2018kkw,Alday:2019nin}. Here we will explain the latter approach because it is easier to see why the method works. Recall for the $\langle\mathcal{O}_2\mathcal{O}_2\mathcal{O}_2\mathcal{O}_2\rangle$ correlator, the reduced correlator has the following Mellin representation 
\begin{equation}
\mathcal{H}=\int \frac{ds dt}{(4\pi i)^2}U^{\frac{s}{2}}V^{\frac{t}{2}-2}\widetilde{\mathcal{M}}(s,t)\Gamma^2\left(\frac{4-s}{2}\right)\Gamma^2\left(\frac{4-t}{2}\right)\Gamma^2\left(\frac{4-\tilde{u}}{2}\right)\;,
\end{equation}
and $s$, $t$, $\tilde{u}$ transform into each other, satisfying $s+t+\tilde{u}=4$. To produce the $\log^2U\log^2V$ singularity in (\ref{LLS}) from this integral, we must have cubic poles in the Mellin integrand at $s=4,6,\ldots$ and $t=4,6,\ldots$. Note that the Gamma function factor $\Gamma^2({\frac{4-s}{2}})\Gamma^2({\frac{4-t}{2}})$ already contributes double poles at these locations. Therefore, the reduced Mellin amplitude only needs to have single poles. Together with crossing symmetry, we can write down an ansatz of the form
\begin{equation}\label{oneloopMtilde}
\begin{split}
\widetilde{\mathcal{M}}(s,t)={}&\sum_{m,n}\left(\frac{c_{mn}}{(s-4-2n)(t-4-2m)}+\frac{c_{mn}}{(s-4-2n)(\tilde{u}-4-2m)}\right.\\
{}&\left.+\frac{c_{mn}}{(t-4-2n)(\tilde{u}-4-2m)}\right)\;.
\end{split}
\end{equation}
Here we will make a seemingly radical assumption that the numerators $c_{mn}$ are just constants independent of $s$, $t$ and $\tilde{u}$. However, this is an assumption which we can explicitly test against the leading logarithmic singularity (\ref{LLS}) once we have solved the Mellin amplitude ansatz. Therefore, we have the following strategy. 
\begin{itemize}
\item Taking the residues in $s$ and $t$ and isolating the $\log^2U\log^2V$ term, we can find the $c_{mn}$ coefficients in a closed form by comparing with $F_2(U,V)$ in (\ref{LLS}) in a small $U$ and $V$ expansion.
\item We then resum (\ref{oneloopMtilde}) and take the residues again\footnote{More precisely, one needs to regularize the one-loop amplitude because (\ref{oneloopMtilde}) is naively divergent.}. This time we focus on just $\log^2U$ but now with the full $V$ dependence. It turns out that we can reproduce the full singularity (\ref{LLS}) including even the subleading terms with $F_1$ and $F_0$. This proves that the ansatz (\ref{oneloopMtilde}) is correct and there are no other single poles in addition to the simultaneous poles.
\item The only ambiguities in the reduced Mellin amplitude are now regular terms which have no poles. However, such regular terms are constrained by the flat-space limit. In fact, one can show that each double sum of the simultaneous poles in (\ref{oneloopMtilde}) becomes a 10d box diagram in flat space. The only allowed ambiguity is a constant term in Mellin space which corresponds to a contact UV counter term. 
\end{itemize}
Here we outlined the calculation without giving the explicit details. The reader can consult \cite{Alday:2018kkw} or Section 12 of \cite{Bissi:2022mrs} in order to reproduce the details of the computation.

\subsection{Stringy corrections}\label{Subsec:stringycorrections}
Up until now, we have been focusing on the supergravity regime where we set the 't Hooft coupling $\lambda=\infty$. In the previous subsection, we considered the first $1/N$ correction while still keeping $\lambda=\infty$ after the large $N$ expansion. It is also interesting to keep $N$ large but consider the $1/\lambda$ corrections which come from stringy effects. In this subsection we will discuss how to compute these tree-level string theory corrections and we will focus on the correlator $\langle \mathcal{O}_2\mathcal{O}_2 \mathcal{O}_2 \mathcal{O}_2\rangle$.

Recall that in applying the position space algorithm in Section \ref{Subsec:posispamethod}, we found that the superconformal Ward identities fix the correlator completely up to an overall constant. However, this is actually only true after we make the important assumption that in the contact part of the ansatz there are no more than two derivatives in the contact Witten diagram vertices. If we consider more derivatives, we will find additional solutions to the superconformal Ward identities which are purely contact. In fact, it is useful to consider the following exercise to work out the new solutions for the first few cases as we increase the derivative number cutoff. These higher-derivative terms can be understood as arising from integrating out the massive stringy states and we are looking at their contributions from an effective field theory perspective in a low energy expansion.
\begin{exercise}\label{ex:counthdsol}
Write down an ansatz as a sum of contact Witten diagrams with no more than $2L$ derivatives and allow them to have the most general R-symmetry dependence. Then impose the superconformal Ward identities, and show that the number of the new solutions $M$ for different $L$ goes as follows
\begin{center}
    \begin{tabular}{|c|c|c|c|c|c|}
    \hline
    Derivative cutoff $L$ & \makebox[1cm]{2} & \makebox[1cm]{3} & \makebox[1cm]{4} & \makebox[1cm]{5} & \makebox[1cm]{6}  \\
    \hline
    \#(New solutions) $M$ & \makebox[1cm]{1} & \makebox[1cm]{0} & \makebox[1cm]{1} & \makebox[1cm]{1} & \makebox[1cm]{1}\\
    \hline
    \end{tabular}
\end{center}
\vspace{0.3cm}

{\bf Hint:} for small $L$ you can just perform the bootstrap calculation in position space; but for larger $L$ it is more convenient to use the tricks we discussed in Section \ref{Subsec:mellinspacemethod} and find the solutions in Mellin space. 
\end{exercise}
These higher-derivative contact solutions are particularly simple to write down in terms of the reduced correlator expressed in the Mellin representation. For solutions with no more than $2L$ derivatives, the reduced Mellin amplitudes are just polynomials of degree $L-2$. We can enumerate the new solutions for different derivative cutoffs as 
\begin{equation}\label{newsolforMtilde}
\begin{split}
{}&4\partial\,:\quad 1\;,\\
{}&6\partial\,:\quad s+t+\tilde{u}=4\;,\quad\text{no new solutions}\;,\\
{}&8\partial\,:\quad s^2+t^2+\tilde{u}^2\;,\\
{}&10\partial\,:\quad st\tilde{u}\;,\\
{}&12\partial\,:\quad  (s^2+t^2+\tilde{u}^2)^2\;,
\end{split}
\end{equation}
which agrees with the counting in Exercise \ref{ex:counthdsol}. Defining $\sigma_2=s^2+t^2+\tilde{u}^2$ and $\sigma_3=st\tilde{u}$, we can write the reduced Mellin amplitude in a $1/\lambda$ expansion in terms of the independent solutions as
\begin{equation}\label{MtildestringLE}
\begin{split}
\widetilde{\mathcal{M}}(s,t)={}&\frac{8}{(s-2)(t-2)(\tilde{u}-2)}+\frac{1}{\lambda^{\frac{3}{2}}}\tilde{\alpha}_{0,0}+\frac{1}{\lambda^2}\tilde{\beta}_{0,0}+\frac{1}{\lambda^{\frac{5}{2}}}(\tilde{\alpha}_{1,0}\sigma_2+\tilde{\gamma}_{0,0})\\
{}&+\frac{1}{\lambda^3}(\tilde{\alpha}_{0,1}\sigma_3+\tilde{\beta}_{1,0}\sigma_2+\tilde{\delta}_{0,0})+\frac{1}{\lambda^{\frac{7}{2}}}(\tilde{\alpha}_{2,0}\sigma_2^2+\tilde{\beta}_{0,1}\sigma_3+\tilde{\gamma}_{1,0}\sigma_2+\tilde{\epsilon}_{0,0})+\ldots\;,
\end{split}
\end{equation}
where the solutions at the previous derivative cutoff orders also carry over to the next one. Note that the power of $\lambda$ changes by $\frac{1}{2}$ between neighboring orders for higher derivative corrections. This makes sense because we know from (\ref{parameterrelation}) that $\lambda^{\frac{1}{2}}=R^2/\ell_s^2$ and a factor of $\lambda^{\frac{1}{2}}$ exactly compensates the difference of the length dimension coming from adding two more derivatives.\footnote{The leading $\lambda^{-\frac{3}{2}}$ correction corresponds to the $R^4$ term in the bulk effective action ($R$ is curvature in this vertex, not to be confused with the radius of AdS).} This expansion of the reduced Mellin amplitude can also be written in an equivalent but more compact form
\begin{equation}\label{MtildestringLEnew}
\widetilde{\mathcal{M}}(s,t)=\frac{8}{(s-2)(t-2)(\tilde{u}-2)}+\sum_{a,b=0}^\infty \frac{\sigma_2^a\sigma_3^b}{\lambda^{\frac{3}{2}+a+\frac{3}{2}b}}\left(\tilde{\alpha}_{a,b}+\frac{1}{\lambda^{\frac{1}{2}}}\tilde{\beta}_{a,b}+\frac{1}{\lambda}\tilde{\gamma}_{a,b}+\ldots\right)\;.
\end{equation}
\begin{exercise}
Prove that the solutions you found in Exercise \ref{ex:counthdsol} for the full correlator translate to the polynomial reduced Mellin amplitudes in (\ref{newsolforMtilde}). Also convince yourself that the expansion (\ref{MtildestringLE}) is the same as (\ref{MtildestringLEnew}).
\end{exercise}

We can already fix an infinite subset of these coefficients by using the flat-space limit formula (\ref{Mellinflatspacelim}). Adapting it to the current case, we have\footnote{We should take $\Delta_i=4$ because we are working with the reduced correlator. Moreover, here we have inverted the relation and written it as a contour integral.}
\begin{equation}\label{Mtildeflatspacelimitf}
\frac{A_{\rm V.S.}(S,T)}{STU}=\lim_{R\to\infty} \frac{R^6c}{32}\int_{\kappa-i\infty}^{\kappa+i\infty}\frac{d\beta}{2\pi i}e^\beta\beta^{-6}\widetilde{\mathcal{M}}\left(\frac{R^2S}{2\beta},\frac{R^2T}{2\beta}\right)\;.
\end{equation}
Here $A_{\rm V.S.}(S,T)$ is the flat-space Virasoro-Shapiro amplitude
\begin{equation}\label{VSflatspace}
\begin{split}
\frac{A_{\rm V.S.}(S,T)}{STU}={}&-\frac{(\alpha')^3}{64}\frac{\Gamma(-\frac{1}{4}\alpha'S)\Gamma(-\frac{1}{4}\alpha'T)\Gamma(-\frac{1}{4}\alpha'U)}{\Gamma(\frac{1}{4}\alpha'S+1)\Gamma(\frac{1}{4}\alpha'T+1)\Gamma(\frac{1}{4}\alpha'U+1)}\\
={}&\frac{1}{STU}\exp\left(2\sum_{n=1}^\infty\frac{\zeta(2n+1)}{2n+1}\left(\frac{\alpha'}{4}\right)^{2n+1}(S^{2n+1}+T^{2n+1}+U^{2n+1})\right)\;,
\end{split}
\end{equation}
where in the second line we have expanded it into polynomials of the flat-space Mandelstam variables $S$, $T$, $U$. We can then perform the Borel transform on the reduced Mellin amplitude (\ref{MtildestringLE}) term by term and then compare with the flat-space result. Note that we need to take the large $R$ limit in the flat-space limit formula (\ref{Mtildeflatspacelimitf}), and only the leading terms at each order of the $1/\lambda$ expansion is selected. Therefore, as is most clear in (\ref{MtildestringLEnew}), the flat-space limit only fixes the $\tilde{\alpha}_{a,b}$ coefficients but does not tell us about the subleading $\tilde{\beta}_{a,b}$ and $\tilde{\gamma}_{a,b}$ coefficients \cite{Goncalves:2014ffa}. 

\begin{figure}[h]
    \centering
        \includegraphics[width=0.92\textwidth]{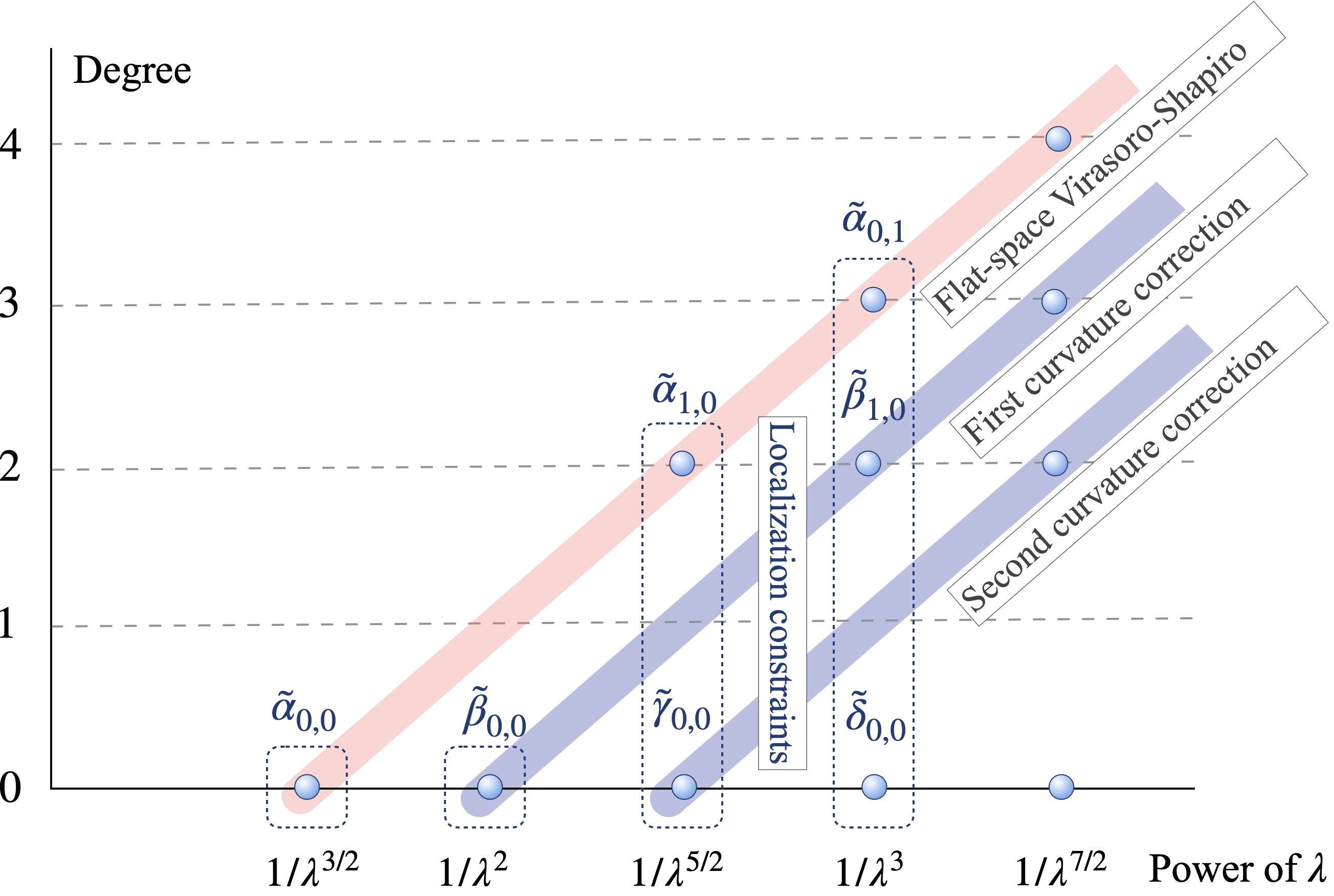}
    \caption{Stringy corrections. The horizontal axis denotes the order in the $1/\lambda$ expansion, while the vertical axis denotes the polynomial degree in $s$, $t$, $\tilde{u}$ of the monomial multiplying each Wilson coefficient. The flat-space limit fixes the leading layer of coefficients marked by the red diagonal strip. Localization constrains the coefficients in the same vertical row by fixing particular linear combinations of them. The worldsheet approach instead organizes successive diagonal layers parallel to the leading one, as curvature corrections to the flat-space Virasoro-Shapiro amplitude.}
    \label{fig:stringycorrection}
\end{figure}

An orthogonal set of constraints comes from supersymmetric localization which allows us to fix more coefficients \cite{Binder:2019jwn,Chester:2019jas,Chester:2020dja}. Schematically, the localization method allows us to compute the value of certain integrated reduced correlators exactly for general values of the coupling
\begin{equation}\label{intcorr}
I(N,\lambda)=\int dzd\bar{z}\,\mathcal{K}(z,\bar{z})\mathcal{H}(z,\bar{z})\;.
\end{equation}
Here $\mathcal{K}(z,\bar{z})$ is a special integration kernel determined by the localization construction. In fact, there are two independent integrated correlators with different kernels \cite{Binder:2019jwn,Chester:2020vyz,Chester:2025kvw}. We will not need the explicit forms of these kernels and refer the reader to the original papers for the details. However, once the integrated correlator $I(N,\lambda)$ is computed using localization, we can expand it in $1/\lambda$ in the 't Hooft limit. Note that both leading and subleading solutions to the superconformal Ward identity in the flat-space limit can contribute to the integrated correlator, and their contributions can be explicitly evaluated by using (\ref{intcorr}). This gives rise to linear relations for the coefficients in (\ref{MtildestringLE}) which can be used to solve for the subleading ones. While this fixes more coefficients, it is also important to note that the progress we are allowed to make using integrated correlators is limited. At higher orders in $1/\lambda$ there are more and more independent solutions to the superconformal Ward identities. However, the presently known localization constructions provide only two independent integrated constraints, including the squashed sphere ones. Thus they eventually become insufficient as the number of contact solutions grows. The constraints of the flat-space limit and localization are illustrated in Figure \ref{fig:stringycorrection}.

To go beyond, we can consider a different organization of the higher-derivative solutions. We first rescale the Mellin-Mandelstam variables by $s,t,\tilde{u}\to \sqrt{\lambda} s, \sqrt{\lambda} t, \sqrt{\lambda} \tilde{u}$, so that all the $\tilde{\alpha}_{a,b}$ terms (and similarly the $\tilde{\beta}_{a,b}$, $\tilde{\gamma}_{a,b}$, etc) in (\ref{MtildestringLEnew}) contribute at the same order. In the absence of an independent definition of string amplitudes in AdS, we use the RHS of (\ref{Mtildeflatspacelimitf}), before taking the $R\to\infty$ limit, as the working definition of its curvature expansion. After a Borel transformation, which is needed for the convergence of the sums, we have the following $1/\lambda$ expansion
\begin{equation}
A_\lambda(S,T)=A^{(0)}(S,T)+\frac{1}{\sqrt{\lambda}}A^{(1)}(S,T)+\frac{1}{\lambda}A^{(2)}(S,T)+\ldots\;.
\end{equation}
Here $A^{(0)}(S,T)$ only involves the $\tilde{\alpha}$ coefficients and  is the flat-space string amplitude described by the $R\to\infty$ limit of (\ref{Mtildeflatspacelimitf}) 
\begin{equation}
A^{(0)}(S,T)=\frac{A_{\rm V.S.}(S,T)}{STU}\;.
\end{equation}
The other terms $A^{(k)}(S,T)$ with $k>0$ describe successive curvature corrections: $A^{(1)}$ involves the $\tilde{\beta}$ coefficients, $A^{(2)}$ involves the $\tilde{\gamma}$ coefficients, and so on. So far this is just a reorganization of the terms in the expansion, which we also illustrate in Figure \ref{fig:stringycorrection}. However, the power of the approach comes from the observation that these corrections $A^{(k)}(S,T)$ themselves admit worldsheet representations \cite{Alday:2023jdk,Alday:2023mvu}. It is a well known fact that the leading flat-space string amplitude $A^{(0)}(S,T)$ can be written as a worldsheet integral 
\begin{equation}
A^{(0)}(S,T)=\frac{1}{U^2}\int d^2z |z|^{-2S-2}|1-z|^{-2T-2}\;.
\end{equation}
But interestingly $A^{(k)}(S,T)$ with $k>0$ also have such a worldsheet structure. They can be written in a similar form
\begin{equation}
A^{(k)}(S,T)=\frac{1}{U^2}\int d^2z |z|^{-2S-2}|1-z|^{-2T-2}G^{(k)}(S,T;z)\;,
\end{equation}
where the trivial unit integrand is replaced by a nontrivial integrand $G^{(k)}(S,T;z)$. The latter can be expanded in a finite dimensional ansatz of single-valued multiple polylogarithms of appropriate transcendental weights, with coefficients rational in $S$ and $T$. The unknown coefficients can then be fixed by imposing various physical conditions, see \cite{Alday:2022uxp,Alday:2022xwz,Alday:2023jdk,Alday:2023mvu} for the details. Crucially, although the ansatz at each fixed order $k$ contains only finitely many unknown, the resulting integrand determines infinitely many Wilson coefficients in the higher-derivative expansion.

\section{Holographic Defect CFTs}\label{Sec:defects}

\subsection{Why defects?}
So far in these lectures we have only discussed holographic CFTs in infinite empty $\mathbb{R}^d$ and we considered correlation functions of local operators. However, there are also nonlocal objects which we can add to the theories. Familiar examples include Wilson loops, surface defects and boundaries. In this section we will collectively call them defects. There is a large body of works studying defects in CFTs and these defect objects are interesting for a number of reasons. 
\begin{itemize}
\item Defects are interesting for their own sake. For example, Wilson loops are important diagnostics for phases of gauge theories. Defects also provide useful descriptions for many phenomena in condensed matter systems. 
\item Adding defects provides a controlled setting to break part of the conformal symmetry (superconformal symmetry) of the theories. This gives us a convenient window to look into setups which are less symmetric. 
\item The inclusion of defects gives rise to more observables and more data, and therefore enriches the theories. For example, there are new operators which are localized to the defect and describing them requires additional new CFT data.
\item Defects are also interesting from the perspective of the AdS amplitude program. Holographically, defects on the boundary are dual to extended objects in the bulk. The previous holographic correlators we considered are dual to scattering amplitudes of particles in AdS. Correlators of local operators in the presence of defects further extend this program by including AdS form factors of particles scattering off extended objects.
\end{itemize}
We will be particularly interested in computing the correlation functions of local operators in the holographic limit of CFTs when defects are present. It turns out that many of the bootstrap techniques we have developed for the defect-free case can be straightforwardly adapted to defect CFT setup. 

In this section, we will first demonstrate how this adaptation works in a simple setup where we are in the supergravity limit and the defect is an ordinary line defect. Then we will explain how we can extend the notion of defects to include seemingly exotic defect dimensions of $0$ and $-1$, which allows us to apply the bootstrap techniques to study systems usually not viewed as defects. Finally, we comment on how various other tools and techniques, such as Mellin representation and unitarity methods, can be generalized to the defect case. 

\subsection{Wilson loop two-point function from bootstrap}\label{Subsec:WL2ptfun}
Let us demonstrate the bootstrap method in the example of $\mathcal{N}=4$ SYM with an infinite and straight $\frac{1}{2}$-BPS Wilson line \cite{Gimenez-Grau:2023fcy}. The presence of the line defect breaks half of the superconformal symmetry. Let us first focus on the conformal group factor and as before the symmetry breaking pattern is easiest to describe in embedding space. The 6d embedding space gets divided into two groups
\begin{equation}\label{PWLseparation}
P^A=\bigg(\underbrace{P^{-1}, P^0,P^1}_{\mathbb{R}^{2,1}\text{ embedding}},\underbrace{P^2,P^3,P^4}_{\text{transverse}}\bigg)=\left(P^a,P^i\right)\;,
\end{equation}
where the first group is the embedding space of the $\mathbb{R}$ occupied by the Wilson line and the second group contains the directions transverse to the first group. Therefore, the original $SO(5,1)$ conformal group is broken to $SO(2,1)\times SO(3)$. 

The R-symmetry is also partially broken after inserting the Wilson line. To see this, we note that holographically the Wilson line is dual to a fundamental string in AdS$_5$ sitting at a point on the internal S$^5$. This point can be described by a fixed six dimensional unit vector $\theta$ satisfying $\theta\cdot\theta=1$. It is then clear that R-symmetry is broken by $\theta$ from $SO(6)$ to $SO(5)$. 

In the presence of the Wilson line, the simplest observable is the one-point function of a bulk local operator inserted away from the defect. For the $\frac{1}{2}$-BPS operators $\mathcal{O}_k(x,t)$ in $\mathcal{N}=4$ SYM, the only structure which preserves all the residual symmetries and has the right scalings is
\begin{equation}\label{WL1pt}
\llangle \mathcal{O}_k \rrangle=a_k\frac{(t\cdot\theta)^k}{|x^i|^k}\;.
\end{equation}
Here we have separated $x=(x^i,x^a)$ into the transverse components $x^i$ and the parallel components $x^a$. The one-point function coefficient $a_k$ is a new piece of CFT data and is not fixed by symmetry. 

The first observable with nontrivial kinematics is the two-point function of two bulk operators inserted away from the defect.\footnote{We can also consider two-point functions where at least one operator is a defect operator. These two-point functions are also fixed by symmetry.} We can define two conformal cross ratios
\begin{equation}\label{crossratiosxichi}
\xi=\frac{(x_1-x_2)^2}{|x_1^i||x_2^i|}\;,\quad \chi=\frac{2x_1^jx_2^j}{|x_1^i||x_2^i|}\;,
\end{equation}
and one R-symmetry cross ratio
\begin{equation}
\sigma=\frac{(t_1\cdot t_2)}{(t_1\cdot \theta)(t_2\cdot \theta)}\;.
\end{equation}
Focusing on the case of $k_1=k_2=2$, the two-point defect correlator can then be written as a function of cross ratios as
\begin{equation}\label{WL2ptdefcalF}
\llangle \mathcal{O}_2(x_1,t_1) \mathcal{O}_2(x_2,t_2) \rrangle = \frac{(t_1\cdot \theta)^2(t_2\cdot \theta)^2}{|x_1^i|^2|x_2^i|^2}\mathcal{F}(\xi,\chi;\sigma)\;,
\end{equation}
after extracting a kinematic factor of one-point functions. The fermionic generators of the unbroken superconformal symmetry impose the following superconformal Ward identity \cite{Liendo:2016ymz}
\begin{equation}\label{defect2ptscfWardid}
\left(\partial_z+\frac{1}{2}\partial_\alpha\right)\mathcal{F}(z,\bar{z};\alpha)\big|_{\alpha=z}=0\;,
\end{equation}
where we have made the change of variables
\begin{equation}\label{defectzzbar}
\xi=\;\frac{(1-z)(1-\bar{z})}{\sqrt{z\bar{z}}},\quad \chi=\frac{z+\bar{z}}{\sqrt{z\bar{z}}}\;,\quad \sigma=-\frac{(1-\alpha)^2}{2\alpha}\;.
\end{equation}

\begin{exercise}
Show that (\ref{WL1pt}) is the only possible form of the defect one-point function. Also use the embedding space to convince yourself that (\ref{crossratiosxichi}) are the correct defect cross ratios. 
\end{exercise}

\begin{exercise}
Just like the $z$, $\bar{z}$ variables for the four-point functions, $z$ and $\bar{z}$ in (\ref{defectzzbar}) have a similar meaning as the operator coordinates. Convince yourself that you can use conformal symmetry to put both operators on a same 2d plane so that the line defect intersect this plane perpendicularly at $0$ and $\infty$. You can further move one operator to $1$, and the other operator has complex coordinate $(z,\bar{z})$ on the plane. Note that this interpretation is not limited to the line defect. We can use the same argument as long as the defect has co-dimension more than 1.
\end{exercise}

\begin{figure}[htbp]
    \centering
    \begin{subfigure}[b]{0.25\textwidth}
        \centering
        \includegraphics[width=\textwidth]{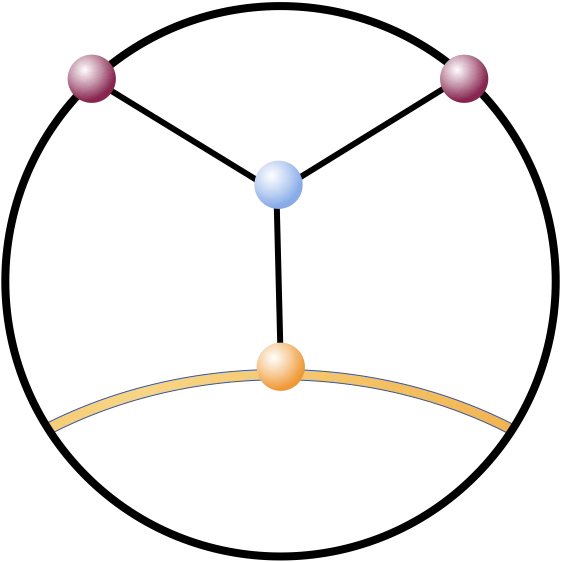}
        \caption{bulk exchange}
        \label{fig:defectWDbulkexch}
    \end{subfigure}
     \hspace{0.05\textwidth}
    \begin{subfigure}[b]{0.25\textwidth}
        \centering
        \includegraphics[width=\textwidth]{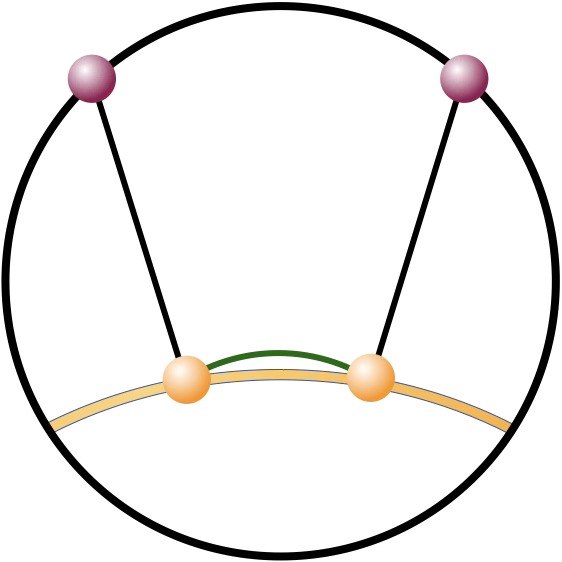}
        \caption{defect exchange}
        \label{fig:defectWDdefectexch}
    \end{subfigure}
     \hspace{0.05\textwidth}
    \begin{subfigure}[b]{0.25\textwidth}
        \centering
        \includegraphics[width=\textwidth]{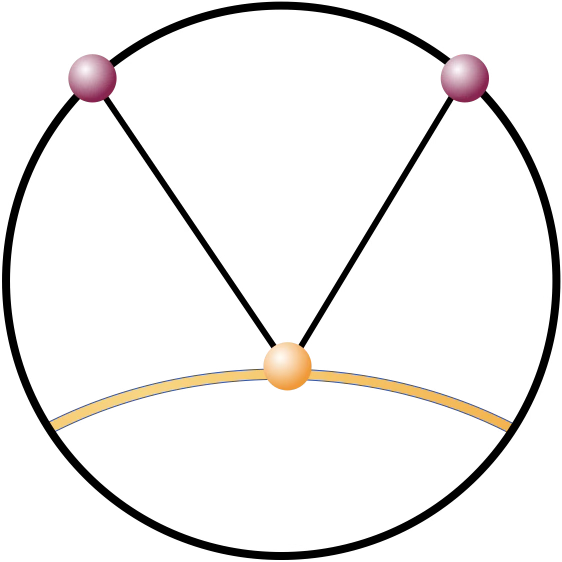}
        \caption{contact}
        \label{fig:defectWDcon}
    \end{subfigure}
    \caption{Three types of defect Witten diagrams at tree level. The yellow curve represents the defect in AdS which is a subspace AdS$_{p+1}$.}
    \label{fig:defectWDtree}
\end{figure}

Let us now set out to bootstrap this simplest two-point function in the supergravity limit. We will adapt the position space method discussed in Section \ref{Subsec:posispamethod} to the defect case. For this to work, we will also need a bit of input from the supergravity picture. We have mentioned that the Wilson line is dual to a fundamental string in AdS$_5$ sitting at a point on S$^5$. The worldsheet of the string occupies an AdS$_2$ subspace which is embedded into AdS$_5$ by setting all but one coordinates parallel to the conformal boundary to zero in terms of the Poincar\'e coordinates. The fluctuations on the worldsheet are dual to defect operators that only live on the CFT line defect, and their spectrum can be obtained by expanding the worldsheet action to the quadratic order. The spectrum of the fluctuations in the bulk remains the same as in Table \ref{tab:kkexch} because the fundamental string does not back-react to the geometry. In this supergravity limit, it is easy to see there are only three types of defect Witten diagrams which contribute to the two-point function and they are depicted in Figure \ref{fig:defectWDtree}.

To imitate the defect-free position space method, we also write down a general ansatz for the defect two-point function as a sum of all the possible defect Witten diagrams
\begin{equation}
\mathcal{A}=\mathcal{A}_{\rm bulk}+\mathcal{A}_{\rm defect}+\mathcal{A}_{\rm contact}\;.
\end{equation}
For the bulk channel exchange part of the ansatz, there are only two fields which can appear
\begin{equation}
\mathcal{A}_{\rm bulk}=\lambda_s h_s E_s+\lambda_{\varphi} h_{\varphi} E_{\varphi}\;.
\end{equation}
They are the supergraviton field $s_2$ and the graviton field $\varphi_{\mu\nu,2}$ in Table \ref{tab:kkexch} from the $k=2$ bulk multiplet. Here $E_s$ and $E_\varphi$ are the corresponding bulk channel exchange Witten diagrams (Figure \ref{fig:defectWDbulkexch}). For example, $E_s$ is explicitly defined by
\begin{equation}
\frac{E_s}{|x_1^i|^2|x_2^i|^2}=\int_{{\rm AdS}_5} dZ \int_{{\rm AdS}_2} dW G^2_{B\partial}(P_1,Z) G^2_{B\partial}(P_2,Z) G^2_{BB}(Z,W)\;,
\end{equation}
where one end of the bulk-to-bulk propagator is integrated over AdS$_5$ and the other end is integrated over the AdS$_2$ worldsheet of the fundamental string. Similarly, $E_\varphi$ is defined with a spin-2 bulk-to-bulk propagator and $\phi_{\mu\nu,2}$ couples to the worldsheet via $\int_{{\rm AdS}_2} g^{ij}\phi_{ij,2}$ where the orthogonal indices are contracted. These diagrams can be evaluated using the same trick as in Section \ref{Subsubsec:exchinposi}. Using (\ref{cubicinttocontact}) and its spin-2 generalization, the bulk channel exchange Witten diagrams can be written as finite sums of contact Witten diagrams (Figure \ref{fig:defectWDcon})
\begin{equation}
\frac{C_{\Delta_1\Delta_2}}{|x_1^i|^{\Delta_1}|x_2^i|^{\Delta_2}}=\int_{{\rm AdS}_2} dZ G^{\Delta_1}_{B\partial}(P_1,Z)G^{\Delta_2}_{B\partial}(P_2,Z)\;.
\end{equation}
These contact Witten diagrams can be evaluated in a closed form (for general AdS$_{p+1}$ defects) as
\begin{equation}\label{defectcontact}
C_{\Delta_1\Delta_2}=\frac{\pi^{\frac{p+1}{2}}\Gamma(\frac{\Delta_1+\Delta_2-p}{2})}{2^{\Delta_1+\Delta_2}\Gamma(\frac{\Delta_1+\Delta_2+1}{2})}{}_2F_1\left(\Delta_1,\Delta_2,\frac{\Delta_1+\Delta_2+1}{2},-\frac{\xi+\chi-2}{4}\right)\;.
\end{equation}
\begin{exercise}
Use Schwinger parameterization to compute the two-point defect contact Witten diagram and show the result is given by (\ref{defectcontact}).
\end{exercise}
Explicitly, we have
\begin{equation}\label{bulkexchange}
\begin{split}
E_s={}&  \frac{\pi  r^2 \log (r)}{4 \left(r^4-r^3 \chi +r \chi -1\right)}\;,\\
E_{\varphi}={}&-\frac{\pi  r^2 \left(\left(r^2-1\right) \left(r^2+2 r \chi +1\right)-2 r \left(r^2 \chi +2 r+\chi \right) \log (r)\right)}{6 \left(r^2-1\right)^3 \left(r^2-r \chi +1\right)}\;,
\end{split}
\end{equation}
where we have introduced the new variable $r$ via
\begin{equation}
\xi+\chi=r+\frac{1}{r}\;.
\end{equation}
\begin{exercise}
Use (\ref{cubicinttocontact}) to reduce scalar bulk channel exchange Witten diagrams to defect contact Witten diagrams. Then use (\ref{defectcontact}) to reproduce $E_s$ in (\ref{bulkexchange}).
\end{exercise}
Similar to the R-symmetry polynomials $Y$ in (\ref{ansatzAs}) for defect-free four-point functions, the information of exchanging an R-symmetry irreducible representation in the bulk channel is also captured by a polynomial of the R-symmetry cross ratio $\sigma$. These can be obtained by solving the Casimir equation and, in our current case, they read
\begin{equation}
h_s=\frac{1}{6} (6-\sigma ) \sigma\;,\quad h_{\varphi}=\sigma^2\;.
\end{equation}
The coefficients $\lambda_s$ and $\lambda_{\varphi}$ are the unknowns which we will need to solve. 

In the defect channel, we similarly have the ansatz in terms of defect channel exchange Witten diagrams (Figure \ref{fig:defectWDdefectexch})
\begin{equation}
\mathcal{A}_{\rm defect}=\lambda_y(\sigma-1)\widehat{E}^{1,0}_{22}+\lambda_x \widehat{E}^{2,1}_{22}\;,
\end{equation}
where we exchange two different defect modes on AdS$_2$. The mode $y^A$ has zero transverse spin with respect to the defect but transforms as a vector under the residual $SO(5)$ R-symmetry. The other mode $x^i$ is an $SO(5)$ singlet but has transverse spin 1. Note that the bulk dual of the Wilson loop is just AdS$_2$ without an internal manifold. Therefore, there are no defect KK modes and the defect channel ansatz has the same exchange spectrum even when the external operators have higher KK weights. In the ansatz, we have also used the explicit R-symmetry polynomials in the defect channel, and $\lambda_y$ and $\lambda_x$ are the unknowns. The defect channel exchange Witten diagram with zero transverse spin is defined as the integral
\begin{equation}
\frac{\widehat{E}^{\widehat{\Delta},0}_{\Delta_1\Delta_2}}{|x_1^i|^{\Delta_1}|x_2^i|^{\Delta_2}}=\int_{{\rm AdS}_2}dW_1dW_2 G^{\Delta_1}_{B\partial}(P_1,W_1)\widehat{G}^{\widehat{\Delta}}_{BB}(W_1,W_2) G^{\Delta_2}_{B\partial}(P_2,W_2)\;,
\end{equation}
where $\widehat{G}^{\widehat{\Delta}}_{BB}(W_1,W_2)$ is the bulk-to-bulk propagator on AdS$_2$. With nonzero transverse spin $s$, the bulk-defect coupling has the form $\int_{{\rm AdS}_{p+1}} \frac{d^{p+1}\hat{z}}{\hat{z}_0^{p+1}}\hat{z}^s_0\widehat{\phi}^{i_1\ldots i_s} \partial_{i_1}\ldots \partial_{i_s}\phi$. One can show  that the diagram with transverse spin $s=1$ can be expressed in terms of the $s=0$ diagram as \cite{Gimenez-Grau:2023fcy} 
\begin{equation}
\widehat{E}^{\widehat{\Delta},1}_{\Delta_1\Delta_2}=2\Delta_1\Delta_2\chi \widehat{E}^{\widehat{\Delta},0}_{\Delta_1+1\Delta_2+1}\;.
\end{equation}
For our particular problem, the defect channel exchange Witten diagrams do not truncate into finitely many contact Witten diagrams as in the bulk channel. But one can still write down equation of motion identities to solve them as an ODE (see \cite{Rastelli:2017ecj,Gimenez-Grau:2023fcy} for details). The explicit results for these defect channel exchange Witten diagrams are
\begin{equation}
\begin{split}
\widehat{E}^{1,0}_{22}={}&\log (r+1)-\frac{r^2 \log (r)}{r^2-1}\;,\\
\widehat{E}^{2,1}_{22}={}&\chi  \left(\frac{-2 r^4+r^3+4 r^2+r-2}{2 \left(r^2-1\right)^2}-\frac{\left(r^4-2 r^2+3\right) r^3 \log (r)}{\left(r^2-1\right)^3}+\left(r+\frac{1}{r}\right) \log (r+1)\right)\;.
\end{split}
\end{equation}

For the remaining contact part of the ansatz, we assume that it has the most general R-symmetry structure but only the zero-derivative contact Witten diagram $C_{22}$ is allowed to appear. The reason is that higher-derivative contact Witten diagrams are more dominant in the high energy limit than the exchange Witten diagrams, in conflict with our flat-space expectation. The precise form of the contact part of the ansatz therefore is
\begin{equation}
\mathcal{A}_{\rm contact}=(\mu_0+\mu_1\sigma+\mu_2\sigma^2)C_{22}\;,
\end{equation}
where 
\begin{equation}
C_{22}=\frac{\pi  r^2 \left(-r^2+\left(r^2+1\right) \log (r)+1\right)}{2 \left(r^2-1\right)^3}\;.
\end{equation}

The final step is to impose the superconformal Ward identity (\ref{defect2ptscfWardid}) on the ansatz. Here it is straightforward to implement it because the ansatz has already reduced to an elementary function. We find that all the unknown coefficients are fixed up to an overall constant
\begin{equation}
\begin{split}
{}&\lambda_{\varphi}=\frac{1}{8}\lambda_s\;,\quad \lambda_y=\frac{\pi}{8}\lambda_s\;,\quad \lambda_x=\frac{\pi}{32}\lambda_s\;,\\ 
{}&\mu_0=\frac{1}{8}\lambda_s\;,\quad \mu_1=-\frac{1}{4}\lambda_s\;,\quad \mu_2=\frac{1}{12}\lambda_s\;.
\end{split}
\end{equation}
The remaining overall coefficient can be fixed in other ways, e.g., by computing the one-point function coefficient of $\mathcal{O}_2$.

\begin{exercise}
Reproduce the bootstrap calculation yourself for the Wilson line two-point function $\llangle \mathcal{O}_2\mathcal{O}_2\rrangle$ and confirm that the superconformal Ward identity fixes all the coefficients in the ansatz. 
\end{exercise}

\subsection{Generalized defects with 0 and -1 dimensions}

Using the bootstrap strategy to compute correlators in other ordinary defect setups, where the defect has at least one dimension, is straightforward, e.g., see \cite{Chen:2023yvw} for the case of $\frac{1}{2}$-BPS surface defects in 6d $\mathcal{N}=(2,0)$ theories. But in the meantime, we can also extend the notion of conformal defects so that the bootstrap techniques can be applied to a greater range of systems. 

To see what these generalized defects are and why the generalization is sensible, let us again go back to embedding space. Let us consider a $p$ dimensional conformal defect in a $d$ dimensional CFT. Just as in (\ref{PWLseparation}), the defect separates the embedding space into two groups
\begin{equation}
\mathbb{R}^{d+1,1}\to \mathbb{R}^{p+1,1}\times \mathbb{R}^{d-p}\;.
\end{equation}
The total conformal group is broken from $SO(d+1,1)$ to the rotation groups of these two factors, namely $SO(p+1,1)\times SO(d-p)$. However, this picture, along with many kinematic results, continue to make sense as we change the defect dimension $p$ as long as the first factor of dimensions contains at least one direction, i.e, $p+2\geq 1$. This gives the lower bound $p\geq -1$. Conventionally the minimal dimension of defects is taken to be 1, and here with the new bound we have added two more values $p=-1$ and $p=0$ to the allowed dimensions. How should we make sense of these seemingly exotic defect dimensions? A natural way to understand them is to use holography.  A general lesson from AdS/CFT is that objects on the boundary gain an extra dimension when moving from the boundary into the bulk AdS. Therefore, the $p=-1$ and $p=0$ defects in the CFT should respectively correspond to a point and a line in AdS. 

It is important to point out that these special defects are not merely abstractions invented for their own sake. We can in fact find concrete realizations in 4d $\mathcal{N}=4$ SYM which allow us to study different aspects of the theory. In what follows, we will explain how $p=-1$ and $p=0$ defects are realized and outline how bootstrap techniques can be used to compute correlators in these systems. 

\subsubsection{$p=-1$: $\mathcal{N}=4$ SYM on real projective space}
One way to realize the $p=-1$ defect is to consider $\mathcal{N}=4$ SYM on $\mathbb{RP}^4$ \cite{Zhou:2024ekb}. Here the 4d real projective space $\mathbb{RP}^4$ is obtained from $\mathbb{R}^4$ by performing a quotient with respect to the conformal inversion\footnote{In Section \ref{Sec:CFTbasics} we also introduced conformal inversion without the minus sign. Here the minus sign is included because we do not want the quotient of $\mathbb{R}^d$ to lead to fixed points. Otherwise, this leads to a boundary CFT.}
\begin{equation}\label{confinvm}
\mathcal{I}:\quad x^\mu\to x'^\mu=-\frac{x^\mu}{x^2}\;.
\end{equation}
In identifying the fields at $x$ and $x'$, we will choose to focus on the case where the $\mathbb{Z}_2$ is accompanied by an additional charge conjugation. For example, the transformations for the bosonic fields in the Lagrangian read \cite{Caetano:2022mus}
\begin{equation}
\Phi^a_I(x'){\rm t}_a=-\widehat{\Phi}^a_I(x){\rm t}_a^{\rm T}\;,\quad A_\mu^a(x'){\rm t}_a=I_\mu{}^\nu A_\nu^a(x) {\rm t}_a^{\rm T}\;,
\end{equation}
where we have explicitly exposed the color indices and ${\rm t}_a$ are the adjoint color matrices. Also we have introduced $\widehat{\Phi}_I=(\Phi_1,\Phi_2,\Phi_3,-\Phi_4,-\Phi_5,-\Phi_6)$, and $I_\mu{}^\nu $ is the inversion tensor
\begin{equation}
I_\mu{}^\nu=\delta_\mu^\nu-\frac{2x_\mu x^\nu}{x^2}\;.
\end{equation}
The $\mathbb{Z}_2$ action on the coordinates  (\ref{confinvm})  breaks the conformal group to $SO(5)\subset SO(5,1)$. Its action on R-symmetry also breaks it to $SO(3)\times SO(3)$. Moreover, one can show that half of the supercharges of the superconformal symmetry are preserved \cite{Wang:2020jgh}.

\begin{figure}[h]
    \centering
        \includegraphics[width=0.6\textwidth]{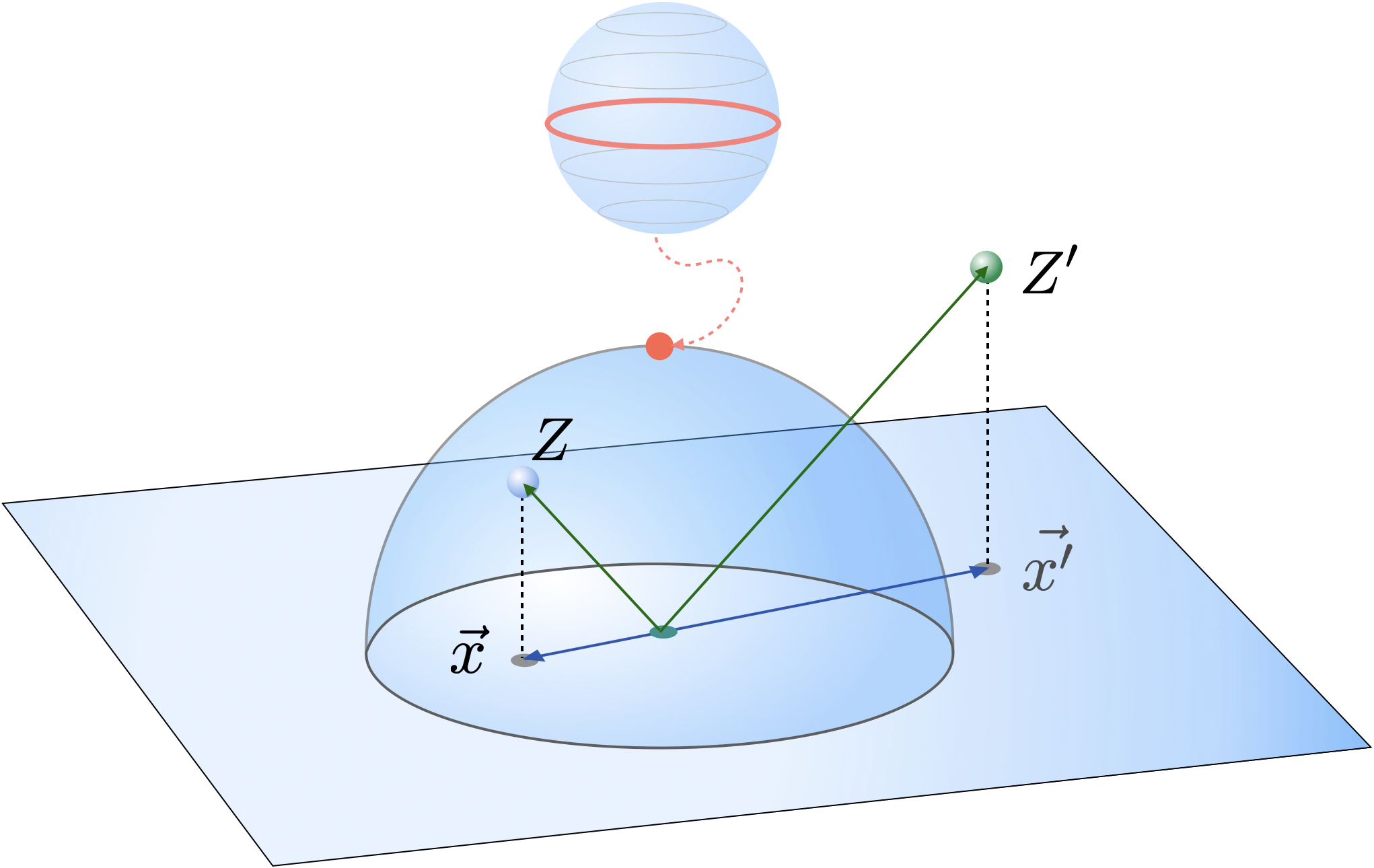}
    \caption{The $p=-1$ dimensional defect realized in $\mathcal{N}=4$ SYM by putting the theory on $\mathbb{RP}^4$ and gauging charge conjugation. On the boundary, $\mathbb{RP}^4$ is obtained from $\mathbb{R}^4$ by taking the $\mathbb{Z}_2$ quotient with respect to the conformal inversion. In the bulk, this is extended into AdS as an inversion with respect to a unit hemisphere. The bulk quotient has a fixed point which is the north pole of the hemisphere. However, the fixed point is resolved in the internal directions as an S$^2$ fixed locus, as is indicated by the red circle.}
    \label{fig:AdSquotientsetup}
\end{figure}

The holographic dual of this setup of $\mathcal{N}=4$ SYM on $\mathbb{RP}^4$ is particularly simple: it is simply the $\mathbb{Z}_2$ quotient of the usual holography on the AdS$_5\times$S$^5$ background \cite{Caetano:2022mus}, and this is illustrated in Figure \ref{fig:AdSquotientsetup}.\footnote{One can also consider putting $\mathcal{N}=4$ SYM on $\mathbb{RP}^4$ without the charge conjugation. However, this setup corresponds to a much more nontrivial bulk geometry \cite{Caetano:2022mus}. The case with charge conjugation is also more interesting because the setup preserves integrability.} The action of $\mathbb{Z}_2$ on the AdS$_5$ factor is obtained by extending the conformal inversion (\ref{confinvm}) into the bulk, and in terms of the Poincar\'e coordinates the action reads
\begin{equation}\label{AdSinversion}
z^\mu\to -\frac{z^\mu}{z_0^2+(z^\mu)^2}\;,\quad z_0\to \frac{z_0}{z_0^2+(z^\mu)^2}\;.
\end{equation}
Geometrically, this corresponds to performing an inversion in the Poincar\'e half space with respect to a unit hemisphere located at the origin. Clearly, going to the boundary at $z_0=0$, we reproduce the boundary conformal inversion (\ref{confinvm}). But unlike the boundary inversion, the inversion in AdS (\ref{AdSinversion}) has a fixed point which is the north pole of the unit hemisphere
\begin{equation}
z_0=1\;,\quad z^\mu=0\;.
\end{equation}
This is the special point singled out by the boundary $-1$ dimensional defect as we discussed. When we further take the internal S$^5$ manifold into consideration, this fixed point is resolved to become a fixed S$^2$ locus. This is because the identification $\Phi\to\widehat{\Phi}$ implies the following $\mathbb{Z}_2$ action on the embedding space coordinates of S$^5$
\begin{equation}
Y^I=(Y^1,Y^2,Y^3,Y^4,Y^5,Y^6)\to \bar{Y}^I=(Y^1,Y^2,Y^3,-Y^4,-Y^5,-Y^6)\;,
\end{equation}
which leaves an S$^2$ invariant. Therefore, the physical picture in the bulk is that we have an O1 orientifold with S$^2$ worldvolume as a defect.\footnote{It is an orientifold because the charge conjugation is related to the orientation reversal on the worldsheet.} This defect is only extended in the internal directions and sits at a point in AdS$_5$.

The simplest nontrivial objects to consider are the two-point functions of $\frac{1}{2}$-BPS operators. We can write them as functions of cross ratios by extracting a factor of one-point functions
\begin{equation}
\llangle \mathcal{O}_{k_1}\mathcal{O}_{k_2}\rrangle=\frac{(t_1\cdot \bar{t}_1)^{\frac{k_1}{2}}(t_2\cdot \bar{t}_2)^{\frac{k_2}{2}}}{(1+x_1^2)^{k_1}(1+x_2^2)^{k_2}}\mathcal{G}_{k_1k_2}(\eta;\sigma,\bar{\sigma})\;,
\end{equation}
where $\eta$ is the conformal cross ratio
\begin{equation}
\eta=\frac{x_{12}^2}{(1+x_1^2)(1+x_2^2)}\;,
\end{equation}
and $\sigma$, $\bar{\sigma}$ are the R-symmetry cross ratios
\begin{equation}
\sigma=\frac{t_1\cdot t_2}{(t_1\cdot \bar{t}_1)^{\frac{1}{2}}(t_2\cdot \bar{t}_2)^{\frac{1}{2}}}\;,\quad \bar{\sigma}=\frac{t_1\cdot \bar{t}_2}{(t_1\cdot \bar{t}_1)^{\frac{1}{2}}(t_2\cdot \bar{t}_2)^{\frac{1}{2}}}\;.
\end{equation}
Similar to the Wilson loop case, the fermionic generators also impose the superconformal Ward identity
\begin{equation}
\left(\partial_{w_1}+\frac{1}{2}\partial_z\right)\mathcal{G}_{k_1k_2}(z;w_1,w_2)\bigg|_{w_1=z}=0\;,
\end{equation}
and with $w_1\leftrightarrow w_2$, after making the change of variables
\begin{equation}
\eta=-\frac{(z-1)^2}{4z}\;,\quad \sigma=\frac{(1-w_1)(1-w_2)}{4\sqrt{w_1w_2}}\;,\quad \bar{\sigma}=\frac{(1+w_1)(1+w_2)}{4\sqrt{w_1w_2}}\;.
\end{equation}

\begin{figure}[h]
    \centering
    \begin{subfigure}[b]{0.45\textwidth}
        \centering
        \includegraphics[width=\textwidth]{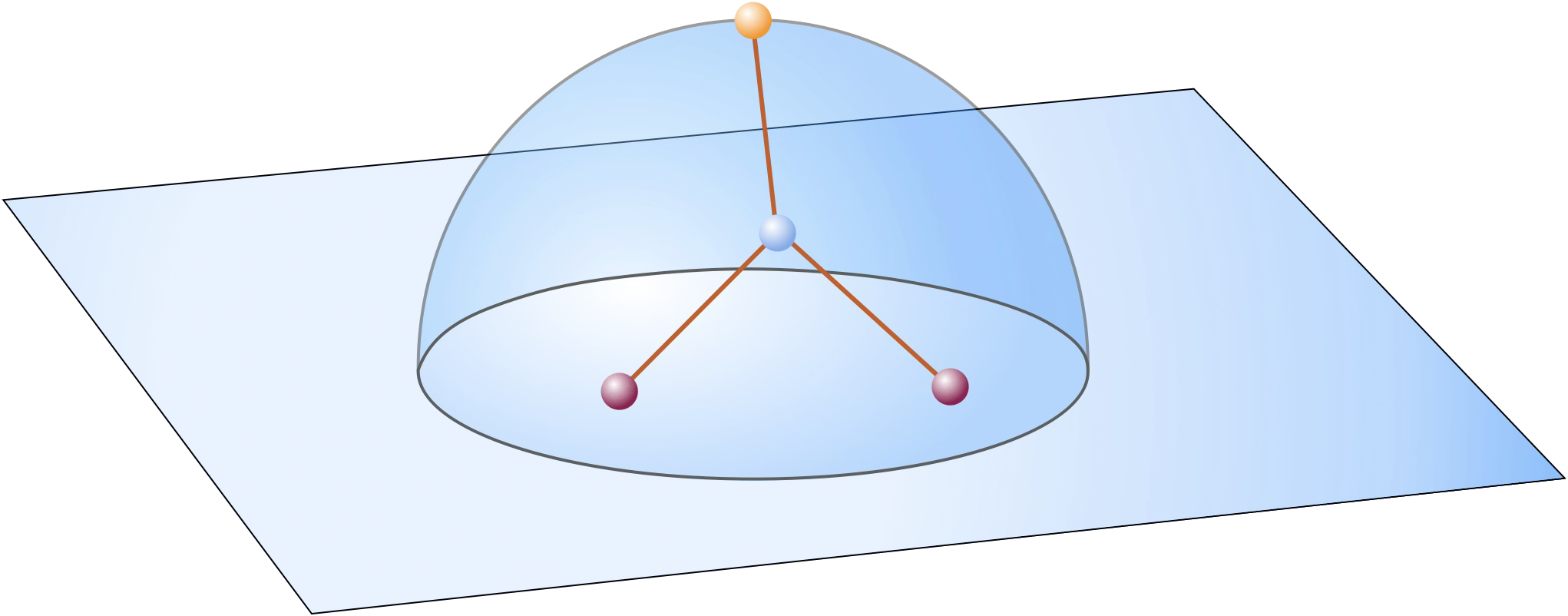}
        \caption{exchange}
        \label{fig:AdSquotientWDexch}
    \end{subfigure}
     \hspace{0.05\textwidth}
    \begin{subfigure}[b]{0.45\textwidth}
        \centering
        \includegraphics[width=\textwidth]{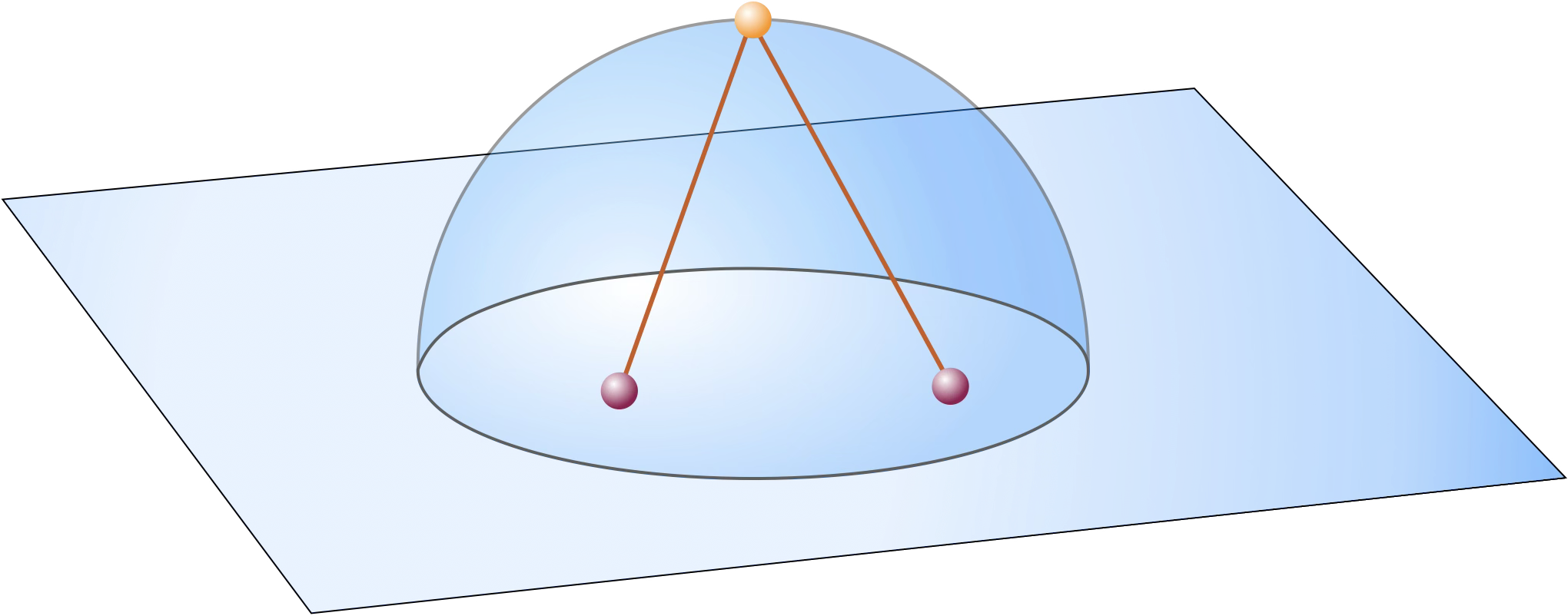}
        \caption{contact}
        \label{fig:AdSquotientWDcon}
    \end{subfigure}
    \caption{Two-point tree-level Witten diagrams for the AdS$_5\times$S$^5/\mathbb{Z}_2$ holographic dual background. The hemisphere surface is not physical and is only drawn here for reference. The propagators in the Witten diagrams are for the quotient space and the defect vertices are anchored at the fixed point in AdS.}
    \label{fig:AdSquotientWD}
\end{figure}

We can now try to bootstrap the two-point functions in the supergravity limit using the same method which we used in the Wilson loop case. The supergravity picture of an O1 orientifold sitting at a point in AdS$_5$ leads us to the conclusion that the leading connected contribution at large $N$ comes from the tree-level diagrams as shown in Figure \ref{fig:AdSquotientWD}. Note that the role of the O1 is to provide vertices localized at the fixed point. Compared to the Wilson loop case, the situation here is a bit simpler because in addition to the contact Witten diagrams we only have the bulk channel Witten diagrams. In particular, there is no analogue for the defect channel. The Witten diagrams should be computed using propagators in the quotient AdS space. However, we can use the method of images to write them in terms of Witten diagrams in the AdS space without the quotient, see \cite{Giombi:2020xah} for the relation and for how to compute the Witten diagrams. The R-symmetry polynomials which capture the information of the exchanged $SO(6)$ representations can also be computed using the Casimir equation. 
Finally, we also need to impose the superconformal Ward identities on the position space ansatz. The difference here is that these Ward identities do not fix the ansatz uniquely because the contact Witten diagram with the singlet R-symmetry structure, i.e., a constant contribution to $\mathcal{G}_{k_1k_2}$, trivially satisfies the superconformal constraint. However, this ambiguity can still be fixed by requiring analyticity of the correlators in the KK levels. We refer the reader to \cite{Zhou:2024ekb} for details. A feature perhaps worth pointing out for this setup is that the tree-level correlators are extremely simple: they are just rational functions of $\eta$ with polynomial dependence on $\sigma$, $\bar{\sigma}$.\footnote{Recall that the defect-free four-point function in the supergravity limit contains ${\rm Li}_2$ functions.} This indicates from the computational point of view that $\mathcal{N}=4$ SYM on $\mathbb{RP}^4$ is the simplest defect system. In fact, unlike the other defects such as the Wilson loop, the real projective space CFT does not have new degrees of freedom living on the defect since the $p=-1$ defect has no defect locus from the boundary perspective. This is therefore a much cleaner setup and gives a nice playground for studying $\mathcal{N}=4$ SYM.

\subsubsection{$p=0$: giant graviton correlators}
\begin{figure}[h]
    \centering
        \includegraphics[width=0.55\textwidth]{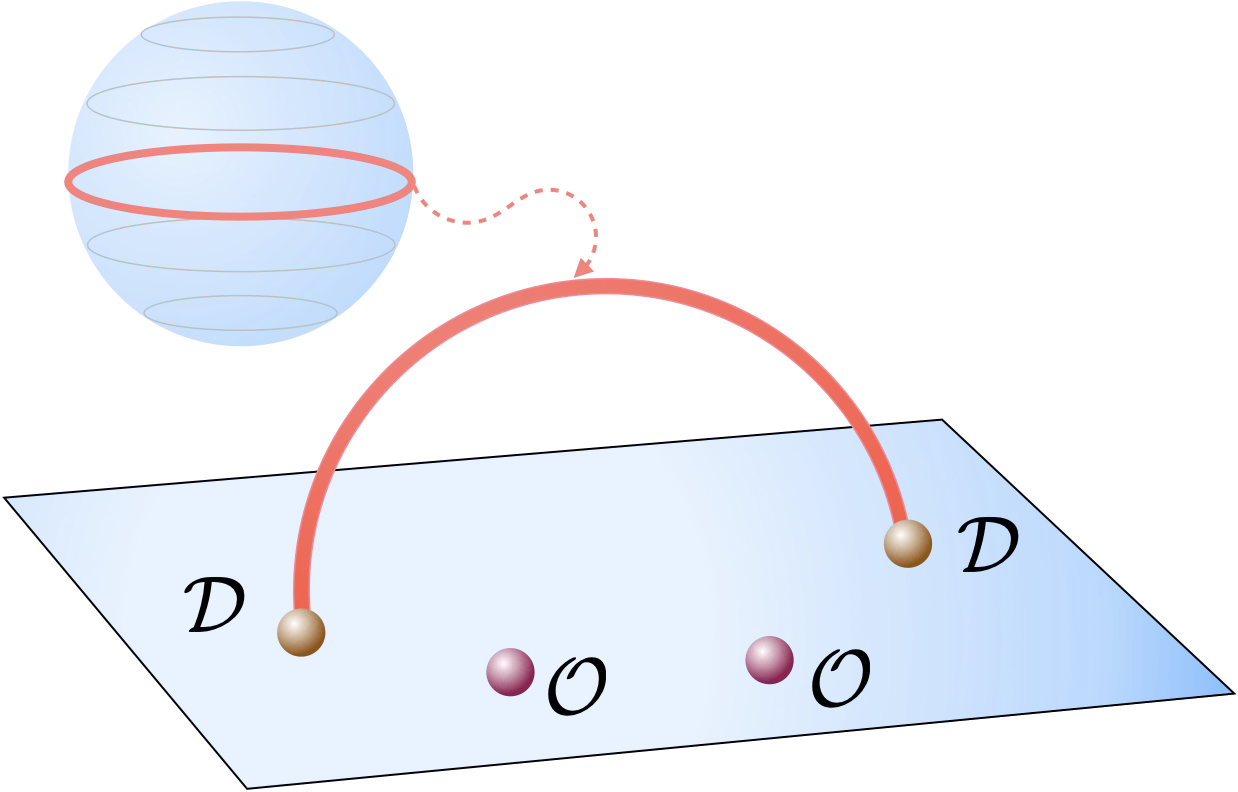}
    \caption{The $p=0$ dimensional defect realized in $\mathcal{N}=4$ SYM as a pair of giant gravitons. The giant graviton has an S$^3$ volume indicated by the red circle which is inside the internal S$^5$ manifold represented by the blue sphere. The remaining dimension is in AdS$_5$ and is a geodesic line connecting the two insertion points of the two giant graviton operators.}
    \label{fig:giantgravitonsetup}
\end{figure}

So far in these lectures we have focused only on correlators of light local operators. In particular, the single-trace $\frac{1}{2}$-BPS operators $\mathcal{O}_k$ dual to supergravitons have protected conformal dimensions $k$ which do not depend on the gauge group rank $N$. Therefore, these operators $\mathcal{O}_k$ are light excitations in the supergravity limit. However, $\mathcal{N}=4$ SYM also contains heavy operators such as the giant gravitons. They are defined as
\begin{equation}
\mathcal{D}(x,t)=\det\left(\Phi(x)\cdot t\right)\;,
\end{equation}
which has the form of a determinant in color space.\footnote{These are the maximal giant graviton, but there are also non-maximal giant gravitons and dual giant gravitons in the theory. Here we will focus on the maximal giant graviton for concreteness.} The giant graviton operator is also $\frac{1}{2}$-BPS and has protected conformal dimension $N$. This makes $\mathcal{D}$ a heavy operator in the large $N$ limit, and we can use these giant gravitons to probe genuinely nonplanar physics. This local operator perspective for heavy operators is of course valid and has been used as the standard viewpoint in the literature. However, it turns out that it is more natural, both conceptually and computationally, to view the giant gravitons as $p=0$ dimensional defects \cite{Chen:2025yxg,Chen:2026ium}, and the advantage is especially clear at strong coupling. This defect perspective can be motivated from the bulk picture where the giant graviton $\mathcal{D}$ is not dual to a supergravity field, but to a D3 brane wrapping an S$^3$ inside S$^5$. This leaves one dimension in the AdS directions, which is a geodesic line connecting the two insertion points of the giant graviton operators (Figure \ref{fig:giantgravitonsetup}). This 1d geodesic can be naturally identified with a line defect, and equivalently a $p=0$ defect on the boundary.

Let us now discuss the translation from the local operator picture to the defect one, focusing on the case where we have two giant graviton operators. The defect correlators are simply defined by dividing by the heavy operator two-point function
\begin{equation}\label{ratioheavy}
\llangle \mathcal{O}_a\ldots \mathcal{O}_n\rrangle=\frac{\langle \mathcal{O}_a\ldots \mathcal{O}_n\mathcal{D}\mathcal{D}\rangle}{\langle \mathcal{D} \mathcal{D}\rangle}\;,
\end{equation}
where the other operators $\mathcal{O}$ are light. The correlator ratio eliminates the explicit dependence on the quantum numbers of the heavy operators, and this agrees with our usual expectation for defect correlators. Importantly, in this defect picture the heavy operators only enter via projectors. For spacetime and R-symmetry, we have respectively the projectors\footnote{For purely bosonic systems, these projectors are sufficient for the description of the heavy operator defect. In the giant graviton case, we also need their ``square roots'', see \cite{Chen:2026ium} for details.}
\begin{equation}
\mathbb{N}_{AB}=\frac{P_{3,A}P_{4,B}+P_{4,A}P_{3,B}}{P_3\cdot P_4}\;,\quad\quad \mathbb{M}_{IJ}=\delta_{IJ}-\frac{t_{3,I}t_{4,J}+t_{4,I}t_{3,J}}{t_3\cdot t_4}\;,
\end{equation}
which are defined in terms of the embedding space vectors and R-symmetry polarizations of the heavy operators. Constructing correlators becomes straightforward thanks to the linearized symmetries and it needs to obey the following rules: for building blocks, we are allowed to use the embedding space vectors and R-symmetry polarizations from the light operators, but only the projectors from the heavy operators; only scalar combinations can appear and the light operators need to have the correct scalings. For one-point functions, it is easy to see we can only write down one structure
\begin{equation}
\llangle O_k(P_i,t_i) \rrangle = a_k \delta_{k,{\rm even}}\frac{(\frac{1}{2}t_i\cdot \mathbb{M}\cdot t_i)^{\frac{k}{2}}}{(P_i\cdot \mathbb{N}\cdot P_i)^{\frac{k}{2}}}\;.
\end{equation}
Writing it out explicitly, we find
\begin{equation}
\llangle O_k(P_i,t_i) \rrangle = a_k \delta_{k,{\rm even}} \left(\frac{(t_i\cdot t_3)(t_i\cdot t_4)x_{34}^2}{(t_3\cdot t_4)x_{3i}^2x_{4i}^2}\right)^{\frac{k}{2}} = \frac{\langle \mathcal{O}_k(P_i,t_i) \mathcal{D}(P_3,t_3) \mathcal{D}(P_4,t_4)\rangle }{\langle \mathcal{D}(P_3,t_3) \mathcal{D}(P_4,t_4)\rangle}\;,
\end{equation}
which confirms the prescription (\ref{ratioheavy}) and identifies the defect one-point function coefficient with the $\langle \mathcal{O}\mathcal{O}\mathcal{D} \rangle$ three-point function coefficient.

\begin{figure}[h]
    \centering
    \begin{subfigure}[b]{0.31\textwidth}
        \centering
        \includegraphics[width=\textwidth]{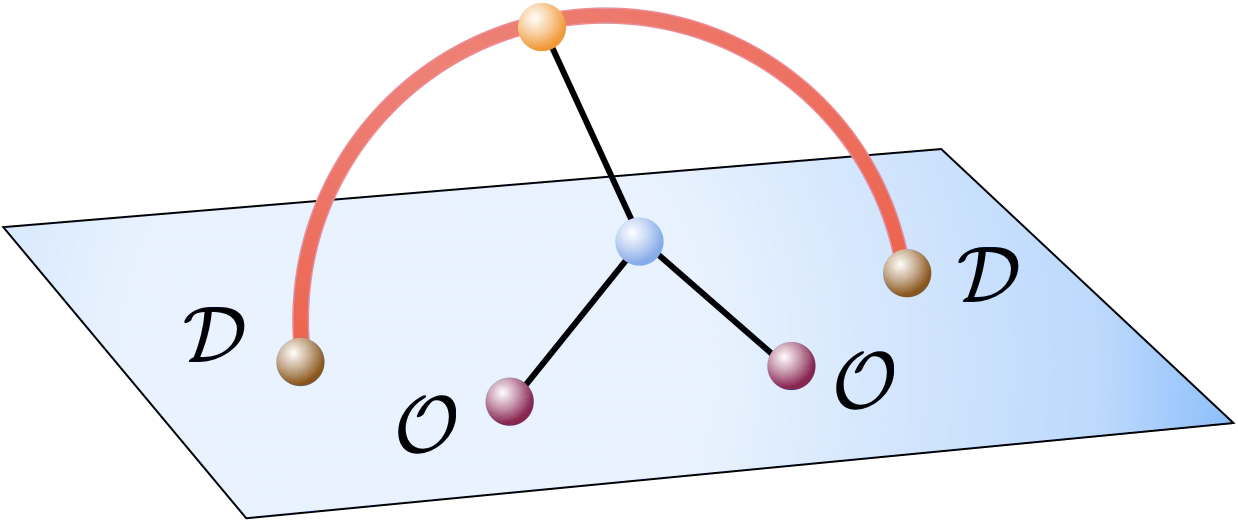}
        \caption{bulk exchange}
        \label{fig:giantgravitonbulkexch}
    \end{subfigure}
     \hspace{0.01\textwidth}
    \begin{subfigure}[b]{0.31\textwidth}
        \centering
        \includegraphics[width=\textwidth]{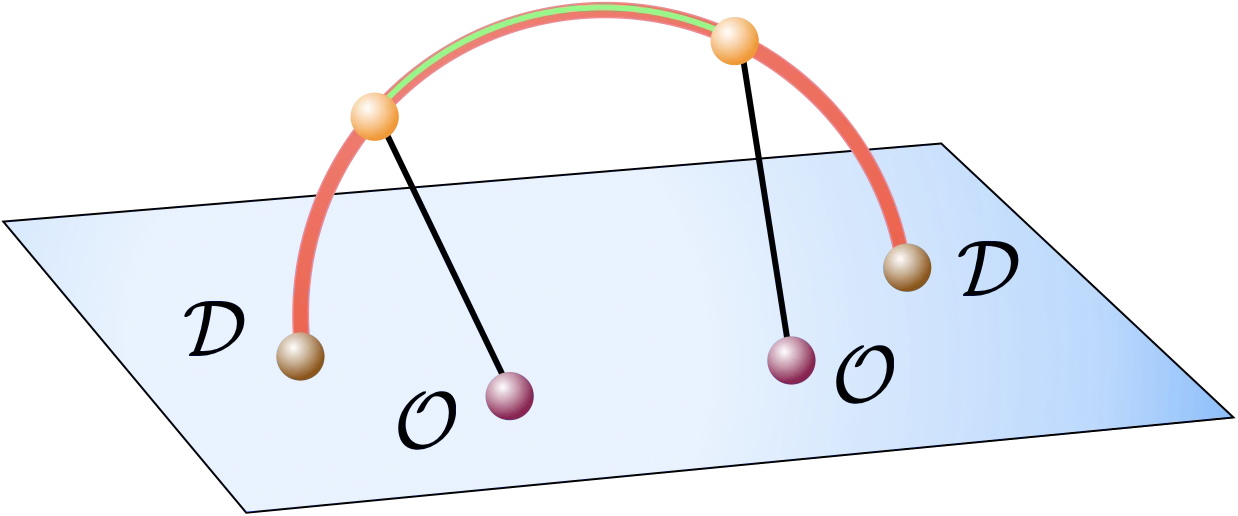}
        \caption{defect exchange}
        \label{fig:giantgravitondefectexch}
    \end{subfigure}
     \hspace{0.01\textwidth}
    \begin{subfigure}[b]{0.31\textwidth}
        \centering
        \includegraphics[width=\textwidth]{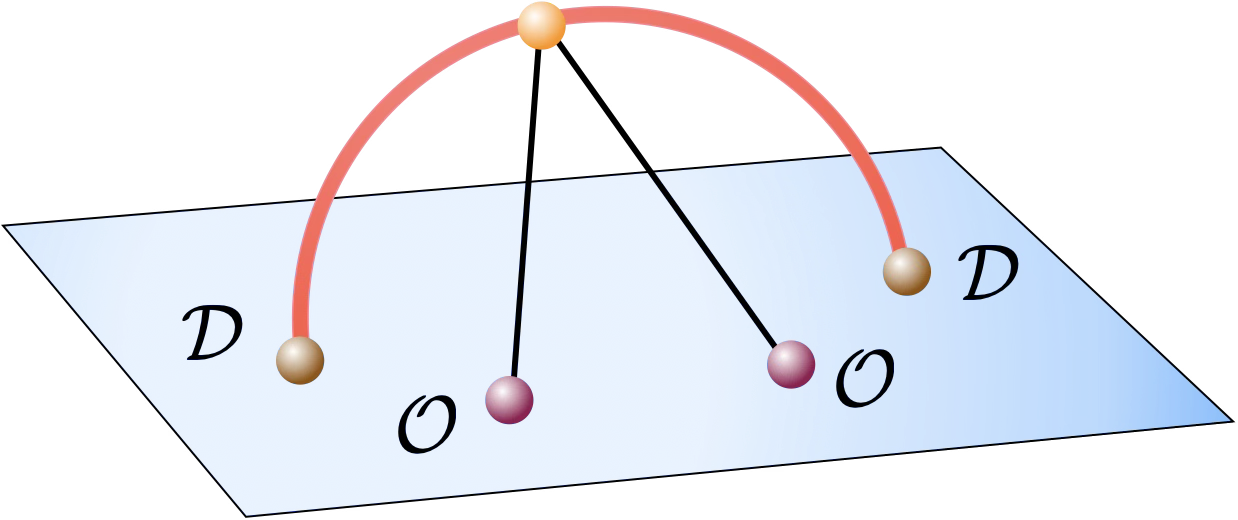}
        \caption{contact}
        \label{fig:giantgravitoncon}
    \end{subfigure}
    \caption{Tree-level Witten diagrams which give the leading connected contribution to the giant graviton correlators. These are the same defect Witten diagrams in Figure \ref{fig:defectWDtree}. Here we have redrawn them differently to manifest the fact that the giant graviton defect is one dimensional in AdS.}
    \label{fig:treelevelgiantgravitonWD}
\end{figure}

Defect two-point functions are no longer fully fixed by symmetry, as we have also expected from the local operator four-point function perspective. In fact, we can use the same four-point function cross ratios to parameterize the defect correlator. Furthermore, we can recycle the superconformal Ward identity (\ref{scfWardidNeq4}) and use the partial non-renormalization theorem result (\ref{solscfWardidNeq4}) to write the defect correlator as
\begin{equation}\label{solscfWardidgiant}
G_{k_1k_2}=G_{k_1k_2,{\rm free}}+\frac{(t_1\cdot t_2)^2x_{13}^4x_{24}^4}{x_{34}^4}R\, H_{k_1k_2}\;.
\end{equation}
Here we have divided by the giant graviton two-point function according to (\ref{ratioheavy}). $G_{k_1k_2,{\rm free}}$ is the defect two-point function in the free theory and $H_{k_1k_2}$ is the defect reduced correlator.

To bootstrap these giant graviton correlators at strong coupling, we turn to the bulk again for the basic input where supergravity gives the description of an effective defect theory. Very schematically, we can start from the D3 brane effective action and integrate out the internal S$^3$.\footnote{More precisely, one also needs to integrate over the moduli of the classical giant graviton solution.} Then we get a 1d effective action which we expand with respect to the number of fluctuations
\begin{equation}\label{S1d}
S_{\rm 1d}=-\frac{N}{2\pi^2}\int\left(1+L^{(2,0)}+L^{(1,0)}+L^{(1,1)}+L^{(0,2)}+\ldots\right)\;.
\end{equation}
Here $L^{(m,n)}$ contains $m$ defect fluctuations and $n$ bulk fluctuations. The term $L^{(2,0)}$ allows us to read off the defect spectrum. The bulk spectrum remains the same because the D3 brane does not back-react to the geometry. The other terms in (\ref{S1d}) correspond to the interacting vertices. However, their explicit details are not needed for the bootstrap calculation. Bootstrapping the defect two-point functions at the leading connected order in the supergravity limit then becomes straightforward and is similar to the previous defect cases. We first write down an ansatz in position space as the sum of all the possible tree-level defect Witten diagrams which are shown in Figure \ref{fig:treelevelgiantgravitonWD}. These Witten diagrams, along with the associated R-symmetry polynomials, can be evaluated explicitly as elementary functions of cross ratios, see \cite{Chen:2025yxg,Chen:2026ium} for details. Imposing the superconformal Ward identities then allows us to fix all the unknowns in the ansatz and fix the correlator. For example, applying the algorithm to the simplest giant graviton correlator with $k_1=k_2=2$, we find the result can be written in the form of (\ref{solscfWardidgiant}) as
\begin{equation}
    G_{22,{\rm free}}= \frac{ 2 (t_1\cdot t_2)^2 (\tau -\sigma  \tau  U+\sigma  V)}{2  (P_1\cdot \mathbb{N}\cdot P_1)(P_2\cdot \mathbb{N}\cdot P_2) U}\;,
\end{equation}
\begin{equation}\label{H22}
    H_{22}= \frac{2V \left(V^2-2 V \log V-1\right)}{N(P_1\cdot \mathbb{N}\cdot P_1)^2(P_2\cdot \mathbb{N}\cdot P_2)^2U (1-V)^3}\;.
\end{equation}
We note that in terms of the correlator expression complexity this is a case which sits in between: the tree-level supergravity two-point functions of $\mathcal{N}=4$ SYM on $\mathbb{RP}^4$ are rational functions and the defect-free four-point functions involve polylogarithmic functions of at most transcendental degree 2; here the giant graviton correlators contain only logarithms. 

Another interesting feature of these giant graviton correlators is about the hidden higher dimensional conformal symmetry which we mentioned at the end of Section \ref{Subsec:algebraicbootstrap}. Here all infinitely many defect two-point functions can also be packaged into a single generating function \cite{Chen:2025yxg,Chen:2026ium}. However, the rules for uplifting the lowest weight reduced correlator $H_{22}$ are different from the defect-free case. For the light operators, the prescription is identical and we also replace $P_i$ by $Z_i$ in the reduced correlator as in (\ref{PPtoZZ}). But for the heavy giant gravitons, we should not uplift the individual vectors because only the projector $\mathbb{N}$ appears. Instead, we replace $\mathbb{N}$ by the combination $\mathbb{N}+\mathbb{M}$, which makes sense geometrically because the combined projector projects to the four dimensional subspace corresponding to the worldvolume of the giant graviton D3 brane. The new uplifting rules make it clear that we now have a partially broken version of the 10d hidden conformal symmetry. Quite interestingly, the same hidden structure can also be identified at weak coupling at each loop order for the Lagrangian insertion integrands. 

Finally, let us mention that this 0d defect perspective is not limited to giant gravitons but gives a natural description for general light-light-heavy-heavy (LLHH) correlators \cite{Chen:2026ium}. Let us consider a four-point function $\langle \mathcal{O}_{\Delta_1} \mathcal{O}_{\Delta_2} \mathcal{O}_\Lambda \mathcal{O}_\Lambda \rangle$. In the infinitely heavy limit where the dimension of the heavy operators $\Lambda\to\infty$, one can prove that the ordinary four-point function conformal blocks reduce to the defect conformal blocks for a defect with dimension $p=0$. This reduction happens both in the bulk channel (identified with the s-channel of the four-point function) and in the defect channel (identified with the union of the t- and u-channels\footnote{It is the union because the defect channel conformal block has two branches: proportional to $V^{\frac{\widehat{\Delta}}{2}}$ when $0<V<1$ and to $V^{-\frac{\widehat{\Delta}}{2}}$ when $V\geq1$. These two branches are reproduced respectively by the t- and u-channel conformal blocks in the heavy limit, precisely in the regime where the particular channel of OPE is convergent. See \cite{Chen:2026ium} for details.}). This gives a general lesson at the level of conformal blocks: any LLHH correlator in the infinitely heavy limit admits a decomposition into defect conformal blocks. Note that this fact is purely kinematic and therefore applies generally to any CFT independent of supersymmetry. One can further exploit this kinematic fact and build analytic conformal bootstrap tools to study LLHH correlators in various physical setups.

\subsection{More analogues for defects}

In these examples we have presented, the analogy with the defect-free case is manifest when we apply the bootstrap idea. We have mostly focused on demonstrating the position space method. However, many other tools and methods for the defect-free correlators can also be analogously extended to the defect case. In this subsection, we will outline a few.

\subsubsection{Mellin space}

We have seen for correlators of local operators the Mellin space representation is a very useful formalism which greatly simplifies the analytic structure. The Mellin space formalism can be similarly generalized to defects as was done in \cite{Rastelli:2017ecj,Goncalves:2018fwx} and it makes the AdS form factor interpretation of defect correlators more clear. 

For concreteness, let us focus on the two-point function of local bulk operators in the presence of a defect of generic dimensions\footnote{Here generic means the defect dimension satisfies $0\leq p< d-1$. Although it can be similarly defined, the Mellin representation suffers from certain issues for the $p=-1$ case. For the $p=d-1$ case, the kinematics is special and one should define the Mellin space separately.}. The Mellin representation of the defect two-point function is
\begin{equation}\label{defectMellin}
\mathcal{F}(\xi,\chi)=\int\frac{d\delta d\gamma}{(2\pi i)^2}\xi^{-\delta}\chi^{-\gamma+\delta}\mathcal{M}(\delta,\gamma)\Gamma(\delta)\Gamma(\gamma-\delta)\prod_{i=1}^2\Gamma\left(\frac{\Delta_i-\gamma}{2}\right)\;,
\end{equation}
where we have defined $\mathcal{F}(\xi,\chi)$ by extracting one-point functions\footnote{The prime is to denote that we set the one-point function coefficients to be one in this factor.}
\begin{equation}
\llangle \mathcal{O}_{\Delta_1} \mathcal{O}_{\Delta_2} \rrangle=\llangle \mathcal{O}_{\Delta_1}\rrangle' \llangle\mathcal{O}_{\Delta_2} \rrangle'\mathcal{F}(\xi,\chi)\;.
\end{equation}
In analogy with the identification $\delta_{ij}=\vec{p}_i\cdot\vec{p}_j$ in the defect-free case, here we also introduce fictitious momenta and divide them into a parallel part and a transverse part $\vec{p}=(\vec{p}_\parallel,\vec{p}_\perp)$. Then we can identify the Mellin-Mandelstam variables in (\ref{defectMellin}) with
\begin{equation}
\delta=\vec{p}_1\cdot \vec{p}_2\;,\quad \gamma=-(\vec{p}_{1,\parallel})^2=-(\vec{p}_{2,\parallel})^2\;,
\end{equation}
where we note momentum is conserved in the directions parallel to the defect. In this defect Mellin representation, tree-level Witten diagrams enjoy similar simplifications as in the defect-free case. The Mellin amplitude of a defect contact Witten diagram without any derivatives is just a constant. The bulk channel exchange Witten diagrams have poles in $\delta$ with polynomial residues in $\gamma$, while the defect channel exchange Witten diagrams have poles in $\gamma$ with polynomial residues in $\delta$. See \cite{Gimenez-Grau:2023fcy} for a detailed discussion. This analytic structure closely resembles that of the flat-space Feynman diagrams in momentum space where we have two particles scattering with a brane. 

Another useful extension of the defect-free result is the flat-space limit formula, which allows us to extract the flat-space form factor from the holographic defect correlator. This is particularly simple to write down in Mellin space \cite{Alday:2024srr}
\begin{equation}\label{flatspacedefect}
\mathcal{M}(\delta,\gamma)\propto \int_0^\infty d\beta \beta^{\frac{1}{2}(\Delta_1+\Delta_2)-\frac{p}{2}-1}e^{-\beta}\mathcal{A}\left(S=-\frac{2\delta\beta}{R^2},Q=-\frac{2\gamma\beta}{R^2}\right)\;,\quad R\to\infty\;,
\end{equation}
where $S$ and $Q$ are the flat space versions of $\delta$ and $\gamma$ and are constructed from the physical momenta. This is almost the same as in the defect-free case (\ref{Mellinflatspacelim}), except that we should use different Mandelstam variables and replace $d$ by $p$. There is also a position space version of the flat-space limit formula where we need to zoom in around a scaling limit of the defect correlator \cite{Chen:2025cod}.

\subsubsection{Stringy corrections}
For defect correlators, we can also go beyond the supergravity approximation and consider stringy corrections. Instead of the Wilson loop, it is more convenient to consider the 't Hooft loop because the $1/\lambda$ correction structure is most similar to that of the defect-free four-point function considered in Section \ref{Subsec:stringycorrections}. 

On the bulk side, the 't Hooft loop is dual to a D1 brane rather than a fundamental string. However, in the limit of tree-level supergravity, the 't Hooft loop two-point function is proportional to the Wilson loop result from Section \ref{Subsec:WL2ptfun}. This is because in both cases the line defects are $\frac{1}{2}$-BPS and preserve the same amount of symmetry. They are also both dual to an AdS$_2$ subspace inside AdS$_5$ with localized degrees of freedom, sitting at a point in S$^5$. We have the same bulk and defect exchange spectrum and therefore the same ansatz. Recall that the ansatz is uniquely determined by the superconformal Ward identity up to an overall constant. Therefore, the defect two-point functions must be proportional. Let us focus on the two-point function $\llangle \mathcal{O}_2 \mathcal{O}_2\rrangle$ where the correlator $\mathcal{F}(\xi,\chi;\sigma)$ is defined in (\ref{WL2ptdefcalF}). Going beyond supergravity, the defect correlator has the following expansion in $1/\lambda$
\begin{equation}\label{Fexpinlambda}
\mathcal{F}=f(\lambda)\mathcal{F}_{\rm SUGRA}+\lambda^{-1}\mathcal{F}_{\rm h.d.}^{(4)}+\lambda^{-\frac{3}{2}}\mathcal{F}_{\rm h.d.}^{(6)}+\lambda^{-2}\mathcal{F}_{\rm h.d.}^{(8)}+\ldots\;.
\end{equation}
Here $\mathcal{F}_{\rm SUGRA}$ is the supergravity two-point function at $\lambda=\infty$. But because the one-point function $\llangle \mathcal{O}_2\rrangle$ is not protected, this solution contributes nontrivially to the stringy correction via the function $f(\lambda)$. The remaining stringy corrections come as higher-derivative corrections, and $\mathcal{F}_{\rm h.d.}^{(M)}$ denotes contact diagram contributions with at most $M$ derivatives. Note that the derivatives must be contracted to ensure rotation invariance and therefore $M$ is even. Also one can show there is no purely contact Witten diagram solution to the superconformal Ward identity with two derivatives or less, and this leads to the absence of $\mathcal{F}_{\rm h.d.}^{(2)}$ in (\ref{Fexpinlambda}). As in the defect-free case, the number of independent solutions grows as we increase the derivative cutoff $M$. 

We can fix the stringy corrections in (\ref{Fexpinlambda}) at low orders using the flat-space limit formula (\ref{flatspacedefect}) and supersymmetric localization \cite{Pufu:2023vwo,Billo:2023ncz,Dempsey:2024vkf,Billo:2024kri}. In flat space, the 1-to-1 scattering amplitude of closed strings scattering with a D-brane has been computed \cite{Hashimoto:1996bf}, and takes the form of the Veneziano amplitude multiplied with a polynomial factor of Mandelstam variables and polarizations. Expanding this string amplitude in terms of $\ell_s$, the flat-space limit formula then allows us to fix the leading terms on the AdS side in the high energy limit. On the other hand, localization fixes $f(\lambda)$, and also constrains the subleading terms via the integrated correlators. In \cite{Alday:2024srr}, these higher-derivative corrections were fixed in this way to eight derivatives up to a linear constraint on the coefficients. We can also apply this strategy to other defect setups and similarly fix the low-lying higher-derivative corrections. But the more powerful route is to use the worldsheet description for the stringy defect correlators and consider them as the curvature corrections to the flat-space amplitude in a similar way as was outlined for the defect-free case in Section \ref{Subsec:stringycorrections}.

\subsubsection{Unitarity method for loop corrections}
\begin{figure}[h]
    \centering
        \includegraphics[width=\textwidth]{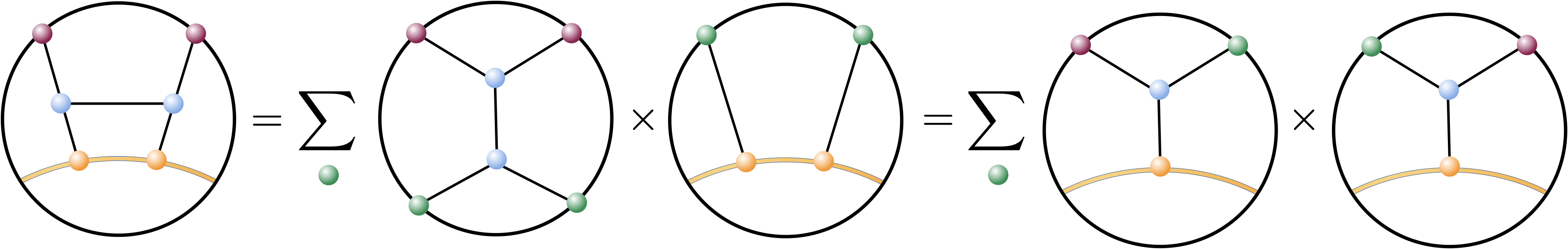}
    \caption{AdS unitarity method for defects. The defect two-point function at one loop can be obtained from gluing in two ways, either as a defect-free four-point function times a defect two-point function, or as the product of two defect two-point functions. A sum over all the intermediate KK modes is also needed.}
    \label{fig:AdSunitaritydefect}
\end{figure}

To conclude the defect discussions, let us mention the computation of loop corrections to the defect correlators. In Section \ref{Subsec:loopsinAdS}, we discussed how to use the AdS unitarity method to construct one-loop corrections from the tree-level correlators. Here we explain how this method can be adapted to defect correlators \cite{Chen:2024orp}. The basic idea is also in terms of correlator gluing and it is schematically illustrated in Figure \ref{fig:AdSunitaritydefect}. On the left, we have a representative diagram at the one-loop level. On the right, we are showing how it can be obtained from two equivalent ways of gluing. The first is in terms of a defect-free tree-level four-point function and a disconnected defect two-point function. The second is in terms of the product of two tree-level defect two-point functions. 

Of course, the rigorous statement of gluing is not at the diagrammatic level, but rather in terms of the conformal block decomposition. Moreover, the gluing is also only for the leading logarithmic singularities of the defect correlator. To state it more precisely, let us introduce the following notation for the new combinations of the cross ratios
\begin{equation}
B=\frac{\xi}{\chi}\;,\quad D=\frac{1}{\chi}\;,
\end{equation}
where $B$ and $D$ go to zero in the bulk and defect channel OPEs respectively. In terms of these variables, the cross ratio factor in the Mellin representation (\ref{defectMellin}) is just $B^{-\delta}D^\gamma$. Considering the reduced correlator $\mathcal{H}$ (which can be defined, e.g., for the case of $\frac{1}{2}$-BPS surface defects in 6d $(2,0)$ theories), the leading logarithmic singularity at the one-loop level is $\mathcal{H}\big|_{\log^2D}$ in the defect channel and $\mathcal{H}\big|_{\log B}$ in the bulk channel. These singularities exhibit the gluing structure illustrated in Figure \ref{fig:AdSunitaritydefect} and can be computed from the lower-order data in a way similar to the defect-free case. Note that the leading singularities in the two channels have the common term $\mathcal{H}\big|_{\log^2D\log B}$. This term can be computed in two ways in these two channels and therefore serves as a nontrivial consistency check of the gluing. The next step is to complete the singularities into the full defect correlator. In \cite{Chen:2024orp} this was considered in Mellin space for the 6d $(2,0)$ surface defect. It turns out that the reduced Mellin amplitude also takes the form of a double infinite sum of simultaneous poles in the defect Mellin-Mandelstam variables
\begin{equation}
\widetilde{\mathcal{M}}(\delta,\gamma)\sim\sum_{m,n}\frac{c_{mn}}{(\delta+n)(\gamma-2m)}\;.
\end{equation}
As in the four-point function case in Section \ref{Subsec:loopsinAdS}, the strategy is also to use $\mathcal{H}\big|_{\log^2D\log B}$ to fix the $c_{mn}$ coefficients and then use the remaining terms in the leading logarithmic singularities to rule out additional single poles. This defect unitarity method has also been used in \cite{Chen:2026fnf} to compute the one-loop correction to the giant graviton correlators, by viewing it as a system with a zero dimensional defect. It turns out that the reduced Mellin amplitudes in both cases have the same analytic structure and the general $c_{mn}$ coefficients can be written down using the same building block ${}_3F_2$ functions. The work \cite{Chen:2026fnf} also obtained the one-loop defect two-point function in position space. One finds similar transcendental functions as in the defect-free case, but with additional new letters in the alphabet.  

\section{Outlook}\label{Sec:outlook}

To end these lectures, let us look out into the future by first looking back at the history. The systematic computation of holographic correlators was once a notoriously difficult problem, and for almost two decades it remained extremely challenging to go beyond the simplest few examples. The new bootstrap strategy, based largely on symmetries and consistency conditions, has significantly lowered this computational hurdle. It has led to many efficient new methods in recent years which are redefining what is computationally possible for holographic correlators. But perhaps more importantly, these bootstrap methods have revealed a remarkable simplicity in the holographic correlators that was largely obscured in the traditional diagrammatic approach. 

It is natural to compare this situation with the development of the flat-space scattering amplitude program, which we have been taking as a major source of inspiration throughout these lectures. In flat space, it was only after the discovery of unexpectedly simple results such as the Parke--Taylor formula that people started to realize that the seemingly complicated perturbative amplitudes are in fact simple and can be dictated by new organizing principles. A flurry of powerful methods and unexpected discoveries then ensued in the following decades, eventually shaping the field into its current form. The recent progress on holographic correlators naturally raises the question of whether we are witnessing the early stages of a similar development in AdS. The basic technologies and initial promising results are there, but there is clearly much more to be found.

There are good reasons for such optimism. The idea of uplifting the flat-space amplitude program into AdS, together with many of its hidden properties and techniques, is very natural, especially if we believe that the general lessons learned from amplitudes are robust enough and are not just coincidences of flat space. The existence of a four-point AdS double-copy relation \cite{Zhou:2021gnu} is one promising piece of evidence in this direction. But this pursuit in AdS should also not be regarded simply as trying to repeat the flat-space story after turning on a curvature ``deformation''. In fact, the bootstrap results are also revealing new structures which have no direct analogues in flat space. Examples include hidden higher-dimensional structures \cite{Caron-Huot:2018kta,Rastelli:2019gtj,Alday:2021odx,Zhou:2021gnu,Chen:2025yxg,Chen:2026ium}, dimensionally reduced structures associated with Parisi--Sourlas supersymmetry \cite{Zhou:2018sfz,Behan:2021pzk,Alday:2021odx}, and Yangian invariance \cite{Rigatos:2022eos}. The physical origins of these structures are currently unknown. It would be very interesting to understand the precise mechanisms behind them, which may reveal new physical lessons as well as provide new computational handles.

 It is of course difficult to predict the future, especially when it comes to identifying which major directions this holographic correlator program should pursue next. Therefore, we will not pretend to outline such a roadmap here. Nevertheless, we can provide the reader of these lecture notes with a list of concrete open problems, some of which are immediate, while others we simply believe are interesting and rewarding to address. 
 
 \begin{itemize}
 \item {\bf BCFW recursion in AdS.} As we emphasized in these lectures, a major bottleneck of the holographic correlator program is to systematically and efficiently go to higher points. It would be highly desirable to find a natural AdS analogue for the flat-space BCFW recursion relation which will enable us to construct higher-point correlators in an on-shell manner. It should also be noted that most studies so far have focused on the kinematically simpler scalar correlators (e.g., $\frac{1}{2}$-BPS operators in $\mathcal{N}=4$ SYM). But perhaps it is worth looking more closely into the correlators of spinning gluons and gravitons in AdS, which despite greater technical complexity have a more direct analogy with the flat-space story.
 
 \item {\bf Geometric structures.} An important lesson from the flat-space amplitude program is that the remarkable simplicity of scattering amplitudes can sometimes be understood geometrically. The discovery of the amplituhedron and related geometric structures \cite{Arkani-Hamed:2013jha,Arkani-Hamed:2017tmz,Arkani-Hamed:2017mur} provides a striking example, where properties of amplitudes which are obscure in the traditional formulation become manifest from a new geometric perspective. It is natural to ask whether the remarkable simplicity we have seen in holographic correlators also has an underlying geometric origin. At present, it is not clear what such a geometric formulation in AdS should look like, or even whether the known flat-space ideas have useful analogues. Exploring this possibility may reveal new organizing principles for holographic correlators. 
 
 \item {\bf Mellin representation for CFT$_{\bf 1}$.} The Mellin space formalism we introduced in these lectures is for CFTs in high enough spacetime dimensions $d$. It has implicitly assumed that the particle number $n$ satisfies $n<d+2$ so that the cross ratios do not have the additional relations from the Gram determinant. However, this condition is violated in CFT$_1$ already in the first nontrivial case of four-point functions, and the counting of cross ratios is $n-3$ for all values of $n$. On the other hand, the situation is also special from the 2d scattering perspective where the special kinematics of $n-3$ Mandelstam variables plays a crucial role in the factorized scattering in integrable theories. Therefore, it would be very interesting to find a Mellin representation tailored to CFT$_1$ that automatically incorporates the Gram determinant constraints while preserving the direct correspondence between cross ratios and Mandelstam variables. Such a formulation would be very useful for studying scattering in AdS$_2$. A particularly interesting application would be to search for integrable QFTs in AdS$_2$, for example by imposing no particle production.
 \item {\bf Spin-1 versions of defects.} In Section \ref{Sec:defects} we  discussed several different setups of holographic defect CFTs. The defect correlators in these setups can be interpreted as AdS form factors describing graviton states scattering from extended objects. However, in Section \ref{Subsec:gluonscattering} we mentioned that one can also study scattering of gluons in AdS, where the simpler spin-1 kinematics makes gluons an ideal playground for exploring the amplitude program in AdS. It would be interesting to construct analogous defect setups involving external gluon states and study gluon form factors in the presence of AdS defects. One natural question is whether the tree-level defect two-point functions exhibit double-copy structures, generalizing the similar observation for defect-free four-point functions. It would also be interesting to understand what the corresponding double copy means for the extended objects themselves.
 
 \item {\bf Defects beyond the probe limit.} Finally, let us mention that in applying the bootstrap strategy to compute holographic defect correlators we have focused on the probe brane regime, where the defects are ``branes'' in AdS  and do not back-react on the geometry. It is clear that there is a whole zoo of ``nonperturbative'' defects whose presence modifies the bulk geometry already at leading order. Examples include defects described by bubbling geometries and holographic interfaces associated with Janus-type solutions. In such cases, the simple picture of particles scattering off a fixed extended object in AdS needs to be generalized to scattering in a nontrivial back-reacted geometry. It would be interesting to understand how the bootstrap philosophy discussed in these lectures can be effectively extended to this regime. In particular, one may ask to what extent defect correlators can be determined directly from symmetries and consistency conditions without explicitly constructing the corresponding back-reacted supergravity solutions, and whether some of the remarkable simplicity found in the probe limit continues to survive.

 \end{itemize}

\acknowledgments
I want to thank the organizers of Pre-Strings 2026 for the invitation to give these lectures. Parts of these lectures were previously also given at the 8th Mandelstam Workshop and School at Stellenbosch University in 2026, Indian Institute of Science in 2025, and at Southeast University, Fudan University and the Institute of Theoretical Physics of the Chinese Academy of Sciences on earlier occasions. I want to thank all the organizers for invitations and hospitality, and all the students and participants for their engaging questions and enthusiasm. This work is supported by the NSFC Grant No. 12275273 and funds from Chinese Academy of Sciences.

\bibliography{refs} 
\bibliographystyle{utphys}
\end{document}